\documentclass[fleqn,usenatbib]{mnras}

\usepackage{newtxtext,newtxmath}

\usepackage[T1]{fontenc}

\DeclareRobustCommand{\VAN}[3]{#2}
\let\VANthebibliography\thebibliography
\def\thebibliography{\DeclareRobustCommand{\VAN}[3]{##3}\VANthebibliography}

\usepackage{graphicx}	
\usepackage{amsmath}	
\usepackage{subfigure}
\usepackage{longtable}
\usepackage{lscape}
\usepackage{adjustbox}
\usepackage{multirow}
\usepackage{mhchem}
\usepackage{xcolor}
\usepackage{booktabs}
\usepackage{caption}

\title[Observations of acetone in hot cores]
{ALMA Observations of Acetone in Hot Cores}

\author[X Zhang et al.]{
Xia Zhang,$^{1,2}$\thanks{E-mail: zhangx@xao.ac.cn}
Xiaohu Li,$^{1,2}$\thanks{E-mail: xiaohu.li@xao.ac.cn}
Zhiping Kou$^{1,3}$
\\
$^{1}$ State Key Laboratory of Radio Astronomy and Technology, Xinjiang Astronomical Observatory, CAS, 150 Science 1-Street, Urumqi, Xinjiang 830011, P.R. China\\
$^{2}$ Xinjiang Key Laboratory of Radio Astrophysics, 150 Science1-Street, Urumqi 830011, P.R. China\\
$^{3}$ University of Chinese Academy of Sciences, Beijing 100049, P.R. China\\}

\date{Accepted XXX. Received YYY; in original form ZZZ}

\pubyear{2015}

\begin{document}
\label{firstpage}
\pagerange{\pageref{firstpage}--\pageref{lastpage}}
\maketitle

\begin{abstract}
Acetone (CH$_3$COCH$_3$) is a ubiquitous interstellar molecule, and serves as an important tracer of hot core chemistry. We conducted a line survey of acetone and its precursor acetaldehyde (CH$_3$CHO) towards 60 hot cores by using the ALMA 3 mm lines observations. We calculated the rotational temperatures and column densities of acetone using the XCLASS software. Acetone was detected in 15 hot cores with rotational temperatures ranging from 89 to 176 K. Its column densities range from (0.9-24) $\times$ 10$^{16}$ cm$^{-2}$. The spatial distributions of acetone exhibit similarities with those of acetaldehyde. The emissions of acetone are concentrated toward the hot core regions and generally exhibit a compact spatial distribution, whereas the emission of acetaldehyde shows a more extended spatial profile. Combined with previous studies, we found a moderately positive correlation between the column densities and rotational temperatures of acetone for the high-mass hot cores ($r = 0.59$). We also found a strong positive correlation between the column densities of acetone and acetaldehyde ($r= 0.82$), indicating a chemical relationship between them. By comparing these observational results with the three-phase model results, we found that the models overpredict the ratio of acetone to methanol relative to the observational data. This discrepancy suggests that current chemical networks may inadequately account for acetone destruction pathways or potential missing physical conditions in the model. Therefore, our large sample observations can provide constraints on chemical models and reinforce the role of acetone as a tracer of complex organic chemistry in warm, dense regions. 
\end{abstract}

\begin{keywords}
astrochemisty -- ISM: molecules -- galaxies: star formation
\end{keywords}



\section{Introduction}
\label{sec:introduction}
Acetone (CH$_3$COCH$_3$), the simplest ketone, is a critically important organic molecule both terrestrially and astronomically, serving as a key participant in atmospheric oxidation processes \citep{harrison2011}. It is also produced in the human body when fat is broken down. Acetone, a ubiquitous molecule in the interstellar medium (ISM), is of substantial interest to radio astronomy due to its rich and dense spectrum—significantly complicated by intramolecular large-amplitude motions. Therefore, it has been detected in a variety of interstellar objects, including high-mass hot cores \citep{combes1987, snyder2002, belloche2013, mcGuire2016, friedel2005, friedel2008, widicus2012, feng2015, suzuki2018, rolffs2011, peng2022, isokoski2013, codella2013, zou2017, mininni2023, palau2017, bogelund2019, baek2022}, and shock positions \citep{peng2013, csengeri2019}, intermediate-mass protostars \citep{fuente2014, palau2011}, low-mass protostars \citep{jorgensen2011, lykke2017, manigand2020, nazari2024, gelder2020, imai2016, chahine2022}, the protoplanetary disk V883 Ori \citep{lee2019}, the cold molecular cloud TMC-1 \citep{agundez2023}. Recently, two large-sample surveys conducted by \citet{chen2025a} and \citet{li2025a} identified acetone in 12 high-mass protostars and 11 massive protostellar clumps using ALMA, respectively. Beyond its predominant detection in the gas phase, acetone has been identified in the ice of the low-mass protostar B1-c with JWST/MIRI \citep{Chen2024}. In addition, it was also found to be present in comet 67P/Churyumov–Gerasimenko and the Murchison meteorite \citep{goesmann2015, altwegg2017, pizzarello2009}. The aforementioned observations of acetone in the ISM and in comets indicate that acetone may be found in both gas and ice phases. These astronomical data are critically important for understanding the formation mechanisms of interstellar acetone. 

Laboratory studies play a crucial role in guiding astronomical observations and interpreting astrophysical data. Extensive experimental work has been conducted on acetone formation in interstellar ice analogs, revealing multiple formation pathways: (1) Acetone forms efficiently in the ice mixtures containing carbon monoxide (CO), methane (CH$_4$), and methanol (CH$_3$OH) upon exposure to ionizing radiation \citep{kaiser2014, maity2015, abplanalp2016}. (2) Pure methanol ice irradiation produces acetone \citep{bennett2007, henderson2015, abou2016, yocum2021}. (3) Binary mixtures (H$_2$O:CH$_3$OH, CH$_4$:CH$_3$CHO) also yield acetone \citep{mvondo2008, singh2022}. Besides, \citet{fedoseev2022} demonstrated ketene (CH$_2$CO) formation from CO and C reactions on H$_2$O-rich ices, proposing its subsequent conversion to acetone via (CH$_2$)$_2$CO intermediate formation and hydrogenation. However, these subsequent steps have not yet been experimentally verified.

Laboratory experiments show that acetone can form through the irradiation of mixed ices containing acetaldehyde (CH$_3$CHO) and methane (CH$_4$), where methyl (\.CH$_3$) and acetyl (CH$_3$\.CO) radicals react to produce it. Astrochemical models further propose acetaldehyde as a direct precursor to acetone \citep{garrod2022, chen2025a}. Beyond its role in acetone formation, acetaldehyde is itself a critical precursor to prebiotic molecules, such as deoxyribose and amino acids, via Strecker synthesis \citep{steer2017}. It is among the most widely detected interstellar molecules, having been observed in diverse environments: cold molecular clouds  \citep[e.g.][]{matthews1985}, prestellar cores \citep[e.g.][]{ vastel2014}, hot cores and hot corinos \citep[e.g.][]{blake1987, cazaux2003}, protostellar shocks \citep[e.g.][] {lefloch2017}, and young disks \citep[e.g.][]{lee2019}. Like acetone, it is also found in comets such as 67P/Churyumov-Gerasimenko \citep{altweg2017} and in meteorites such as Murchison \citep{jungclaus1976}.

In observations, acetone is a very interesting molecule. \citet{friedel2008} and \citet{zou2017} found that large N-bearing molecules and large O-bearing molecules are spatially separated. The presence of acetone in chemically distinct environments - specifically, regions in both N‑bearing and O‑bearing species - indicates that this molecule in Orion KL is co‑spatial with two types of chemistry. Here, N‑bearing molecules typically trace compact, hotter regions where gas‑phase chemistry dominates, while O‑bearing molecules are considered tracers of grain‑surface processes such as radical–radical and photolysis reactions. This co‑spatiality suggests that acetone likely has a unique formation mechanism involving both gas‑phase and grain‑surface processes. The difference in the rotational temperature of acetone in Orion KL and Sgr B2(N) also suggests the possibility of multiple formation pathways for acetone that correspond to different gas temperatures. Therefore, there is a need for large sample sizes and systematic observations to understand the chemistry of acetone, particularly its potential unique formation mechanism in the ISM.

The ALMA Three-millimeter Observations of Massive Star-forming regions (ATOMS) project has observed 146 massive clumps at the 3 mm band with ALMA \citep{liu2020a, liu2020b}. Out of 453 compact dense cores, 60 have been identified as hot core candidates using three molecuels (CH$_3$OH, CH$_3$OCHO, and C$_2$H$_5$CN) \citep{qin2022}. In this work, we conduct a survey using acetone lines based on the ATOMS continuum and line data at the 3 mm band. The primary goal is to systematically study acetone in high-mass hot cores to better understand its formation and destruction pathways. The observations and data reduction are briefly described in Section \ref{sec:obser}, and the observational results, including parameter calculations and spatial distributions are presented in Section \ref{sec:results}. We investigate the environmental impact of H{\sc ii} regions on acetone chemistry, compare the column densities and excitation temperatures of acetone with other sources, discuss the column density correlations of acetone with acetaldehyde and methanol, and compare our observational results with chemical models in Section \ref{sec:discussion}. The main results and conclusions are summarized in Section \ref{sec:conclusions}.

\section{Observations}
\label{sec:obser}
The sample selection for the ALMA observations is described in detail by \citet{liu2020a, liu2020b} and \citet{qin2022}. The ALMA band 3 observations were made towards 146 massive clumps from late September to mid-November 2019 (Project ID: 2019.1.00685.S; PI: Tie Liu), using both the Atacama Compact 7 m Array (ACA; Morita Array) and the 12 m array (C43-2 or C43-3 configurations). The band 3 observations include eight spectral windows (SPWs), in which we utilized SPWs 7 and 8 with a broad bandwidth of 1875 MHz each, located in the upper sideband. These SPWs, with a spectral resolution of $\sim$ 1.6 km s$^{-1}$, were used for continuum imaging and line surveys. Their frequency ranges are 97536–99442 MHz and 99470–101390 MHz, respectively. Due to the smaller source sizes of hot cores and the less missing flux issue for complex organic molecule (COM) lines, we only utilized the 12 m array data for the identification of acetone and its precursor, acetaldehyde in this study. Data reduction was performed using version 5.1.15 of the Common Astronomy Software Applications (CASA) package \citep{mcmullin2007}. Further details regarding the spectral setups, band pass, flux, and phase calibrators are described in \citet{liu2020a}. SPWs 7 and 8 were used to create continuum images from the line-free channels centred at 99.4 GHz (or 3 mm). Continuum-subtracted spectral cubes from the ALMA 12m arrays were cleaned using the CASA 5.3 TCLEAN task. This was done by applying the natural weighting, setting the gridder with the `pblimit' parameter of 0.2, and a multiscale deconvolver. All images were corrected for the primary beam. The self-calibration was performed by running three rounds of phase self-calibration and one round of amplitude self-calibration to improve all the images. We adopted a 15\% flux uncertainty for all data points. 

For two cores in our sample, IRAS 18056-1952 and IRAS 18507+0110, which exhibit particularly rich spectral line emission. The method used to select line-free channels is described below. We first assumed that no line-free channels were present in the SPW and ran tclean in CASA. Spectra were then extracted from a region matching the beam size and centered on the continuum peak. Following \citet{jorgensen2016}, these spectra were fitted using both a symmetric Gaussian and a skewed Gaussian to obtain an initial estimate of the continuum emission. We then applied a three-step iterative selection procedure for line-free channels adapted from the local noise estimation method of \citet{loomis2021}: 1. Calculate the standard deviation $\sigma$ of the entire dataset. Flag channels with intensities exceeding $\sigma$, along with 10 adjacent channels on each side, as NaN. 2. Compute a new $\sigma$ from the remaining data. Flag channels with intensities >$3\sigma$, plus 10 adjacent channels per side, as NaN. 3. Repeat step 2 once more using the updated dataset. The remaining unflagged data were identified as line-free channels. These final line-free channels were then used for imaging with the same tclean parameters described above (natural weighting, pblimit = 0.2, multiscale deconvolver, and primary beam correction).

The spectra of our sample were obtained from a region that matches the beam size and is centered on the peak of the continuum. The resulting continuum images from the 12 m array have angular resolutions, characterized by the synthesized beam size, ranging from $\sim$ 1.5$\arcsec \times$1.2$\arcsec$ - 3.1$\arcsec \times$2.3$\arcsec$ and maximum recoverable angular scales of about 14.5–20.3 arcsec (~0.1 pc in size; or 2 arcsec at 10 kpc distance). The mean 1 $\sigma$ noise level for SPWs 7-8 is below 10 mJy beam$^{-1}$ per channel for lines. Table \ref{tab:spws7-8} lists parameters for SPWs 7 and 8 with the central frequency, bandwidth, spectral resolution, beam size and P.A, and channel rms, respectively. The complete sample of sources in which acetone was detected is presented in Table \ref{tab:physical-parameter}, along with the derived physical parameters.

\begin{table}
 \caption{Key observational parameters of SPWs 7 and 8.}
 \label{tab:spws7-8}
 \begin{tabular}{ccc}
 \hline
  Parameter  &SPW 7   &SPW 8\\
  \hline
 Central Frequency (MHz) &~98489 &~100430\\
 Bandwidth (MHz)	&1875	&1875\\
 Spectral Resolution (MHz)	&~0.488	&~0.488\\
 Beam Size (arcsec$\times$arcsec)	&1.6$\times$1.2 - 3.0$\times$2.4	&1.6$\times$1.2 - 3.1$\times$2.4\\
 Beam P.A. (degree)	&-90 - 90	&-90 - 90\\
 channel rms (mJy beam$^{-1}$)	&2.55 - 7.48	&2.72 - 6.43\\
 \hline
 \end{tabular}
\end{table}

\section{Results}
\label{sec:results}
\subsection{Line identifications and fitting of the spectra}
\label{sec:fits}
We extracted spectra from 60 line-rich cores that are considered hot core candidates \citep{qin2022}. We identified spectral line transitions and performed parameter calculations using the eXtended CASA Line Analysis Software Suite \citep[XCLASS\footnote{\url{http://xclass.astro.uni-koeln.de}};][]{moller2017}. The molecular spectroscopic parameters were obtained from the Jet Propulsion Laboratory molecular databases (JPL\footnote{\url{http://spec.jpl.nasa.gov/}}) for acetone and acetaldehyde \citep{pickett1998}. The spectroscopic data for acetone originate from \citet{groner2002}, and those for acetaldehyde from \citet{kleiner1996}. Assuming that the molecular gas meets the conditions of local thermodynamic equilibrium (LTE), XCLASS solves the radiative transfer equation to generate synthetic spectra for specific molecular transitions, taking into account factors such as source size, beam filling factor, line profile, line blending, excitation, and opacity. In the XCLASS modeling, the input parameters include source size ($\theta$), rotational temperature (T$_{\text{rot}}$), and column density (N), full width at half-maximum of the observed line ($\Delta$V) and velocity offset (V$_{\text{off}}$) \citep{moller2017}. In this work, we used the deconvolved angular sizes of the continuum sources as their source sizes, provided by \citet{qin2022}, and a single line width (FWHM) that fits both acetone and acetaldehyde transitions, as shown in Table \ref{tab:physical-parameter}. The velocity offsets relative to the systemic velocities of the hot cores are determined using the commonly detected CH$_3$OH line at 100.6389 GHz with an upper-level energy of Eu = 233 K as a reference. This high excitation condition ensures that the emission predominantly traces the warm, dense gas of the hot core. Therefore, we designated the rotational temperatures, column densities, and velocity widths as free parameters to replicate the observed spectra, with the details of the column density calculation provided in \citet{moller2017}. To optimize the parameters for rotational temperature and column density, we employed the Modeling and Analysis Generic Interface for eXternal numerical codes \citep[MAGIX;][]{moller2013}, which aids in optimizing the fit, identifying the best parameter solutions, and providing corresponding error estimates through various optimization algorithms or algorithm chains. In this study, we utilized the Levenberg–Marquardt (LM) and the Markov chain Monte Carlo (MCMC) algorithms to obtain the optimized rotational temperatures and column densities. The quality of the fit was checked by visually inspecting both the overall profile and the line shapes of the most isolated transitions

\begin{table*}
 \caption{Physical parameters of hot cores.}
 \label{tab:physical-parameter}
 \setlength{\tabcolsep}{2.8pt}
 \begin{tabular}{lcccccccccc}
 \hline
  Source  &RA   &Dec &Distance &$\theta_{\text{source}}$  &T$_{\text{rot}}$ &CH${_3}$COCH${_3}$ &CH${_3}$CHO &FWHM &N$_{\text{H$_2$}}^a$ &Abundance\\
&($^{h m s}$)  & ($^{\circ ~\prime ~\prime\prime}$) &(kpc) &(arcsec) &(K) &N (cm$^{-2}$) &N (cm$^{-2}$) &(km s$^{-1}$) &(cm$^{-2}$) &CH${_3}$COCH${_3}$\\
  \hline
 I13079–6218 &13:11:13.75 &–62:34:41.55 &3.8 &2.4   &110$\pm$21 & (1.6$\pm$0.8)$\times$10$^{16}$ &(1.1$\pm$0.1)$\times$10$^{16}$ &6.2 &(7.8$\pm$0.7)$\times$10$^{23}$ &(2.0$\pm$1.0)$\times$10$^{-8}$\\
 I13134–6242 &13:16:43.20 &–62:58:32.30 &3.8 &2.2   &94$\pm$23  &(2.0$\pm$1.5)$\times$10$^{16}$ &(5.0$\pm$0.9)$\times$10$^{15}$ &6.5 &(8.2$\pm$0.3)$\times$10$^{23}$ &(2.4$\pm$1.8)$\times$10$^{-8}$\\
 I16065–5158$^*$ &16:10:20.30 &-52:06:07.1 &3.98 &3.1  &101$\pm$18  &(8.7$\pm$4.2)$\times$10$^{15}$ &(5.5$\pm$0.6)$\times$10$^{15}$ &5.0 &(5.0$\pm$1.5)$\times$10$^{23}$ &(1.7$\pm$1.0)$\times$10$^{-8}$\\
 I16164-5046$^*$ &16:10:19.99 &–52:06:07.25 &3.57 &2.8  &123$\pm$21  &(3.5$\pm$1.3)$\times$10$^{16}$ &- &5.5 &(1.9$\pm$0.3)$\times$10$^{24}$ &(1.8$\pm$0.7)$\times$10$^{-8}$\\
 I16272–4837C1 &16:30:58.77 &–48:43:53.57 &2.92 &2.2   &89$\pm$9 &(3.4$\pm$1.1)$\times$10$^{16}$ &(9.0$\pm$0.2)$\times$10$^{15}$ &5.0 &(4.1$\pm$0.7)$\times$10$^{23}$ &(8.3$\pm$3.0)$\times$10$^{-8}$\\
 I16318-4724$^*$ &16:35:33.96 &–47:31:11.59 &7.68 &2.2   &100$\pm$22 &(1.7$\pm$0.8)$\times$10$^{16}$ &(1.1$\pm$0.2)$\times$10$^{16}$ &6.6 &(3.2$\pm$0.3)$\times$10$^{23}$ &(5.3$\pm$0.3)$\times$10$^{-8}$\\
 I16348-4654$^*$ &16:38:29.65 &-47:00:35.67 &12.09 &0.8 &113$\pm$32 &(1.0$\pm$0.9)$\times$10$^{17}$ &(5.0$\pm$0.6)$\times$10$^{16}$ &6.9 &(4.0$\pm$0.3)$\times$10$^{24}$ &(2.5$\pm$0.2)$\times$10$^{-8}$\\
 I17008–4040 &17:04:22.91 &–40:44:22.91 &2.38 &1.7 &115$\pm$17 &(3.0$\pm$1.3)$\times$10$^{16}$ &(1.0$\pm$0.2)$\times$10$^{16}$ &5.0 &(8.3$\pm$0.8)$\times$10$^{23}$ &(3.6$\pm$1.6)$\times$10$^{-8}$\\
 I17233–3606$^*$ &17:26:42.46 &–36:09:17.85 &1.34 &3.9  &104$\pm$22 &(1.9$\pm$1.2)$\times$10$^{16}$ &(7.0$\pm$0.8)$\times$10$^{15}$ &5.0 &(1.8$\pm$0.3)$\times$10$^{24}$ &(1.1$\pm$0.7)$\times$10$^{-8}$\\
 I18056–1952 &18:08:38.23 &–19:51:50.31 &8.55 &1.1  &104$\pm$12 &(1.6$\pm$0.5)$\times$10$^{17}$ &(8.0$\pm$1.5)$\times$10$^{16}$ &7.3 &(7.4$\pm$0.4)$\times$10$^{24}$ &(2.2$\pm$0.1)$\times$10$^{-8}$\\
 I18117–1753 &18:14:39.51 &–17:52:00.08 &2.57 &1.6  &100$\pm$6 &(9.9$\pm$0.2)$\times$10$^{15}$ &(6.0$\pm$0.4)$\times$10$^{15}$ &4.6 &(3.4$\pm$0.6)$\times$10$^{23}$ &(2.9$\pm$0.5)$\times$10$^{-8}$\\
 I18159-1648C1 &18:18:54.66 &–16:47:50.28 &1.48 &1.7  &100$\pm$9 &(1.7$\pm$0.7)$\times$10$^{16}$ &(2.0$\pm$0.2)$\times$10$^{15}$ &5.0 &(8.1$\pm$1.1)$\times$10$^{23}$ &(2.1$\pm$0.9)$\times$10$^{-8}$\\
 I18507+0110$^*$ &18:53:18.56 &+01:14:58.23 &1.56 &1.3  &
 176$\pm$3 &(2.4$\pm$0.3)$\times$10$^{17}$ &(3.6$\pm$0.7)$\times$10$^{16}$ &6.0 &(2.0$\pm$0.2)$\times$10$^{25}$ &(1.2$\pm$0.2)$\times$10$^{-8}$\\
 I18507+0121 &18:53:18.01 &+01:25:25.56 &1.56 &1.7  &128$\pm$8 &(5.8$\pm$0.8)$\times$10$^{16}$ &(2.0$\pm$0.2)$\times$10$^{16}$ &6.4 &(8.3$\pm$0.6)$\times$10$^{23}$ &(7.0$\pm$1.0)$\times$10$^{-8}$\\
 I19095+0930$^*$ &19:11:53.99 &+09:35:50.27 &6.02  &0.5  &100$\pm$1 &(3.7$\pm$0.6)$\times$10$^{16}$ &(1.2$\pm$0.2)$\times$10$^{16}$ &3.5 &(1.2$\pm$0.1)$\times$10$^{25}$ &(3.1$\pm$0.6)$\times$10$^{-9}$\\
 \hline
 \multicolumn{11}{l}{Notes.$^*$ represents these sources associated with H{\sc ii};  $\theta$ source denotes the source size, which discussed in detail in \citet{qin2022}; $^a$ \citet{chen2025b}.}\\
 \multicolumn{11}{l}{All rotational temperatures were derived from the fit of the acetone lines.}
 \end{tabular}
\end{table*}

\begin{figure*}
\centering
\includegraphics[scale=0.46]{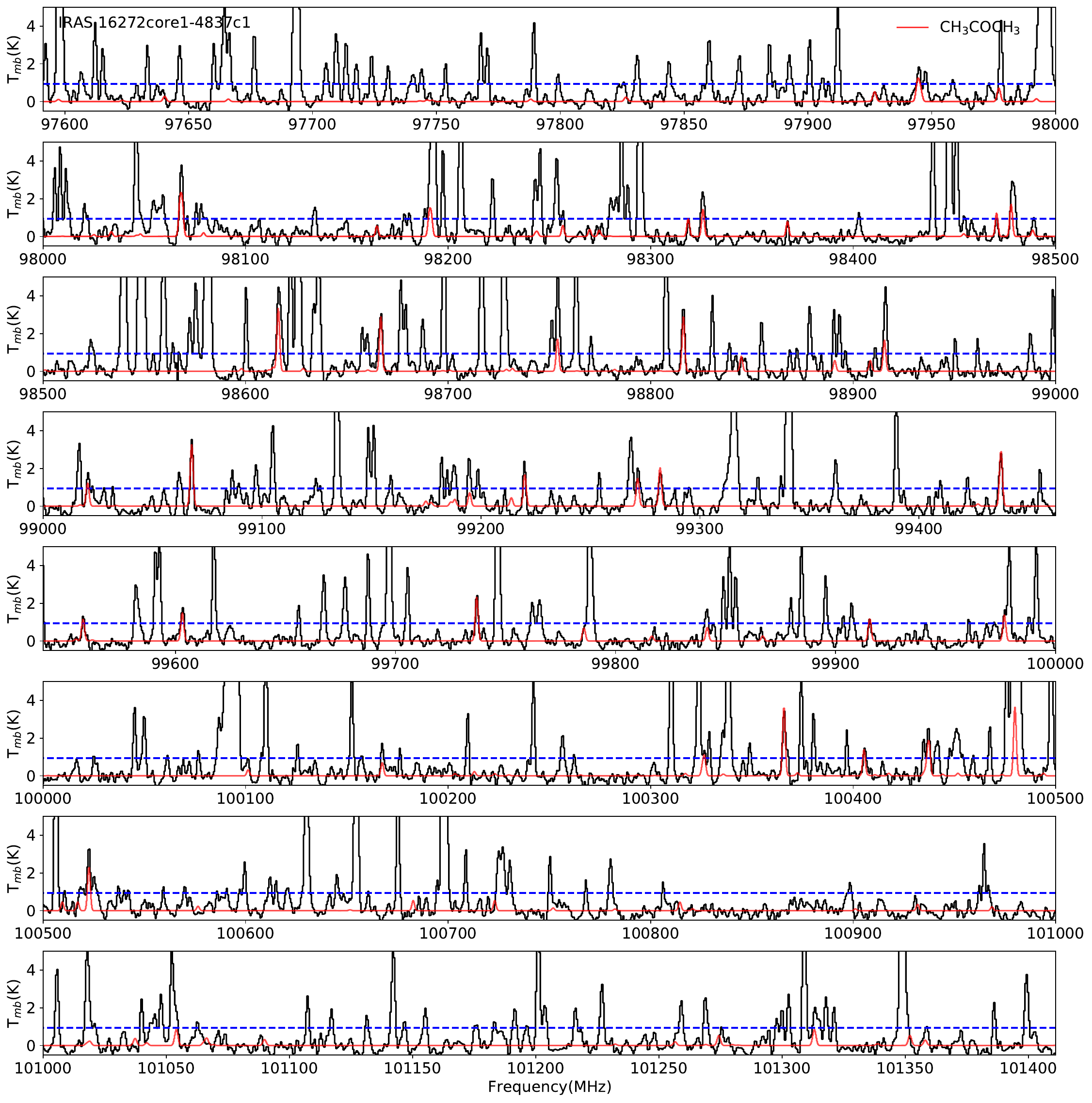}
\caption{Full band spectra and line identification towards IRAS 16272-4837C1. The observed spectrum is shown in black. XCLASS synthesized spectra of CH${_3}$COCH${_3}$ is overlaid in red. The horizontal blue dashed line indicates the 3 $\sigma$ noise level in each window. The spectra of other sources are presented in Fig. \ref{fig:B1}.}
\label{fig:IRAS16272-4837c1}
\end{figure*}

In our work, we considered spectral lines with intensities above the 3$\sigma$ noise level for identification and subsequent analyses. Table \ref{tab:physical-parameter} summarizes the derived parameters for the molecular lines of acetone, which we detected in 15 out of the 60 line-rich hot cores. To investigate potential correlations between acetone detection and source parameters, we examined whether source distance \citep[which spans approximately 1.34–12.09 kpc in our sample,][]{liu2020a} or mass correlate with detection. We find no clear trend, as acetone is detected across the full distance range and in sources with varied masses \citep[8.5-1079.2M$_{\odot}$,][]{chen2025b}. Instead, non-detections appear more closely linked to observational factors such as lower line intensities and/or increased spectral blending in line-rich spectra. Although acetone emission was detected in all observed hot cores, the spectra in several sources are too weak and/or blended to reliably constrain the rotational temperature. In these cases, the limitations arise from either insufficient signal-to-noise ratio (SNR) (peak intensities below the 3$\sigma$ threshold) or severe spectral contamination with fewer than three unblended transitions. The identified acetone and acetaldehyde transition lines above the 3$\sigma$ threshold are listed in Table \ref{tab:Iden-trans-acetone}. These lines are clean and not blended with other lines. We detected a total of 15 spectral features that can be assigned to 28 acetone transitions, and 2 acetaldehyde transitions, all originating from the ground state. Fig. \ref{fig:IRAS16272-4837c1} shows a sample spectra toward IRAS 16272-4837C1 with modeled molecular spectra for acetone overlaid. Additional spectra are presented in Fig. \ref{fig:B1}.

Additionally, we identified emission lines of acetaldehyde, a potential precursor to acetone \citep{garrod2022, chen2025a}, in 14 hot cores where acetone has been detected. For the source I16164-5046, acetaldehyde transitions are too weak to yield reliable fits. However, only two emission lines of CH$_3$CHO were observed in each source. The corresponding frequencies are 98863.314 MHz and 98900.951 MHz, with upper level energies of $\sim$ 16 K making it hard to constrain rotational temperatures of CH$_3$CHO. Therefore, we assume that CH$_3$CHO has the same rotational temperature as acetone in order to estimate its column densities. This approach is based on the assumption that acetone and acetaldehyde originate from the same spatial region. Due to the limited number of detected acetaldehyde transitions and their similar upper-level energies, the MCMC algorithm failed to converge for a robust statistical uncertainty. Therefore, we adopted a systematic uncertainty of 15\% for N(CH$_3$CHO), propagated from the dominant flux calibration error, as the final error estimate.

Since acetone has two isomers, propanal (CH$_3$CH$_2$CHO) and propylene oxide (c-CH$_3$CHOCH$_2$, a chiral molecule), we also attempted to detect these isomers in the 15 line-rich hot cores where acetone was detected. However, neither propanal nor propylene oxide was detected. This non-detection is most likely attributable to the limited number (33 for propanal and 78 for propylene oxide) and weak intensity of their spectral lines within the observed frequency range, which makes their identification challenging.

\subsection{Rotation temperatures, column densities and abundances}
\label{sec:temperatures}
The molecular transitions are clear and cover a large upper-level energy E$_u$ range, making them well-suited for constraining the physical parameters. Based on this approach, we summarized the rotational temperatures and column densities of acetone in Table \ref{tab:physical-parameter}. The rotational temperatures of acetone ranges from 89 to 176 K. Its column densities range from (0.9 to 24) $\times$ 10$^{16}$ cm$^{-2}$. Its abundance range from (0.3 to 8.3) $\times$ 10$^{-8}$. As a precursor to acetone, acetaldehyde exhibits column densities in the range of (0.2 to 18) $\times$ 10$^{16}$ cm$^{-2}$, all of which are lower than the corresponding acetone column densities from the same source. The mean rotation temperature of CH$_3$COCH$_3$ is 110$\pm$15 K, the average column density is 5.3 $\times$ 10$^{16}$ cm$^{-2}$ in our results. The mean abundance of acetone was calculated to be 3.0 $\times$ 10$^{-8}$, based on the H$_2$ column densities from \citet{chen2025b} and the authors used the same source sizes as in this study, which are derived from the deconvolved angular sizes of 3mm continuum images. These findings are consistent with previous studies of high-mass star-forming regions (e.g., \citet{peng2022, li2025a, chen2025a}). Among these 15 sources, seven are associated with ultra-compact H{\sc ii} regions \citep{qin2022, zhang2023}. The data reveal that the acetone abundances in hot cores associated with ultra-compact H{\sc ii} regions are below 2$\times$10$^{-8}$, except for I16318-4724 and I16348-4654, which exhibit lower values compared to hot cores without ultra-compact H{\sc ii} regions. 

Although we did not detect CH$_3$CH$_2$CHO and c-CH$_3$CHOCH$_2$ in our survey, we derived their column densities across 15 line-rich hot cores to be (0.5 to 14.2) $\times$ 10$^{16}$ cm$^{-2}$ and (0.2 to 5.0) $\times$ 10$^{16}$ cm$^{-2}$, respectively. These estimates are derived from the experimentally determined branching ratios of (4.82±0.05):(2.86±0.13):1 for acetone:propanal:propylene oxide in interstellar ice analogs \citep{singh2022}. \citet{lykke2017} reported an observed CH$_3$COCH$_3$/CH$_3$CH$_2$CHO ratio of 8 in the low-mass protostar IRAS 16293-2422, while \citet{belloche2013} found a lower limit of 3.6 in Sgr B2(N), consistent with laboratory results. In contrast, chemical models predict peak fractional abundance ratios of 0.22 (slow), 0.83 (medium), and 0.07 (fast) in gas phase \citep{garrod2013}, favoring propanal over acetone - a trend that is diametrically opposed to both laboratory experiments and astronomical observations \citep{singh2022, lykke2017, belloche2013}. These contradictions highlight the need for refinements in chemical models to reconcile theory with laboratory and observational constraints.

\subsection{Spatial distributions}
\label{sec:spatial}
\begin{figure*}
\subfigure[]{
\includegraphics[width=5.5cm]{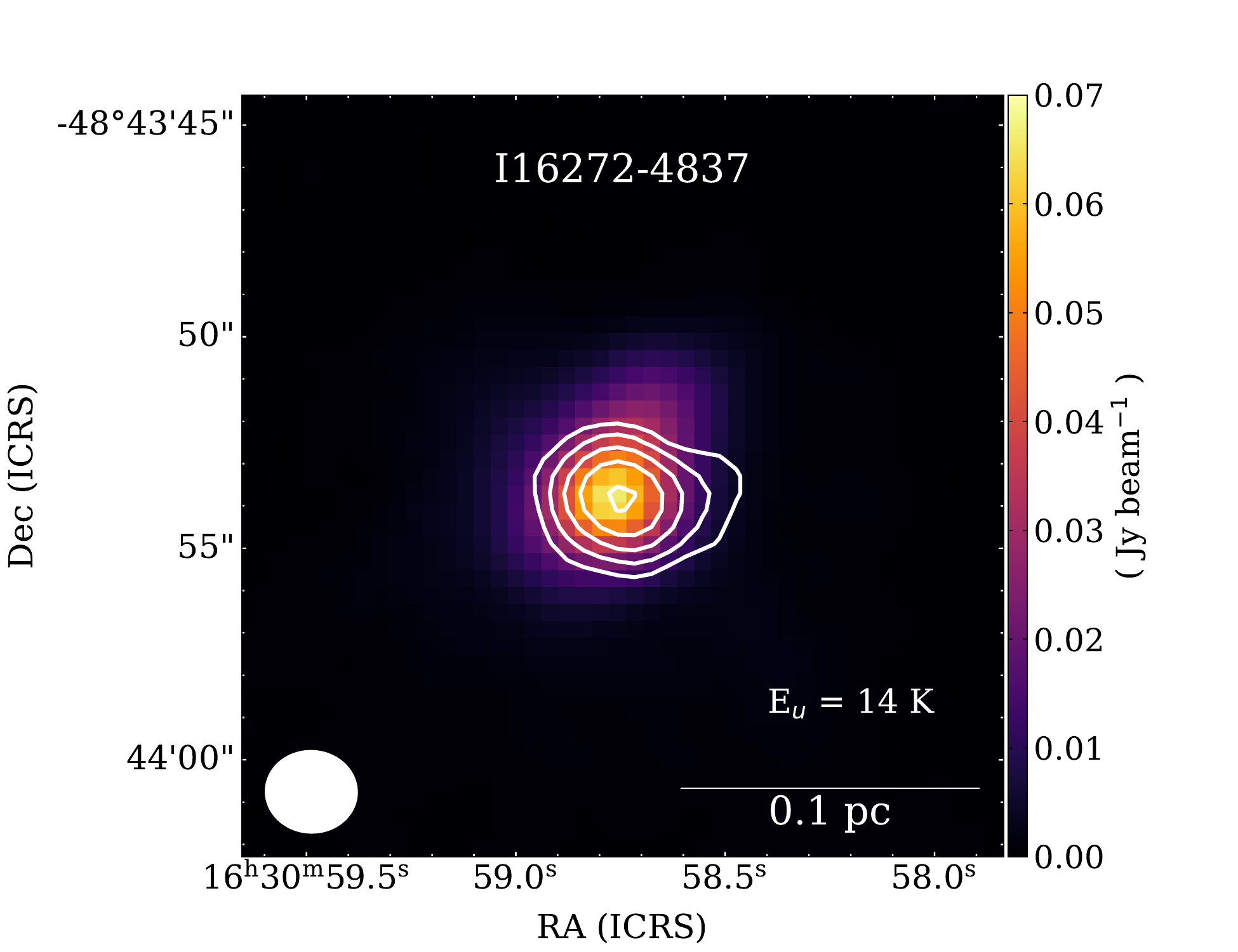}
}
\quad
\subfigure[]{
\includegraphics[width=5.5cm]{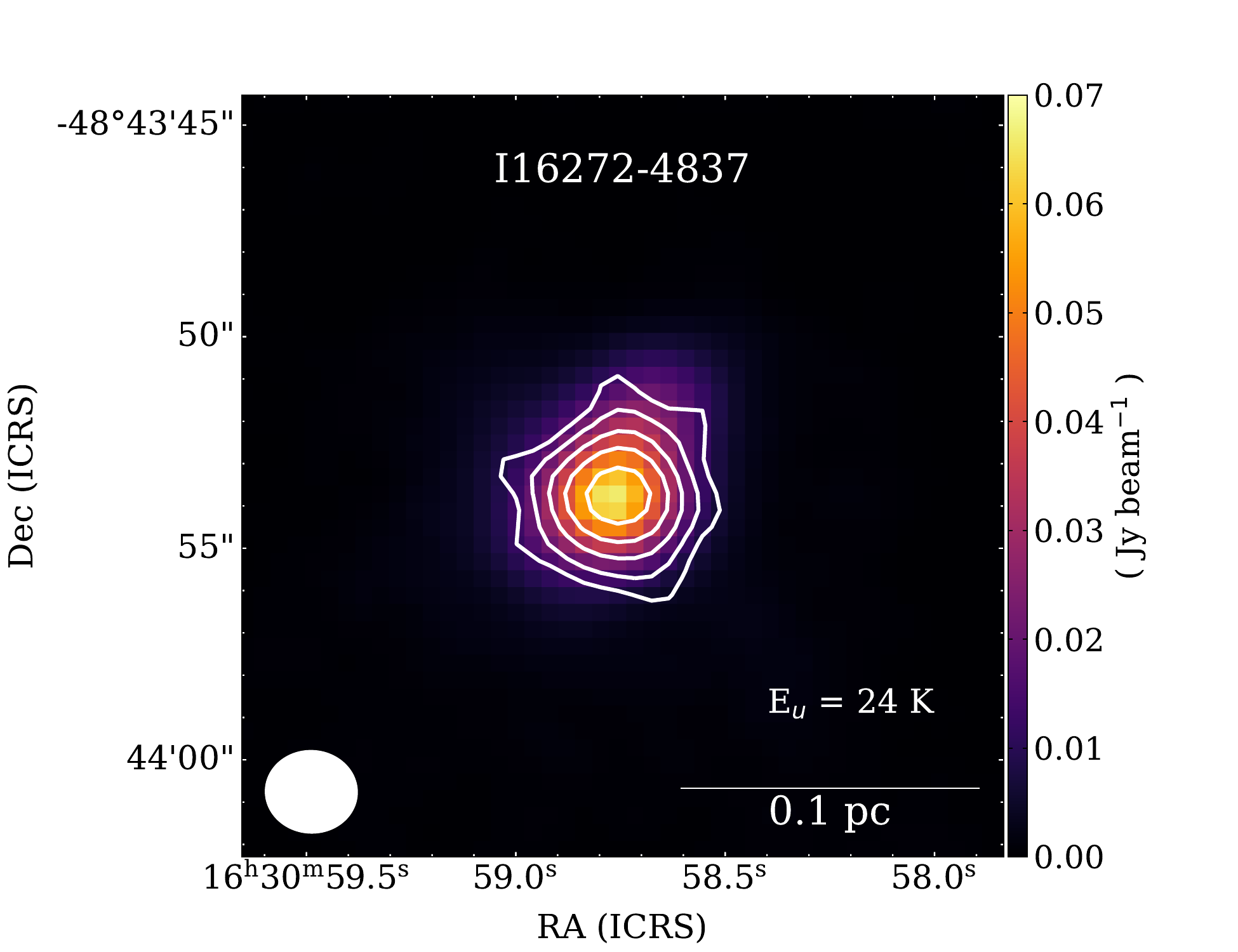}
}
\quad
\subfigure[]{
\includegraphics[width=5.5cm]{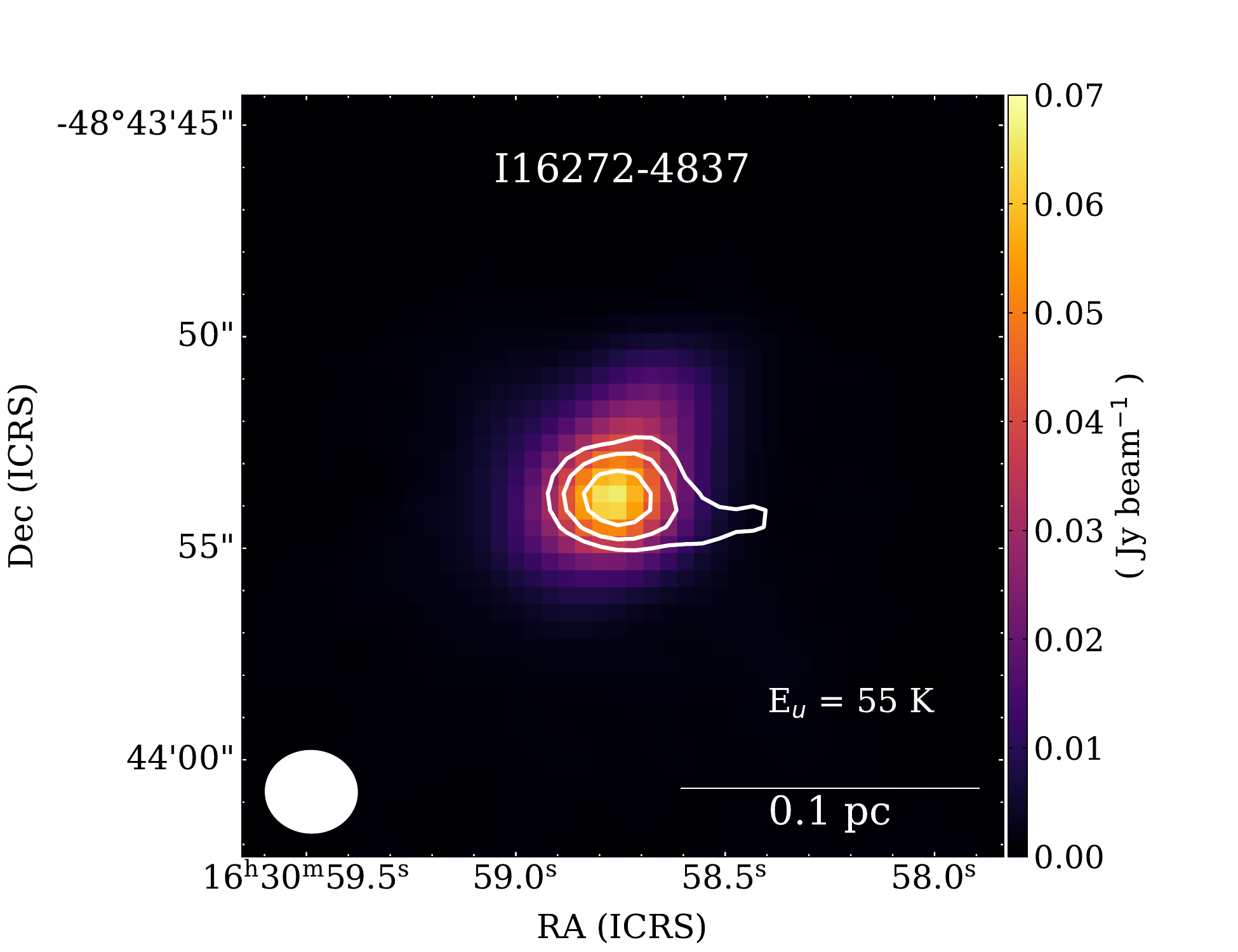}
}
\quad
\subfigure[]{
\includegraphics[width=5.5cm]{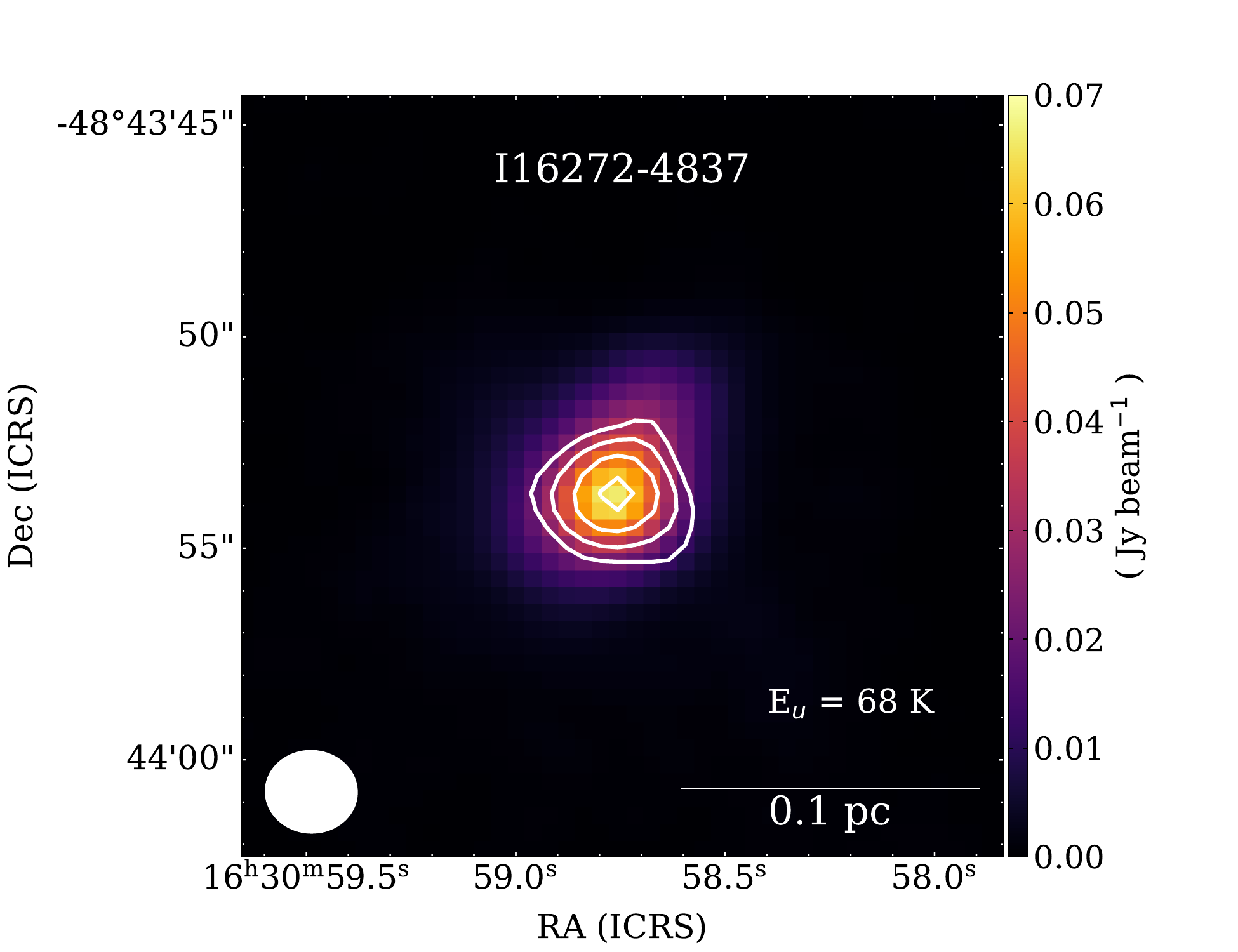}
}
\quad
\subfigure[]{
\includegraphics[width=5.5cm]{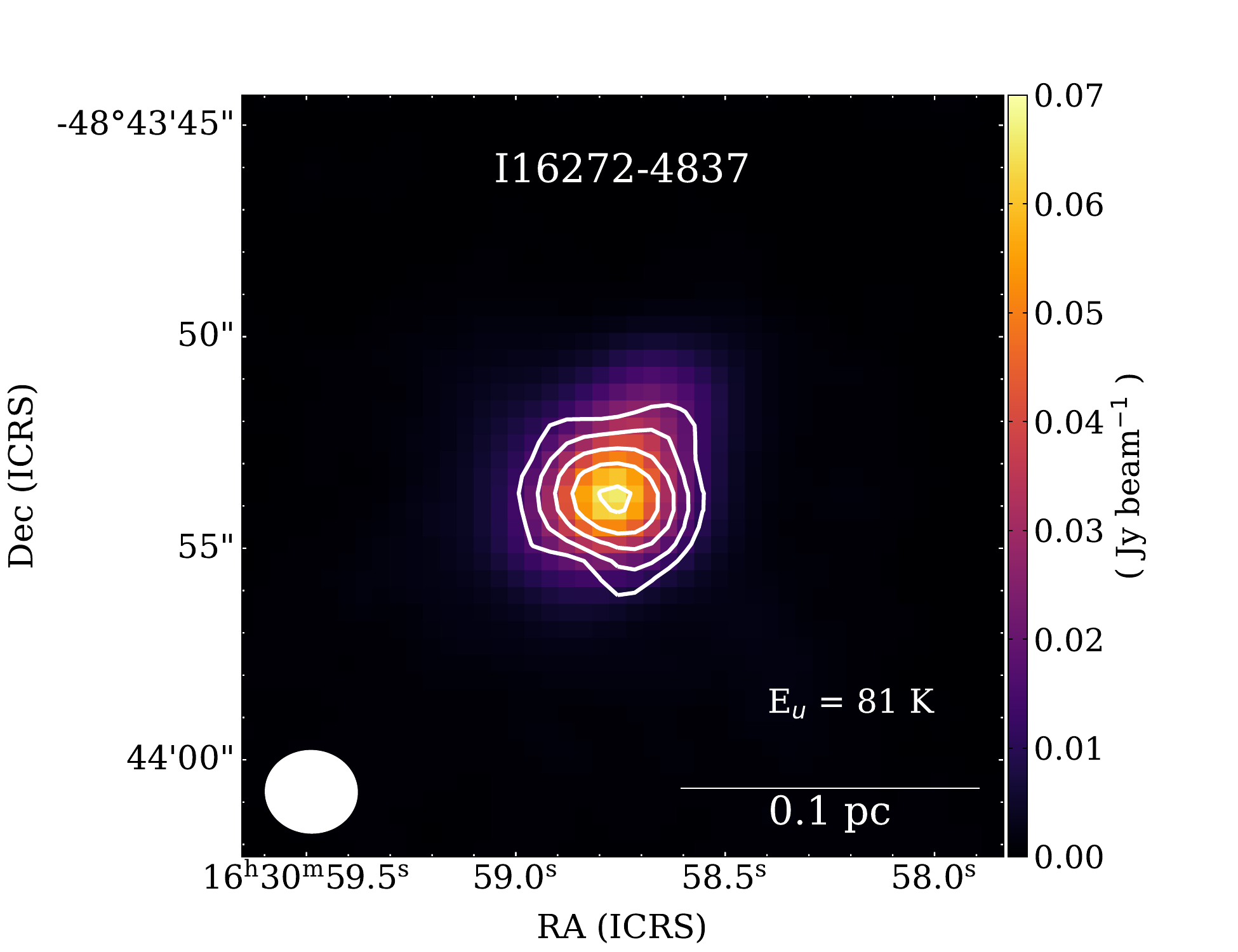}
}
\quad
\subfigure[]{
\includegraphics[width=5.5cm]{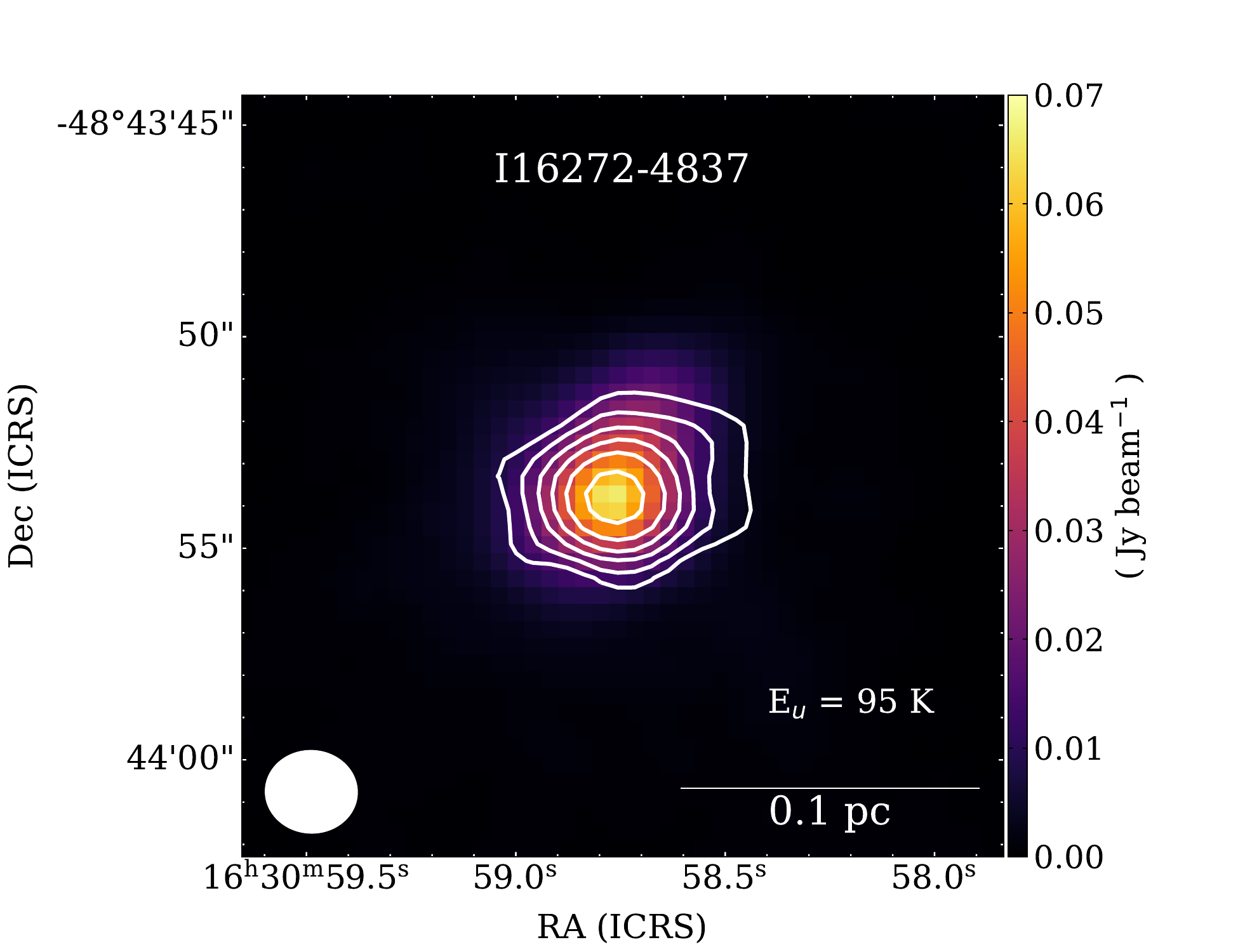}
}
\caption{Integrated intensity maps of representative CH${_3}$COCH${_3}$ unblended lines observed towards I16272-4837C1. The color scale is the continuum at a wavelength of 3 mm. The white contours indicate CH${_3}$COCH${_3}$ at difference upper energies. The contour levels are at the 3, 5, 8, 12, 18, 28$\sigma$. Beam size is shown in the bottom left-hand corner.}
\label{fig:spatial_distribution}
\end{figure*}

We obtained integrated intensity maps of acetone using its six principal transitions toward the source I16272-4837C1, as shown in Fig. \ref{fig:spatial_distribution}. Analysis of the spatial distributions reveals that all peak emissions coincide with the continuum peaks and exhibit no significant differences in their spatial profiles. 

\begin{figure*}
\centering
\subfigure[]{
\includegraphics[width=5cm]{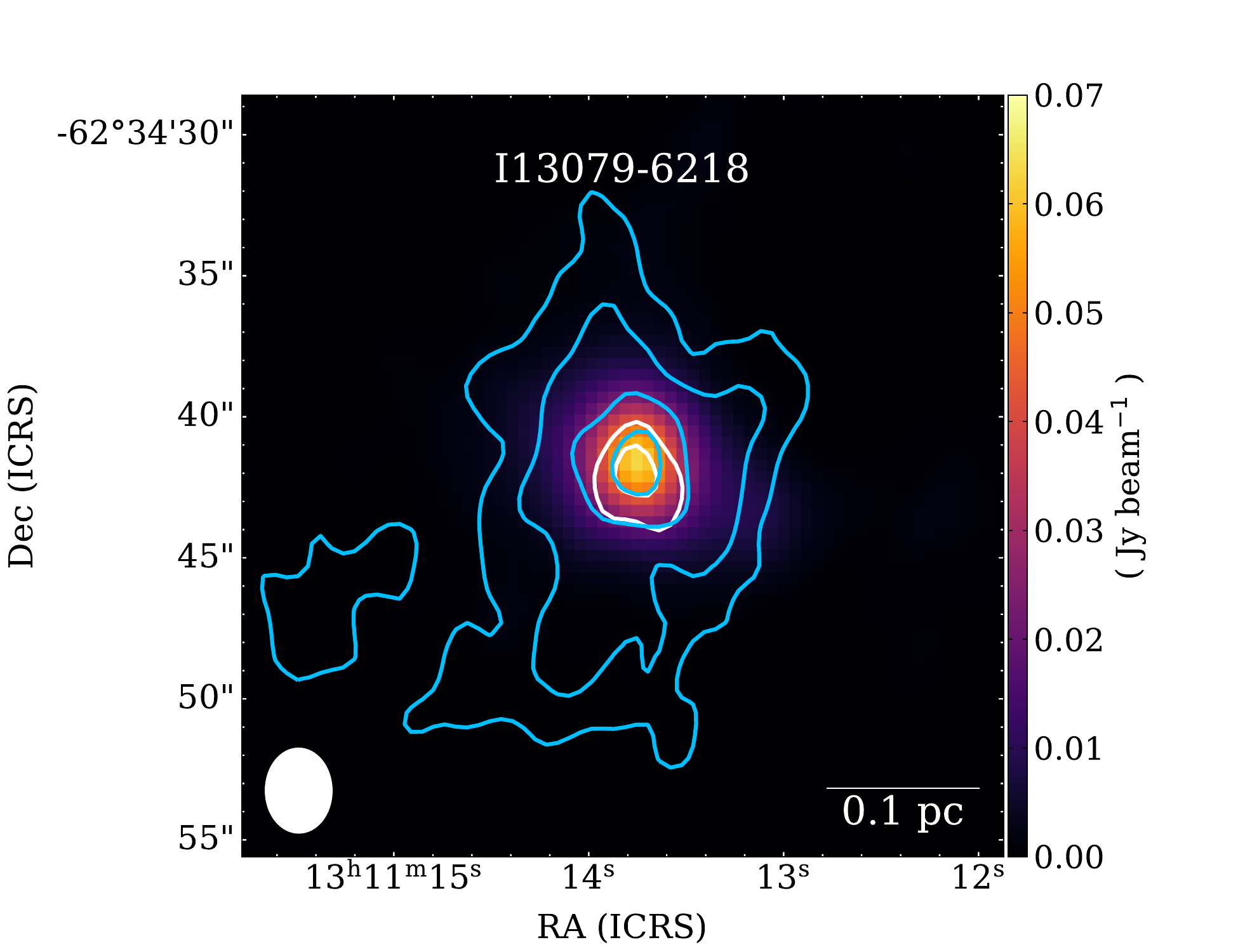}
\label{fig:sub1}
}
\vspace{-2mm}
\subfigure[]{
\includegraphics[width=5cm]{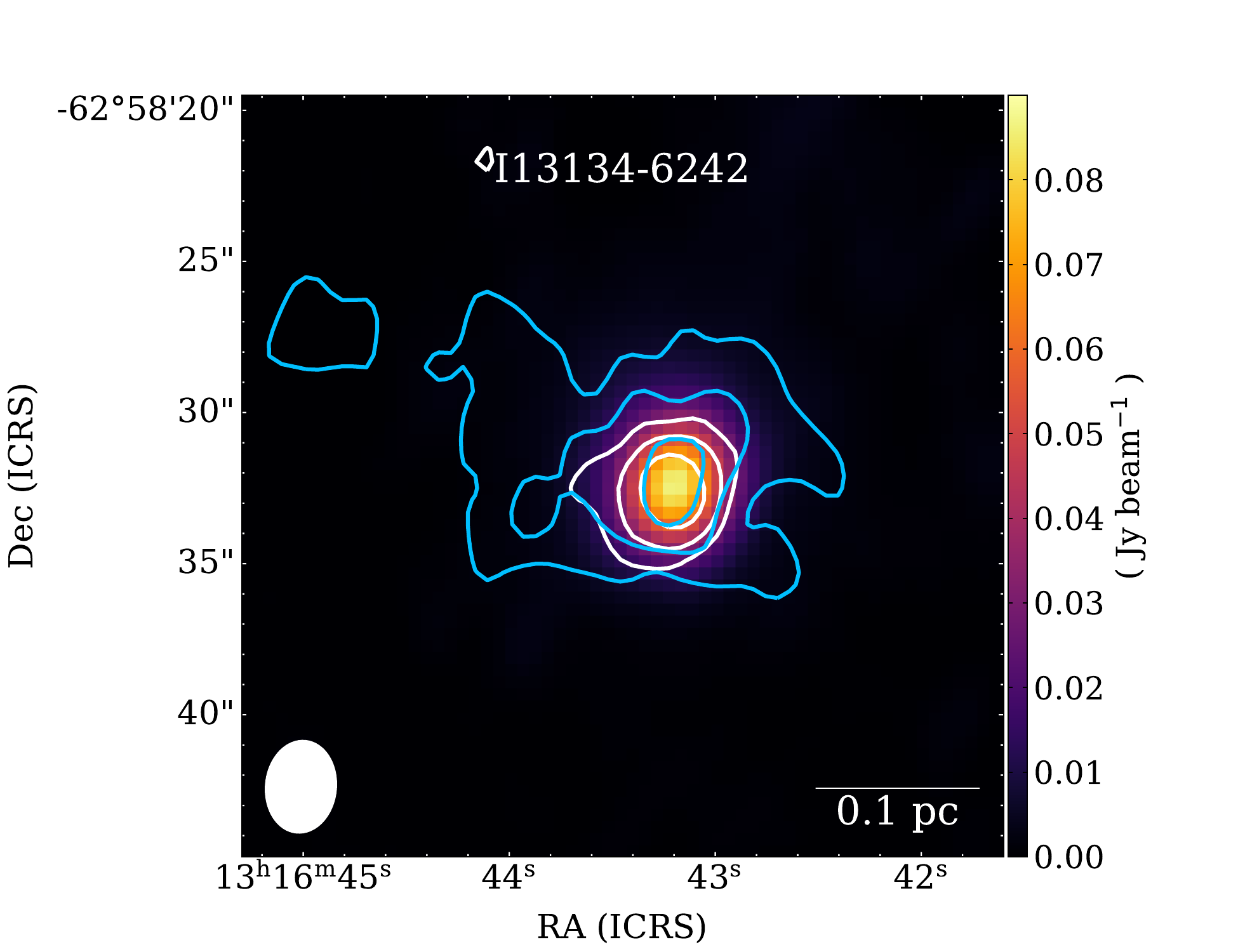}
\label{fig:sub2}
}
\vspace{-2mm}
\subfigure[]{
\includegraphics[width=5cm]{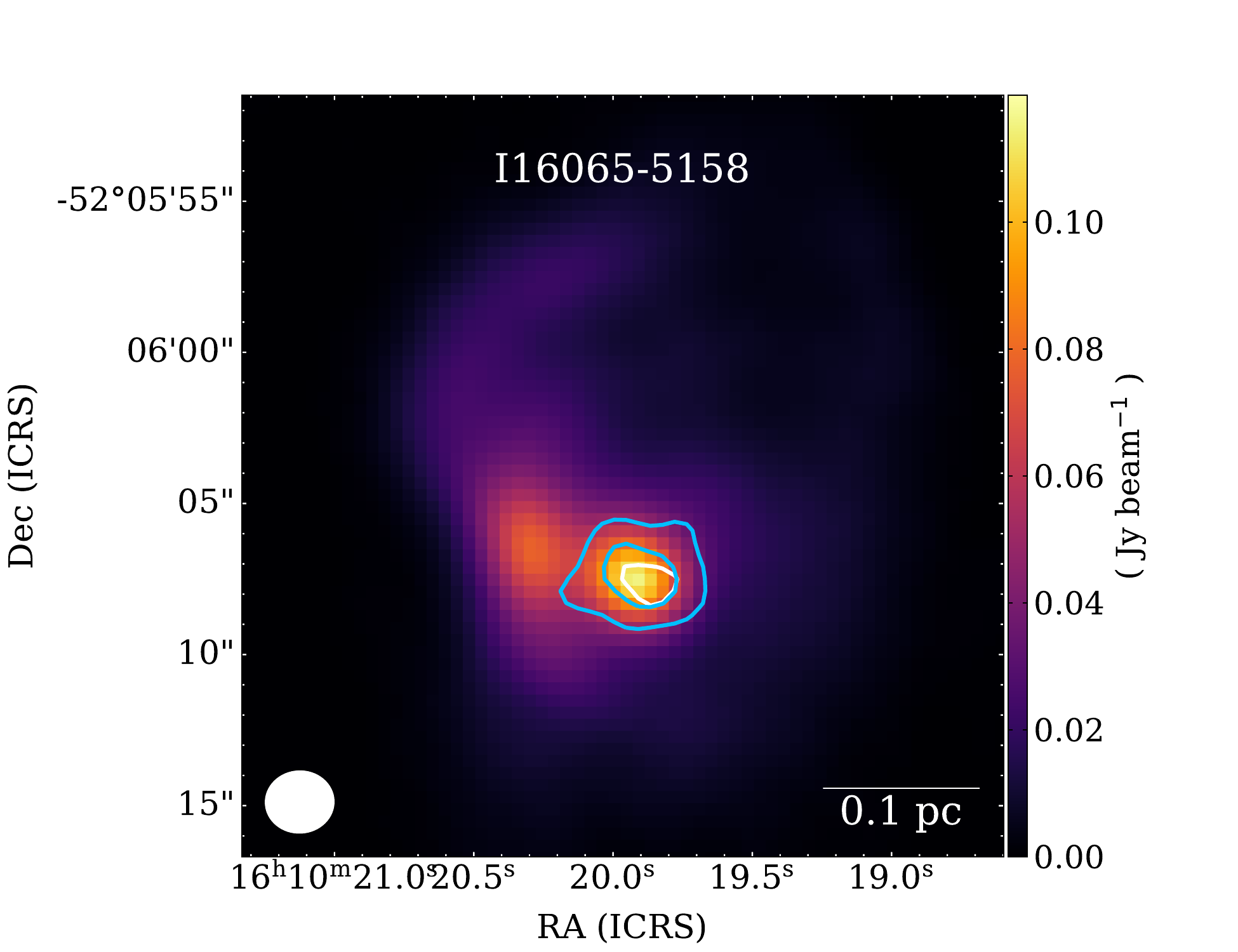}
\label{fig:sub3}
}
\vspace{-2mm}
\subfigure[]{
\includegraphics[width=5cm]{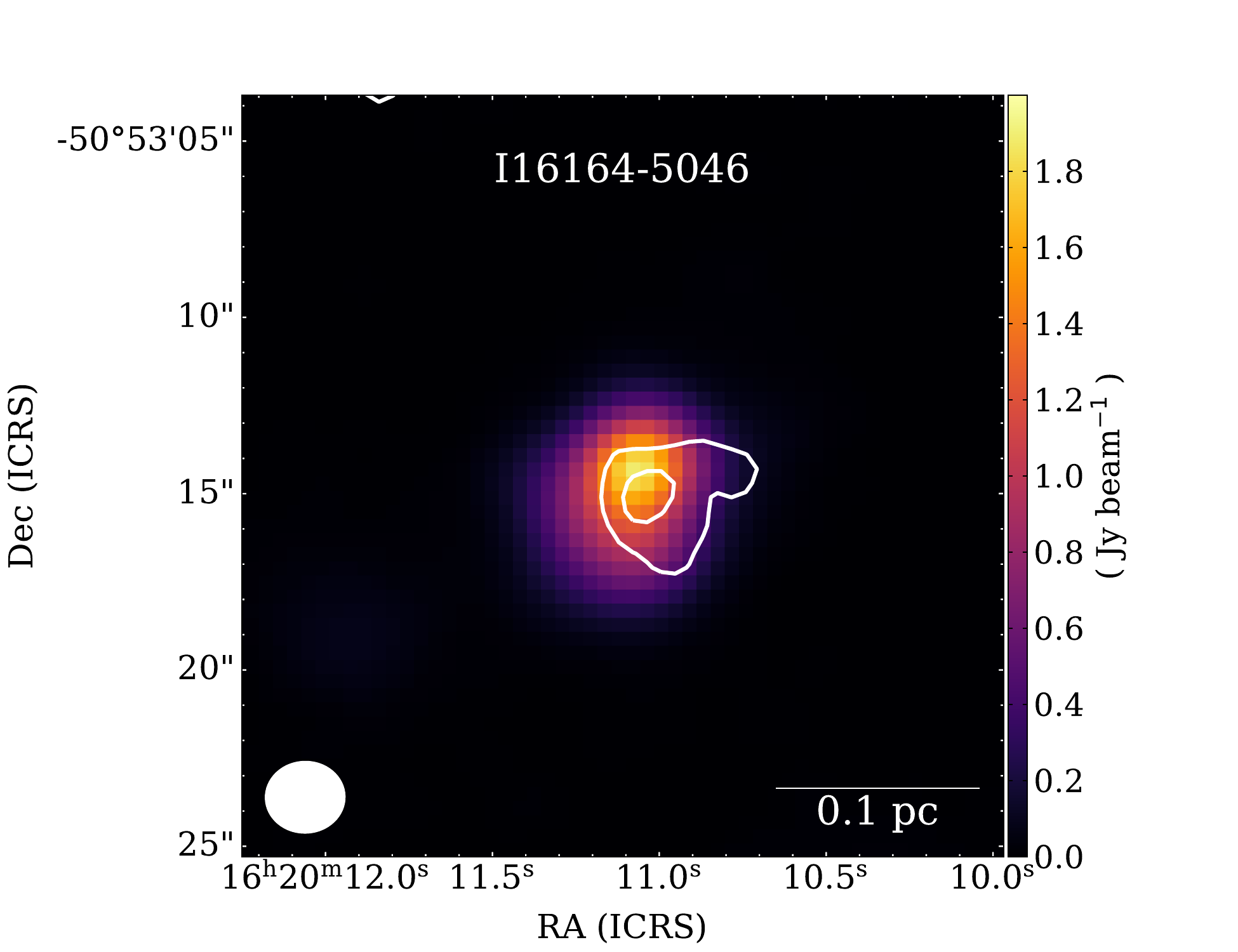}
\label{fig:sub4}
}
\vspace{-2mm}
\subfigure[]{
\includegraphics[width=5cm]{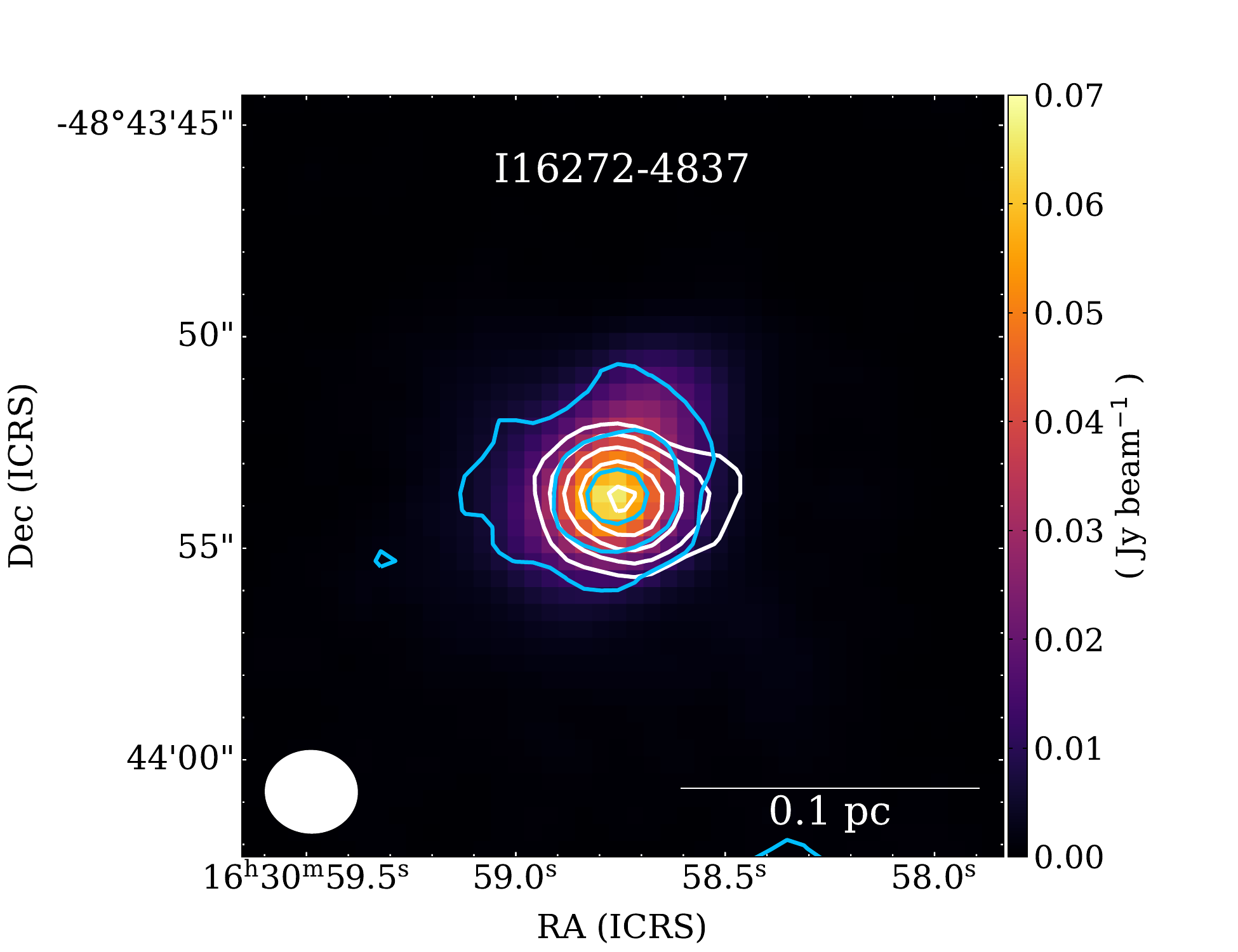}
\label{fig:sub5}
}
\vspace{-2mm}
\subfigure[]{
\includegraphics[width=5cm]{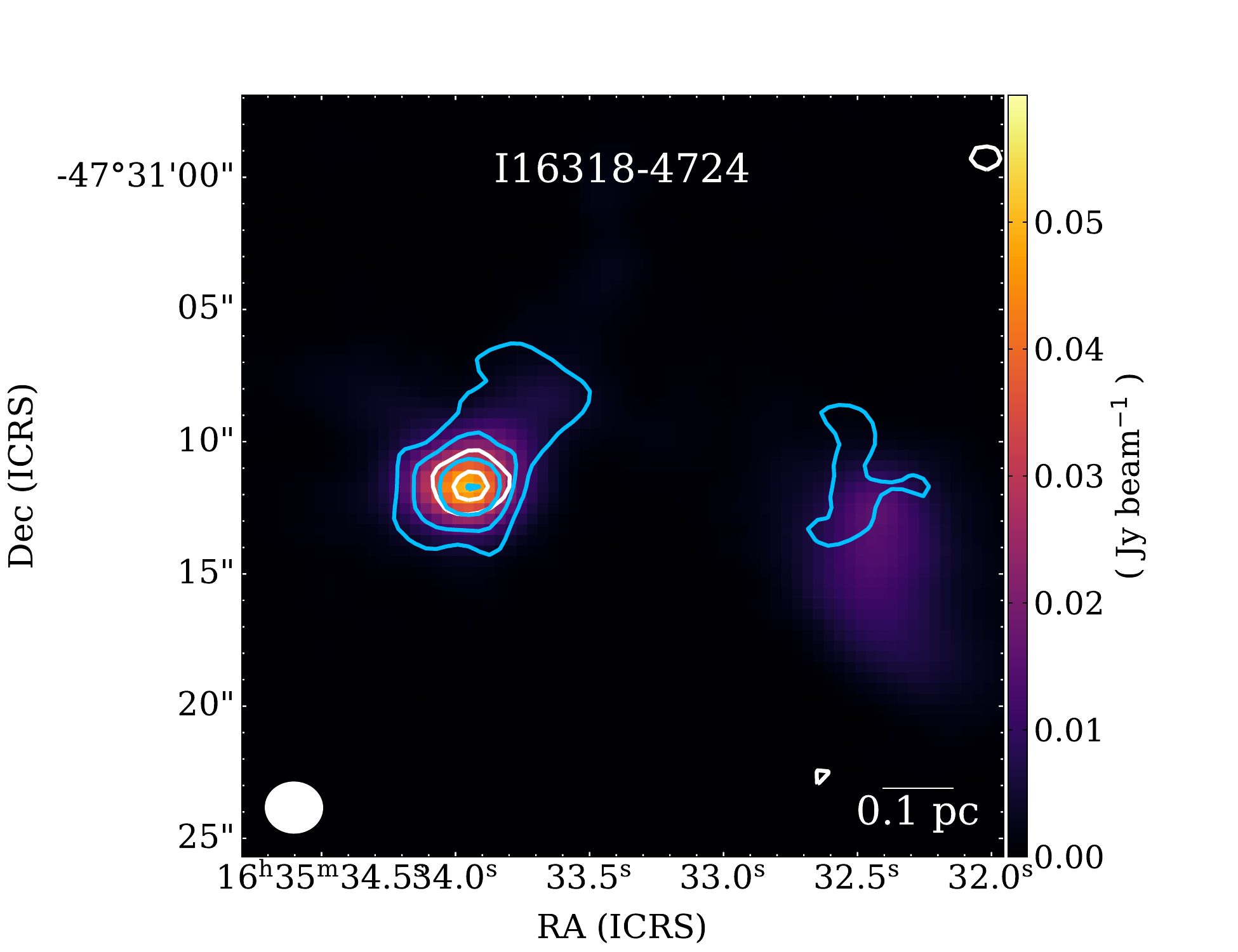}
\label{fig:sub6}
}
\vspace{-2mm}
\subfigure[]{
\includegraphics[width=5cm]{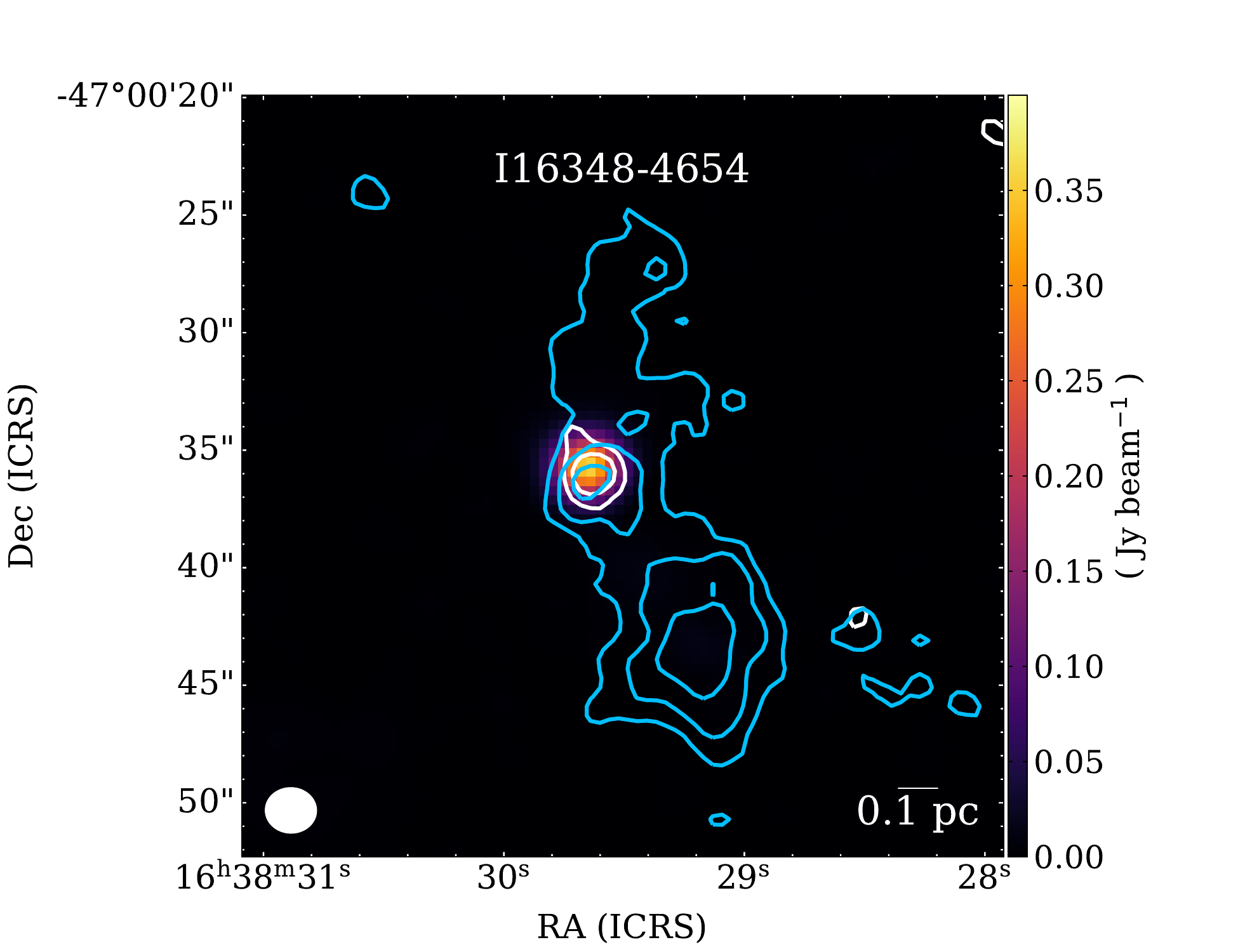}
\label{fig:sub7}
}
\vspace{-2mm}
\subfigure[]{
\includegraphics[width=5cm]{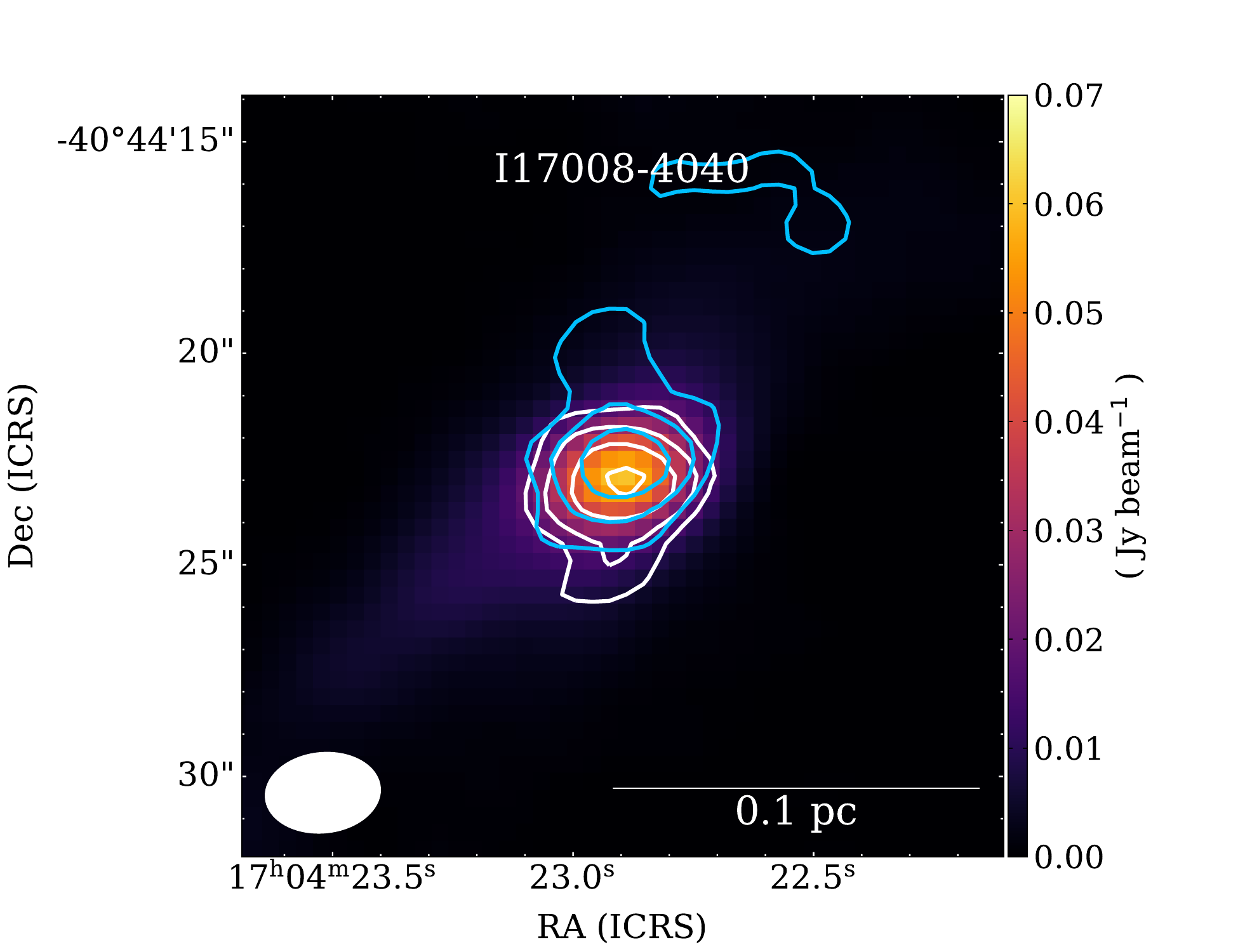}
\label{fig:sub8}
}
\subfigure[]{
\includegraphics[width=5cm]{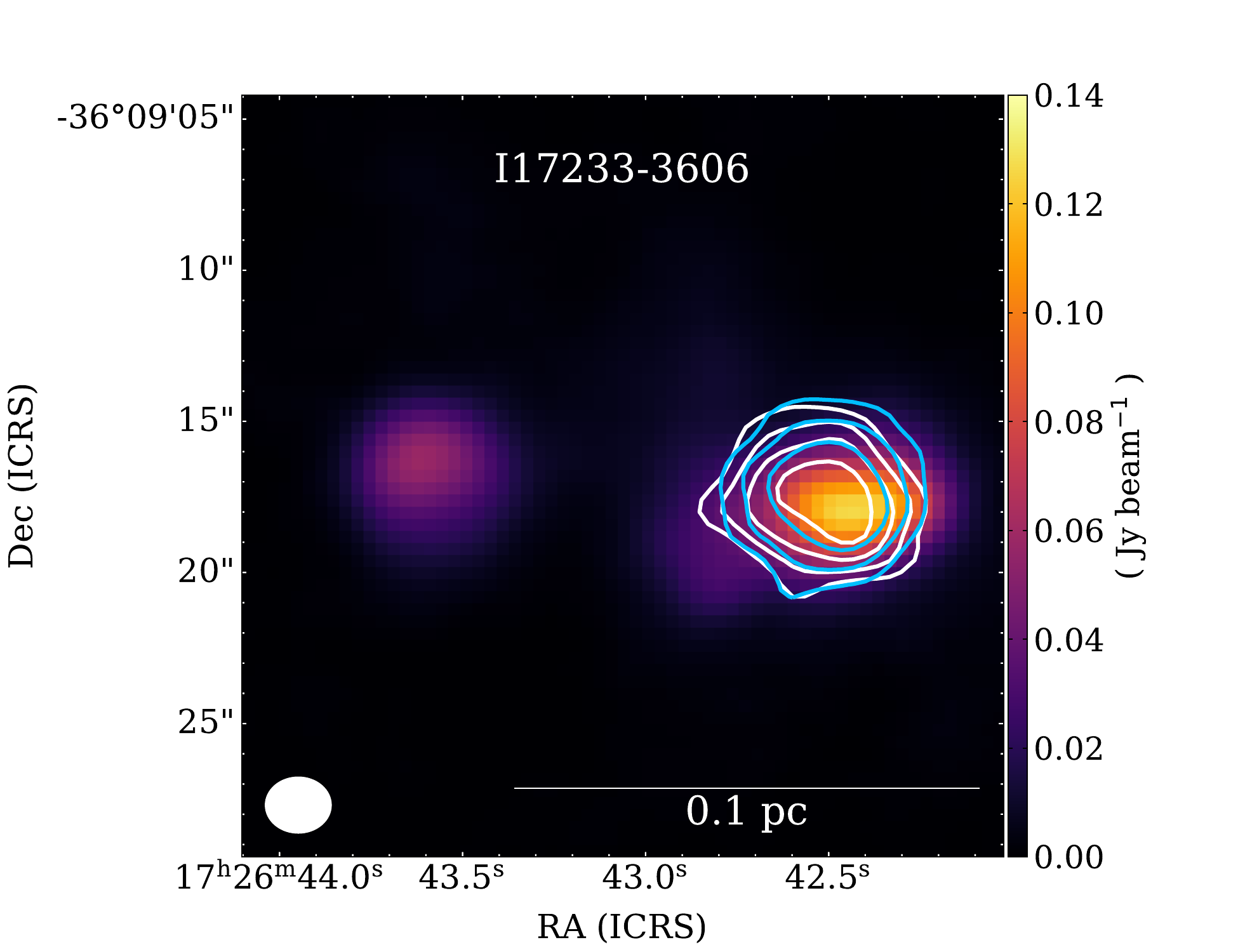}
\label{fig:sub9}
}
\vspace{-2mm}
\subfigure[]{
\includegraphics[width=5cm]{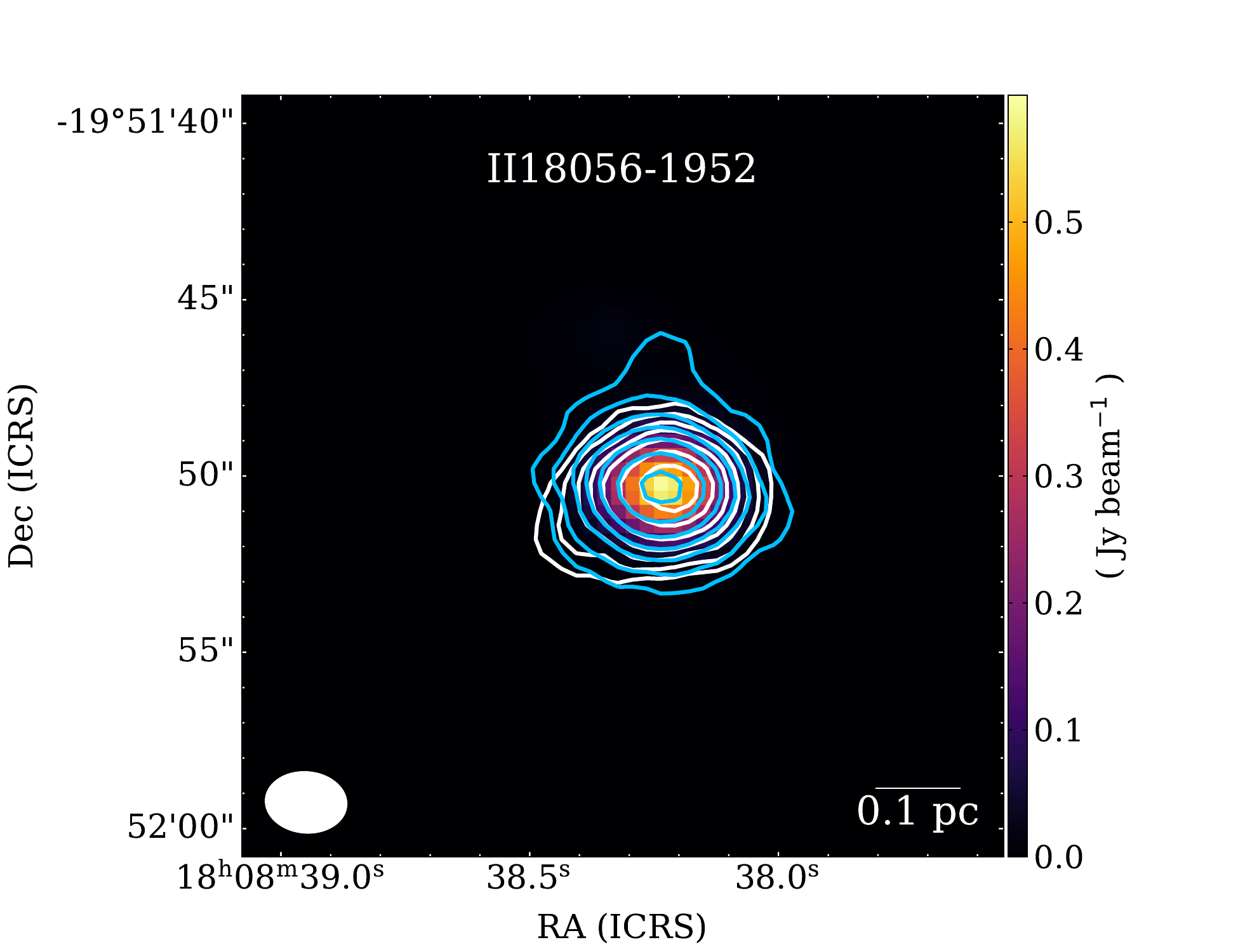}
\label{fig:sub10}
}
\quad
\subfigure[]{
\includegraphics[width=5cm]{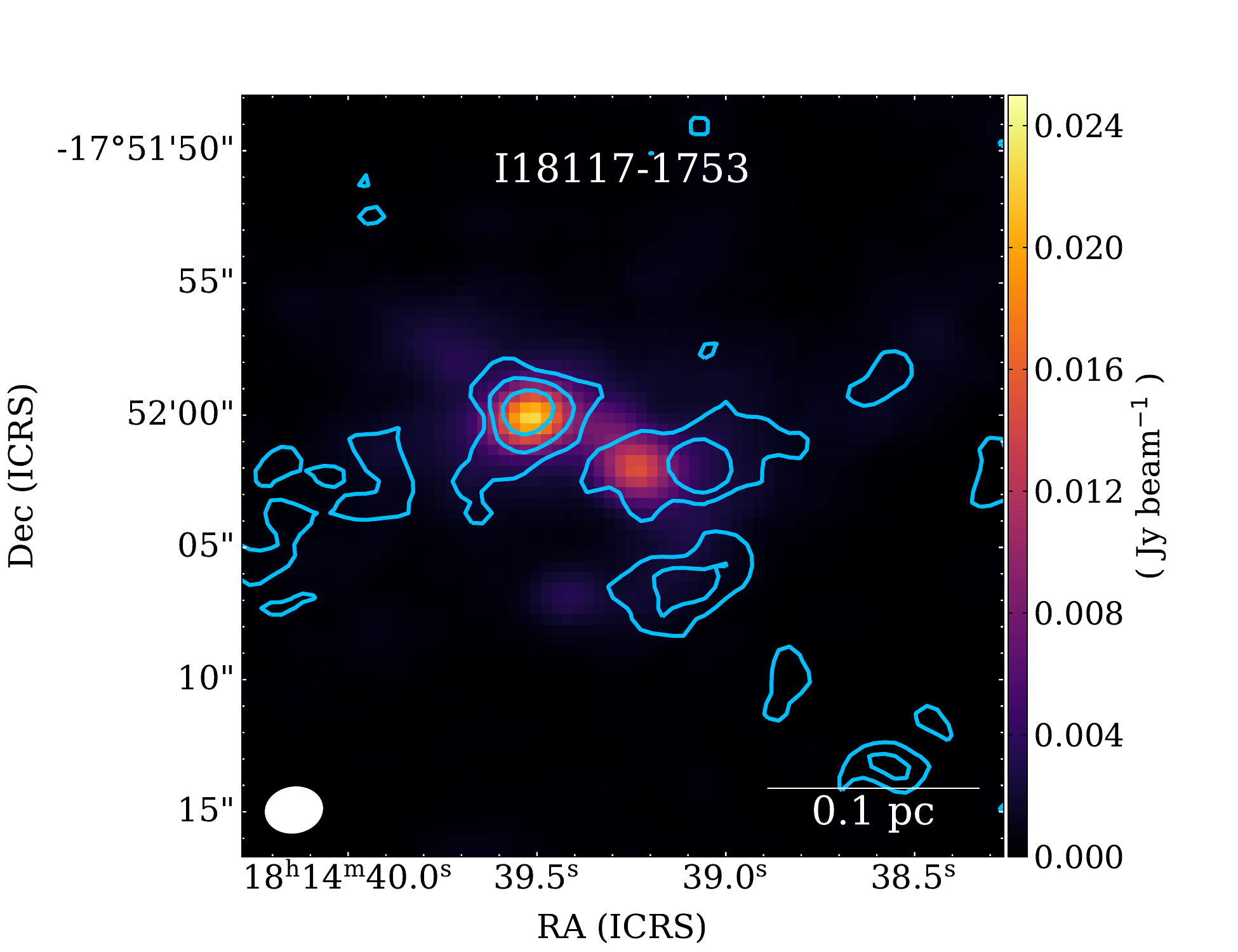}
\label{fig:sub11}
}
\vspace{-2mm}
\subfigure[]{
\includegraphics[width=5cm]{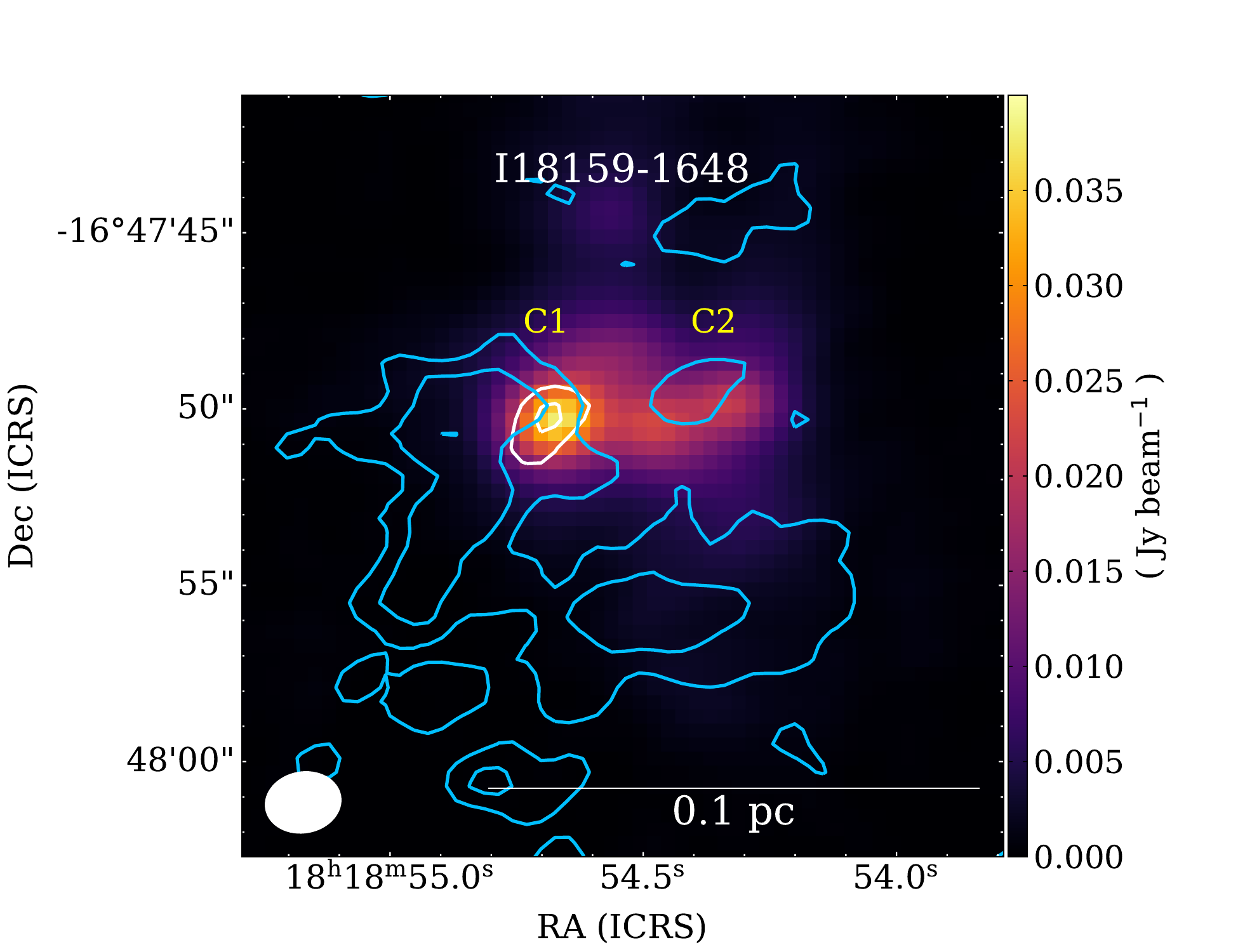}
\label{fig:sub12}
}
\quad
\subfigure[]{
\includegraphics[width=5cm]{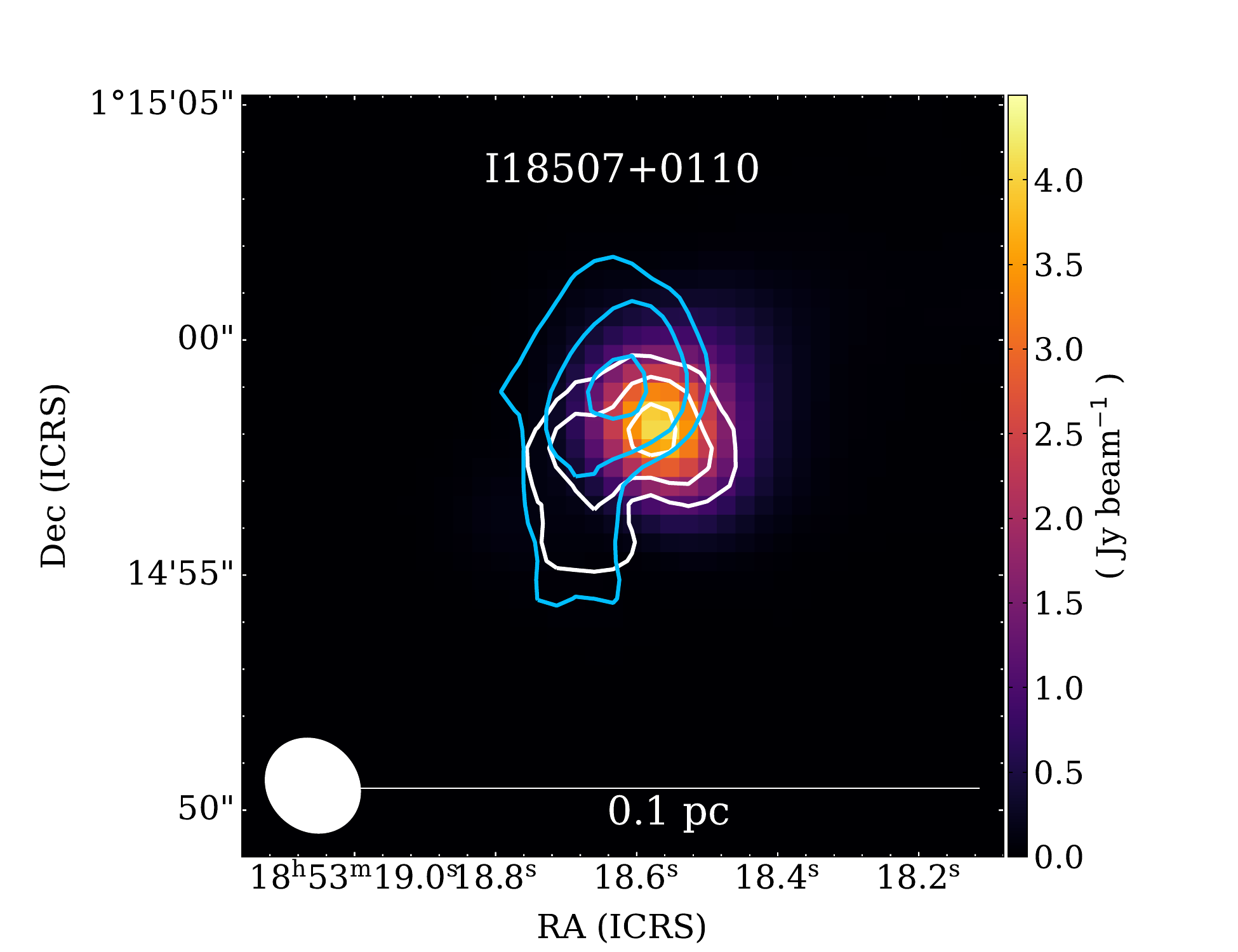}
\label{fig:sub13}
}
\vspace{-2mm}
\subfigure[]{
\includegraphics[width=5cm]{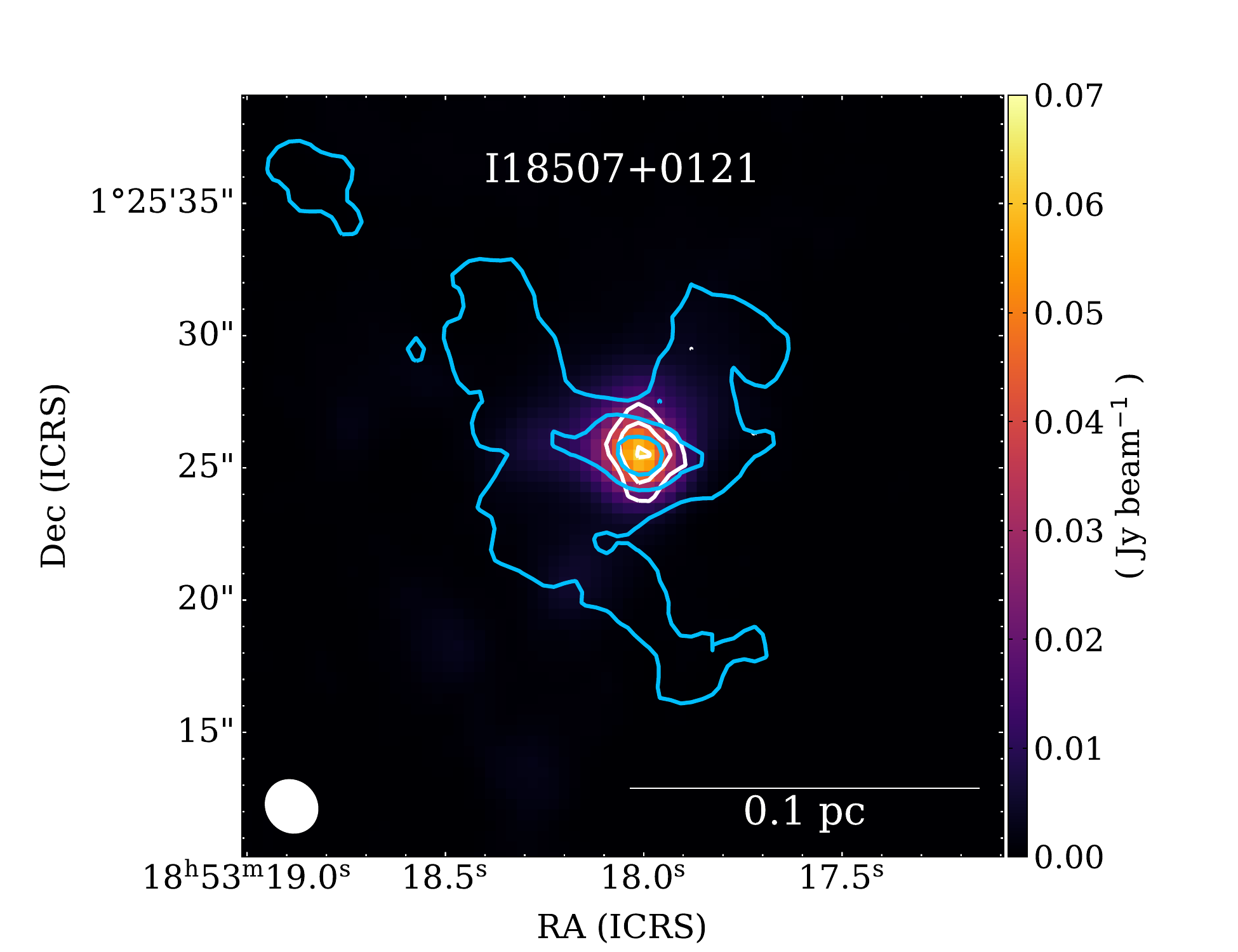}
\label{fig:sub14}
}
\vspace{-2mm}
\subfigure[]{
\includegraphics[width=5cm]{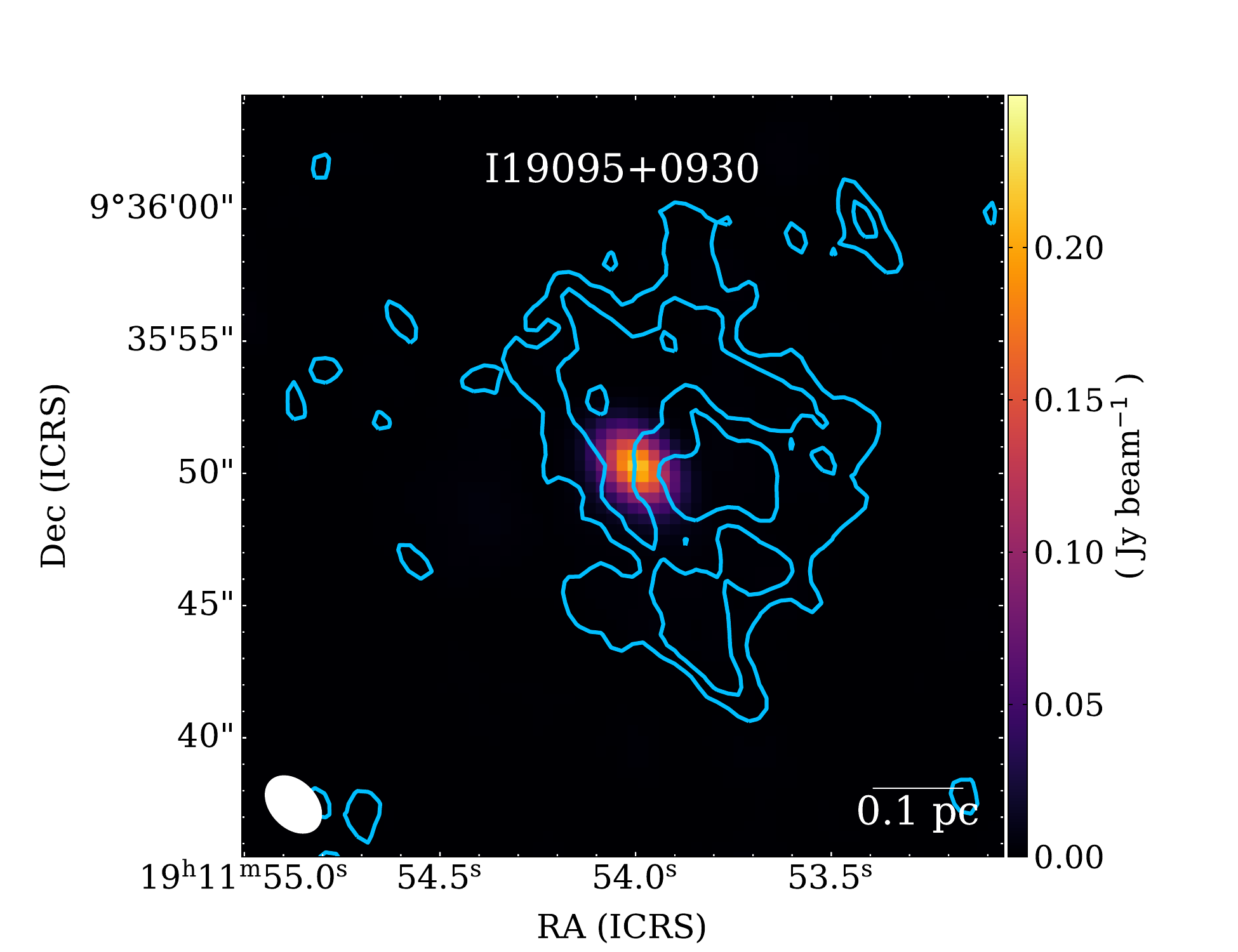}
\label{fig:sub15}
}
\caption{Integrated intensity maps of unblended CH$_3$COCH$_3$ (98800 MHz) and CH$_3$CHO (98863 MHz) lines toward high-mass star forming regions. The color scale represents the 3 mm continuum emission. White and deepskyblue contours indicate CH$_3$COCH$_3$ (upper-level energy E$_u$ = 14.09 K) and CH$_3$CHO (upper-level energy E$_u$ = 16.59 K) emission, respectively. For sources I18117-1753 and I19095-0930, the acetone lines are too weak and are thus not plotted in the figure, which only shows the integrated intensity of CH$_3$CHO. For the source I16164-5046, the acetaldehyde line is too weak and is not shown in the figure. The white contours indicate the CH$_3$COCH$_3$ emission at the 3, 5, 8, 12, 18, 28, and 40$\sigma$ levels ($\sigma$ = 15-55 mJy beam$^{-1}$ km s$^{-1}$), while the deepskyblue contours indicate the CH$_3$CHO emission at the same levels ($\sigma$ = 15-70 mJy beam$^{-1}$ km s$^{-1}$). Beam size is shown in the bottom left-hand corner.}
\label{fig:acetone_acetaldehyde}
\end{figure*}

To investigate the relationships between acetaldehyde and acetone, we generated integrated maps of CH$_3$COCH$_3$ at a rest frequency of 98800 MHz and CH$_3$CHO at a rest frequency of 98863 MHz toward 15 massive protostellar regions, as illustrated in Fig. \ref{fig:acetone_acetaldehyde}. For sources I18117-1753 and I19095-0930, the acetone lines are too weak and are not shown in the figure. For the source of I16164-5046, the acetaldehyde line is also too weak and is not shown in the figure. From the Fig.\ref{fig:acetone_acetaldehyde}, the spatial distributions of acetone exhibit similarities with acetaldehyde. The line emission peaks of two species coincide with the continuum peaks in most regions. The emission of acetone is concentrated in the hot core regions and generally displays a compact spatial distribution, whereas the emission of acetaldehyde shows a more extended spatial profile. This suggests that acetaldehyde formation is dominated by gas-phase processes at relatively low temperatures before the bulk of the species desorb from icy grain mantles \citep{manigand2020}. Furthermore, \citet{li2025b} propose that the formation of acetaldehyde likely occurs in regions affected by widespread shock waves, as supported by the observed spatial correlation between acetaldehyde distribution and the emission patterns of SiO and H$^{13}$CO$^+$. In Section \ref{sec:fits}, in order to obtain acetaldehyde column densities, we assumed the same rotational temperature for acetaldehyde as for acetone. This likely results in overestimated rotational temperatures for acetaldehyde, as it exhibits a more extended spatial distribution than acetone. Consequently, the derived column densities for acetaldehyde are likely overestimated since the assumption of co-spatial distribution does not hold.

\section{Discussion}
\label{sec:discussion}
\subsection{Influence of H{\sc ii} regions on acetone chemistry}
\label{sec:Influence}
The systematically lower acetone abundances in hot cores associated with ultra‑compact H{\sc ii} regions (Section \ref{sec:temperatures}) point to a direct environmental influence on its chemistry, likely through enhanced ultraviolet (UV) radiation. While photons with h$\nu$ > 13.6 eV are absorbed in creating and maintaining the ionized region, the abundant far-UV photons (6–13.6 eV) can penetrate the surrounding molecular envelope. These photons are capable of photodissociating acetone directly or depleting its chemical precursors, thereby suppressing its observed abundance. Thus, the lower acetone abundances in sources associated with H{\sc ii} regions can be attributed to a UV-driven chemical environment where photodissociation processes are significant. In this picture, the presence of an ultra-compact H{\sc ii} region serves as an indicator of a local radiation field that inhibits acetone formation or accelerates its destruction.

\subsection{Comparison with the other sources}
\label{sec:compare}
\begin{figure}
    \includegraphics[scale=0.26]{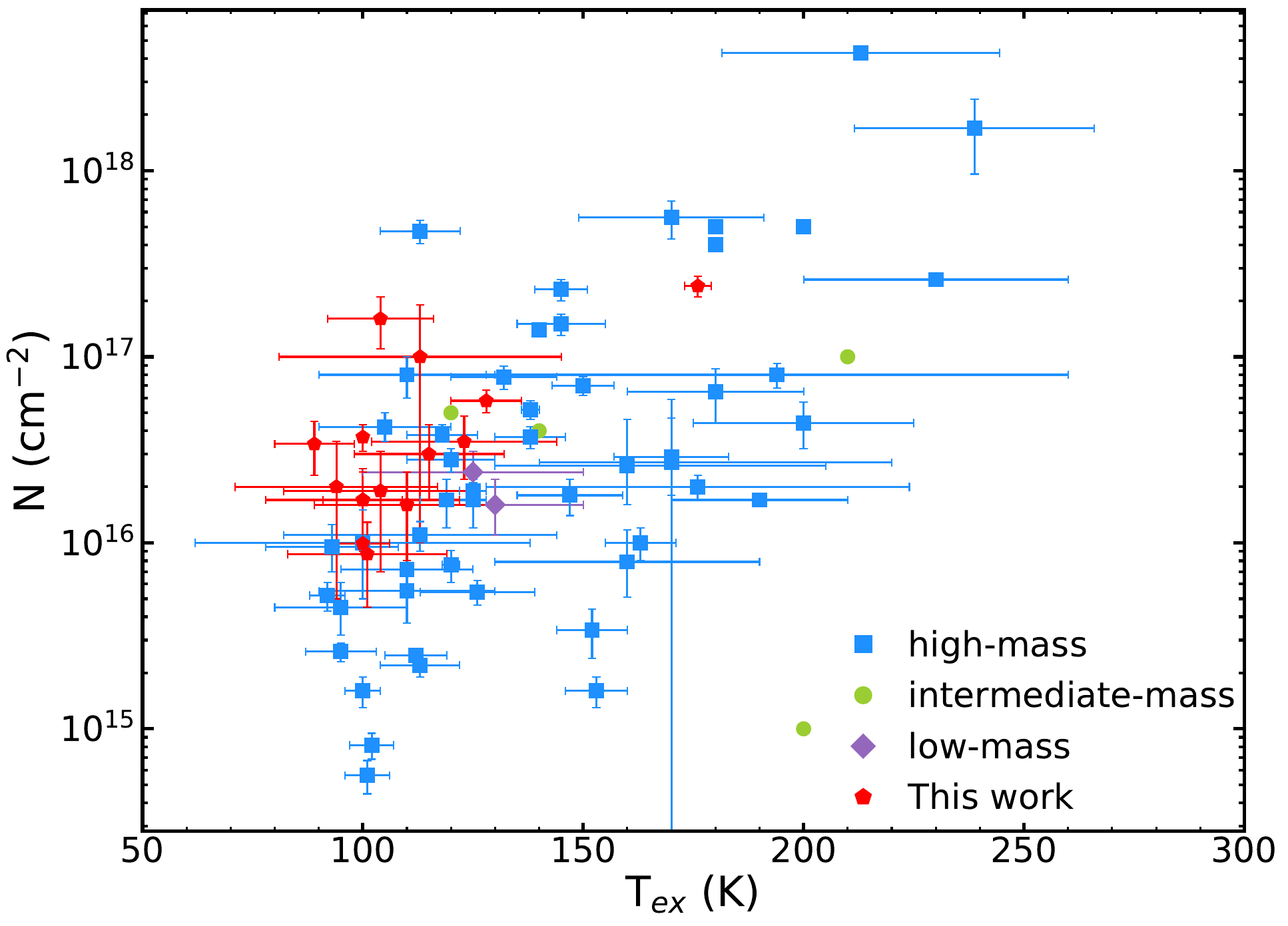}
    \caption{Column density as a function of rotational temperature, categorized by source types. Purple and green represent low-mass hot cores and intermediate-mass hot cores, respectively, while blue and red denote high-mass hot cores.}
    \label{fig:observation}
\end{figure}

To investigate the relationship between the column density and rotational temperature of acetone across different source types, we compared our sample (15 sources) with literature data (52 sources), as listed in Table \ref{tab:acetone-obser}. Fig. \ref{fig:observation} presents all data points color-coded by source type. The linear least‑squares fit was performed exclusively on the high‑mass hot cores. Visual inspection shows that other source types (e.g., low-mass and intermediate-mass hot cores) do not occupy distinctly separate regions in this parameter space compared to the high-mass hot cores. The best-fit linear relationship for high-mass hot cores is described by the equation log(N(CH$_3$COCH$_3$)) = 0.012T$_{rot}$ + 14.767, with a Pearson correlation coefficient of 0.59 (p = 2.17 $\times$ 10$^{-7}$). This indicates a moderately positive correlation between the two variables. However, the relatively modest correlation strength (R$^2 \approx$ 0.35) suggests that rotational temperature alone accounts for only a fraction of the variance in acetone column density, implying the potential influence of additional physical or chemical factors.

\citet{jorgensen2018} identified a correlation between rotational temperatures and desorption temperatures for oxygen-bearing COMs formed on icy grains. In our observations, the mean rotational temperature of acetone (110$\pm$15 K) aligns closely with its experimentally desorption temperature of 133 K \citep{schaff1998}. Although the experimental value is slightly higher than the observational one, actually acetone desorbs at lower temperatures in astrophysical environments, due to differences in desorption timescales between laboratory and interstellar conditions. By combining the spatial distribution of acetone, we can infer that its formation primarily occurs in hot and compact regions, where it is initially formed on grain ice surfaces before being released into the gas phase through thermal desorption. However, the contribution of high-temperature gas-phase chemical reactions to acetone production cannot be ruled out. Detailed chemical pathways will be discussed in Section \ref{sec:chemistry}. 

\subsection{The related chemistry of acetone}
\label{sec:chemistry}
To explain the observed abundance of acetone in space, multiple formation pathways have been proposed. These include gas-phase ion-molecule reactions \citep{combes1987, herbst1990}, surface radical-radical recombination \citep{garrod2008}, the radiolysis of methanol \citep{bennett2007}, and experimental routes involving oxygen atoms \citep{hudson2017}.

In this work, we focus on the pathways that can be directly investigated through our observational data, namely those involving its potential precursor, acetaldehyde, and its indirect relative, methanol. The gas-phase formation route begins with the radiative association between the methyl ion and acetaldehyde, \protect\\

\noindent CH$_3^+$ + CH$_3$CHO $\rightarrow$ CH$_3$COCH$_4^+$ + hv, \hfill(R1) \protect\\

\noindent followed by dissociative recombination with electrons to form acetone, \protect\\

\noindent CH$_3$COCH$_4^+$ + e$^-$ $\rightarrow$  CH$_3$COCH$_3$ + H. \hfill(R2) \protect\\

\noindent However, \citet{herbst1990} found that reactions (R1-R2) are much slower than previously estimated, rendering this pathway insufficient to explain the observed acetone abundance in hot cores. These reactions can only account for the significantly lower abundance ($\sim$10$^{-11}$) observed in cold sources like TMC-1. Consequently, grain-surface chemistry is considered crucial for acetone formation in hot cores \citep{ehrenfreund2000}. Several models and experiments have identified the radical-radical reaction between CH$_3$ and CH$_3$CO on dust grains (R3) as a major pathway \citep{garrod2008, singh2022}, \protect\\ 

\noindent CH$_3$\.CO + \.CH$_3$ $\rightarrow$   CH$_3$COCH$_3$. \hfill(R3) \protect\\

\noindent The CH$_3$CO radical is derived from hydrogen abstraction of acetaldehyde. According to \citet{bennett2007}, acetone can also form via the radiolysis of methanol, potentially explaining cold, extended acetone distributions. Therefore, the observed correlation of acetone with both acetaldehyde and methanol in our sample provides key observational constraints on these interconnected gas‑grain chemical processes.

From Fig. \ref{fig:spatial_distribution} and Fig. \ref{fig:acetone_acetaldehyde}, we note that the acetone emissions are more concentrated near the hot cores, whereas acetaldehyde emissions trace extended regions. This morphological difference demonstrates that acetone is more closely associated with hot cores compared to extended regions. These observational findings strongly suggest that the formation mechanisms of acetone may involve high-temperature gas-phase reactions (e.g., neutral-neutral) or grain surface reactions, as proposed by \citet{herbst1990} and \citet{hudson2017}, rather than ion-neutral reactions followed by dissociative recombination. Our results are further supported by the comprehensive modeling work of \citet{chen2025b}, which identifies the formation of acetone via one gas-phase route \citep{tsang1988, garrod2022} and three solid-phase pathways \citep{tsang1988, belloche2022}. The gas-phase route is Reaction (R4):\protect\\

\noindent 2-C$_3$H$_7$ + O $\rightarrow$ CH$_3$COCH$_3$ + H. \hfill(R4) \protect\\ 

\noindent Three solid-phase pathways are Reactions (R5)-(R7):\protect\\

\noindent CH$_3$CHO + CH$_2$ $\rightarrow$ CH$_3$COCH$_3$, \hfill(R5) \protect\\

\noindent CH$_3$CH(\.{O})CH$_3$ + H $\rightarrow$ CH$_3$COCH$_3$ + H$_2$, \hfill(R6) \protect\\

\noindent CH$_3$\.{C}(OH)CH$_3$ + H $\rightarrow$ CH$_3$COCH$_3$ + H$_2$. \hfill(R7) \protect\\

The combined evidence from astronomical observations, theoretical models, and laboratory experiments strongly supports the significance of both grain-surface and gas-phase processes in the interstellar synthesis of acetone. This multi-pathway formation scenario provides a robust explanation for the observed abundance and spatial distribution of acetone in hot core environments. Nevertheless, observational evidence remains insufficient to confirm potential molecular reactions. Further observations are required to verify whether these reaction pathways represent dominant formation routes.

\begin{figure}
    \centering
    \includegraphics[scale=0.29]{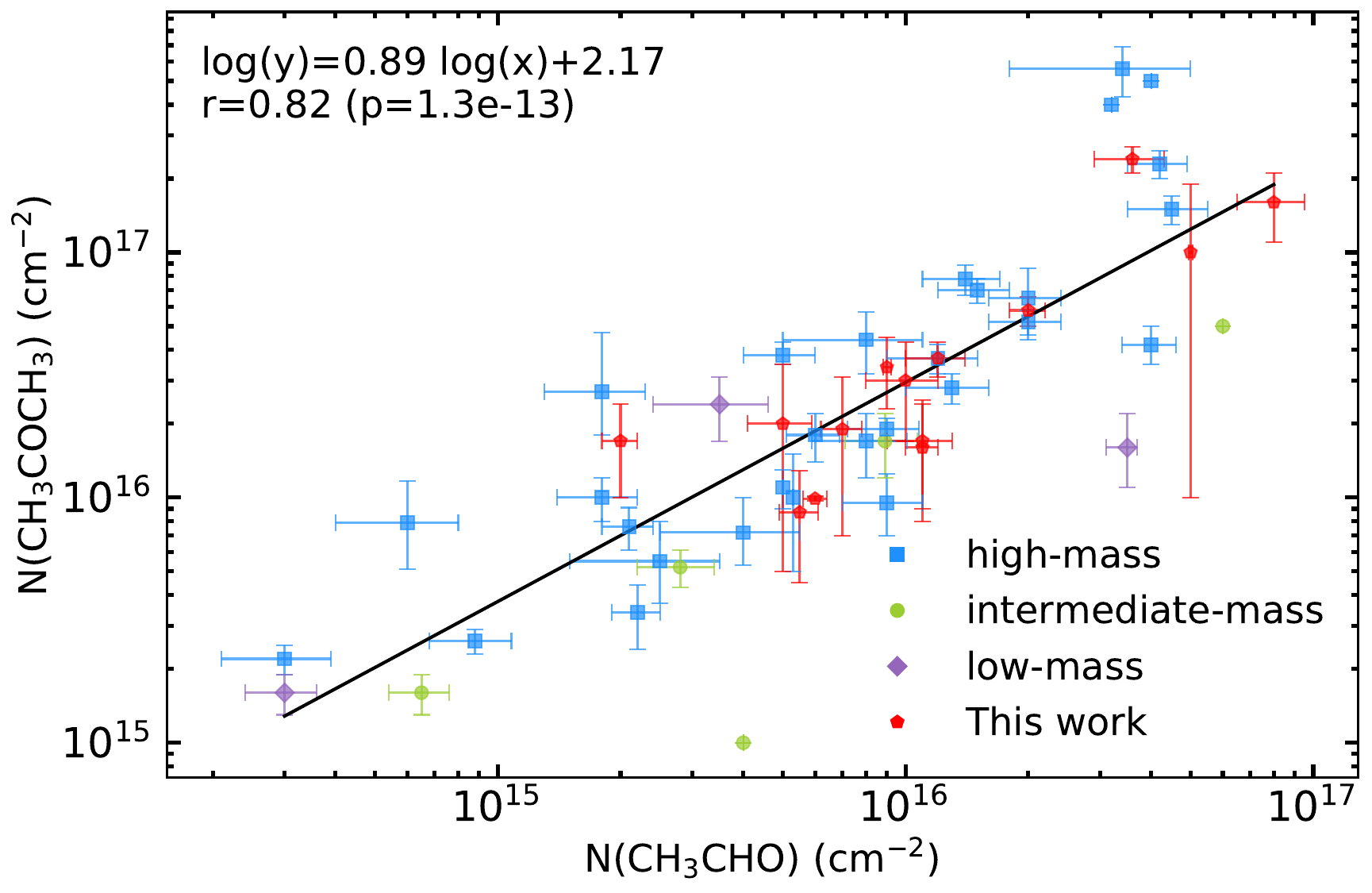}
    \caption{CH$_3$COCH$_3$ column density versus CH$_3$CHO column density, categorized by source types. Purple and green represent low-mass and intermediate-mass hot cores, respectively, while blue and red denote high-mass hot cores. All values are derived from interferometric observations.}
    \label{fig:acetone_aceta}
\end{figure}

Fig. \ref{fig:acetone_aceta} displays the correlation between acetone and acetaldehyde column densities for our sample of 15 sources, along with comparative data from 37 literature sources. Table \ref{tab:acetone-obser} summarizes reported the column densities of acetone, acetaldehyde, and methanol from previous interferometric studies with well-constrained source sizes. Linear least-squares fitting reveals a significant positive correlation between the column densities of acetone and acetaldehyde across different source types (Fig. \ref{fig:acetone_aceta}). The best-fit relationship between the column densities of acetone and acetaldehyde, log(N(CH$_3$COCH$_3$)) = 0.89N(CH$_3$CHO) + 2.17, with a Pearson’s correlation coefficient of 0.82 (p = 1.3$\times$ 10$^{-13}$), strongly suggests a chemical relationship between these species. The derived values align remarkably well with previous large-scale surveys of massive star-forming regions \citep{li2025a} ($r = 0.81$), but are lower than the values reported by \citet{chen2025a} ($r = 0.90$). These findings are supported by recent chemical models that identify multiple formation pathways for both molecules. According to the models, acetone is formed through three grain surface routes and one gas-phase pathway \citep{garrod2022, chen2025a}. Their chemical connection is established through the surface reaction of CH$_3$CHO and CH$_2$. We note that the derived column densities of CH$_3$CHO have inherent uncertainties due to the assumed rotational temperatures in our study. However, in combination with other observational results, we provided valuable constraints for the chemical model and further highlighted the significance of acetaldehyde in acetone formation.

\begin{figure}
    \centering
    \includegraphics[scale=0.29]{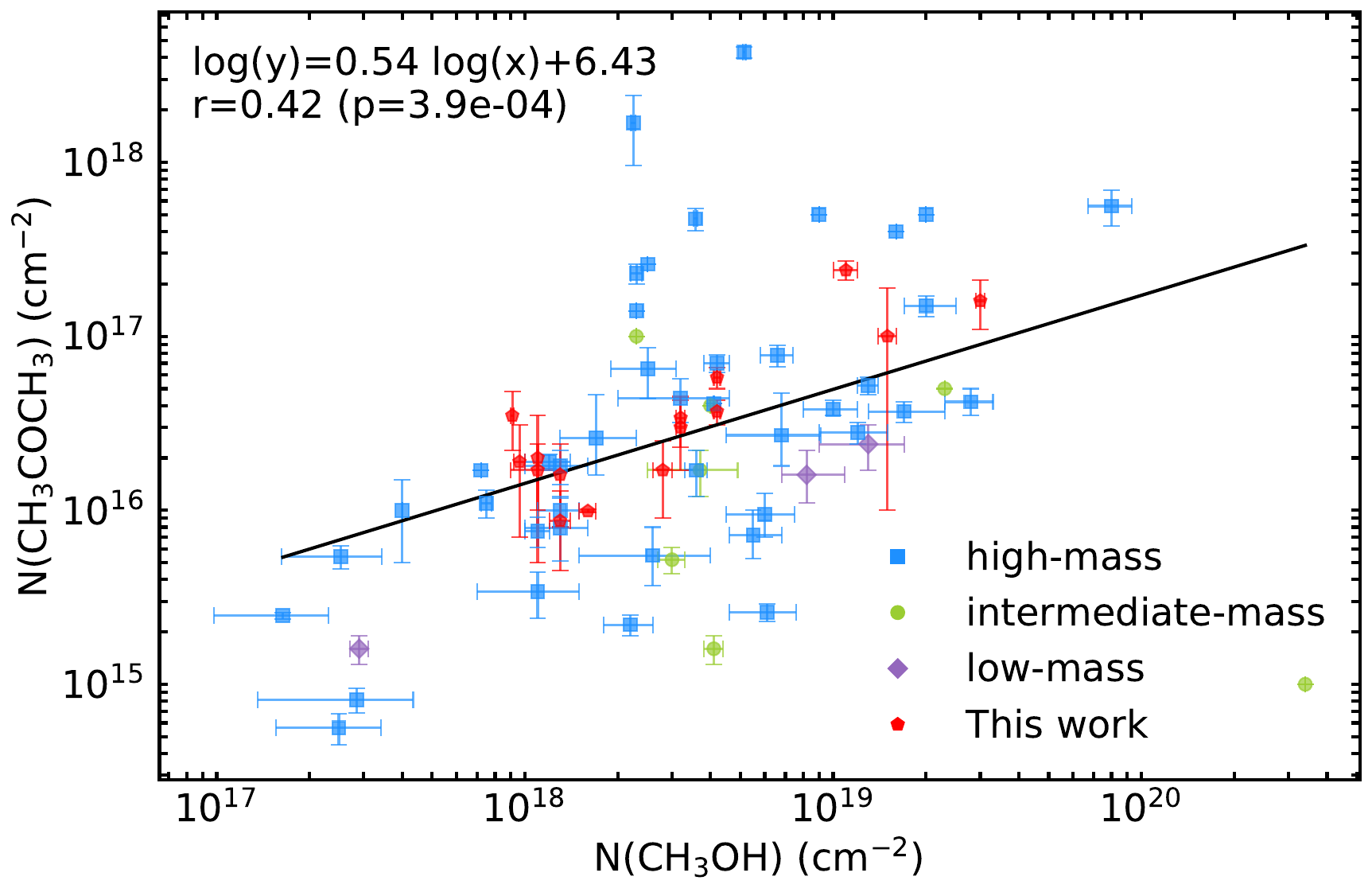}
    \caption{Column density of CH$_3$COCH$_3$ versus column density of CH$_3$OH, categorized by source types. Purple and green represent low-mass and intermediate-mass hot cores, respectively, while blue and red denote high-mass hot cores. All values are derived from interferometric observations.}
    \label{fig:acetone_metha}
\end{figure}

Since acetone can be efficiently formed in pure methanol ice upon exposure to ionizing radiation through experimental research, we investigated the relationship between the column densities of methanol and acetone across our sample (15 sources) and literature data (50 sources), as shown in Table \ref{tab:acetone-obser}. Methanol column densities for our sample were obtained from \citet{qin2022}, who derived them using the main isotopologue, CH$_3$OH though spectral fitting. A potential limitation of this method is that it may yield slightly underestimated methanol column density for optically thick transitions. In other studies, the calculated column densities of methanol in most of the literature were either corrected using isotopologues, including $^{13}$CH$_3$OH and CH$_3^{18}$OH or derived by fitting optically thin transitions, as noted in Table \ref{tab:acetone-obser}. Only the studies by \citet{rolffs2011} and \citet{feng2015} did not apply such corrections, which may lead to an underestimation of their reported methanol column densities. Fig. \ref{fig:acetone_metha} illustrates the correlation between acetone and methanol column densities in different source types. Linear least-squares fitting reveals a weak positive correlation between the two molecules ($r = 0.42$), with the best-fit relationship: log(N(CH$_3$COCH$_3$)) = 0.54Nlog(CH$_3$OH) + 6.43. This suggests that the column densities of methanol and acetone do not exhibit a strong positive correlation, implying that methanol may serve as an indirect precursor for acetone formation through radical-mediated secondary processes. Specifically, via photo-dissociation, methanol yields radical species that subsequently react to form molecular precursors of acetone.

\begin{figure}
    \includegraphics[scale=0.31]{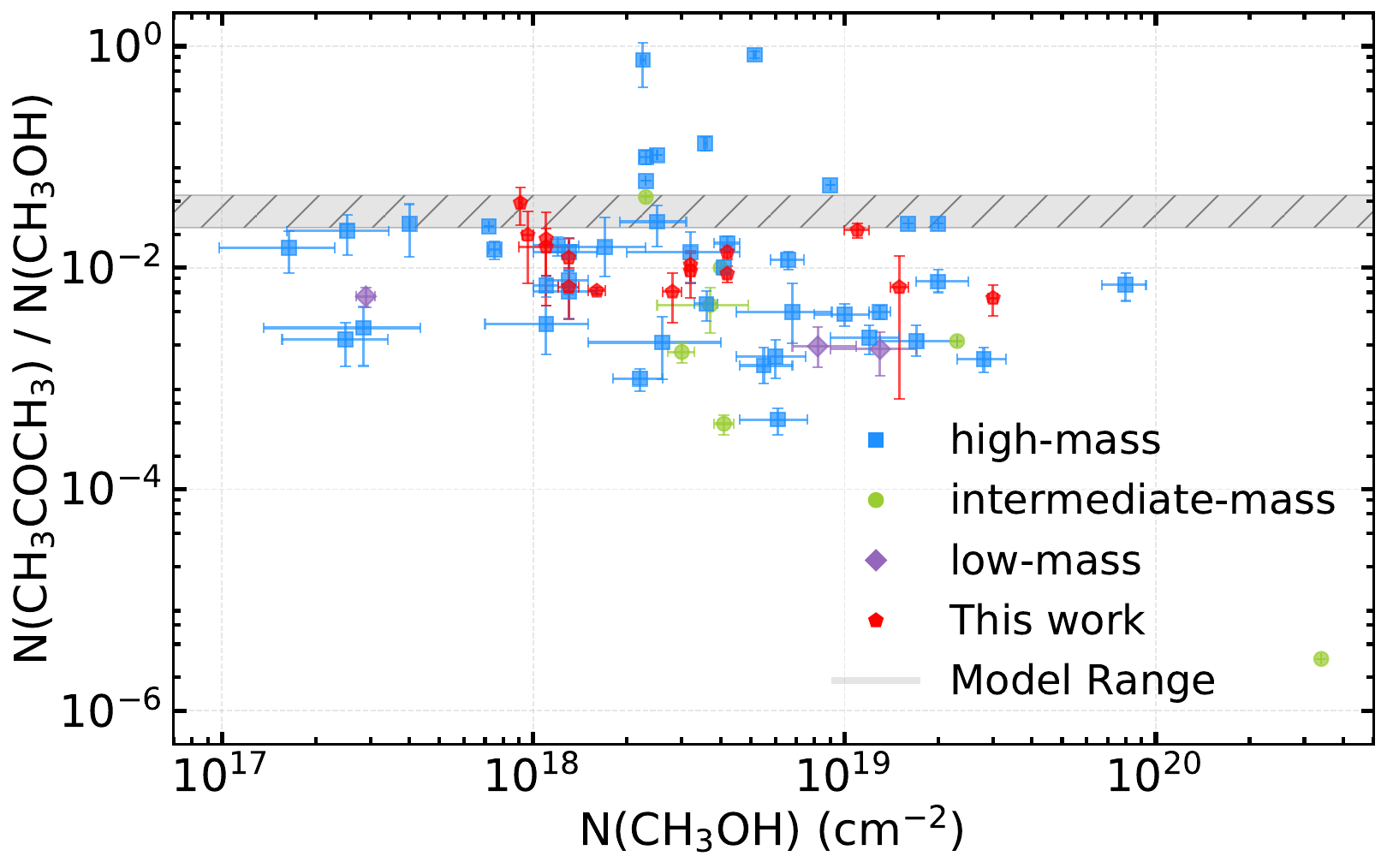}
    \caption{The relationship between the column density ratio of CH$_3$COCH$_3$/CH$_3$OH and the column density of CH$_3$OH, categorized by source types. Purple and green,  represent low-mass and intermediate-mass hot cores, respectively, while blue and red denote high-mass hot cores. The gray hatched region shows the range of gas-phase abundance ratios predicted by the MAGICKAL astrochemical model across three different timescales.}
    \label{fig:acetone_methanol}
\end{figure}

In order to better understand the relationship between the observational data and theoretical predictions, we compared our observations with model results. We analyzed the column density ratios of CH$_3$COCH$_3$/CH$_3$OH and the column densities of CH$_3$OH in our sample of 15 sources, along with literature data comprising 50 sources (see Table \ref{tab:acetone-obser}).When using literature data, it should be noted that the methanol column densities reported by \citet{rolffs2011} and \citet{feng2015} were not corrected for optical depth effects, whereas other studies performed such corrections, typically through isotopologue analysis or by fitting optically thin transitions. Since uncorrected methanol column densities may be underestimated due to optical depth, the acetone-to-methanol ratios in both our sample and the aforementioned literature sources could be systematically higher than their intrinsic values. As illustrated in Fig. \ref{fig:acetone_methanol}, we compared these observational results with predictions from the three-phase MAGICKAL model \citep{garrod2013, garrod2022}, in its updated version used by \citet{belloche2022} and \citet{chen2025a}. This model includes two evolutionary stages for hot cores: (1) a free-fall collapse stage followed by (2) a warm-up stage characterized by three timescales: 5 $\times$ 10$^4$ yr (fast), 2 $\times$ 10$^5$ yr (medium), and 1 $\times$ 10$^6$ yr (slow). The gray hatched region in Fig. \ref{fig:acetone_methanol} denotes the modeled abundance ratio range. Different colors in the figure represent various types of sources. We found that 12 observational points fall within the predicted range, 7 exceed the upper bound, and 48 lie below the lower bound. We observed no significant differences among these source types; however, this may be attributed to the limited sample size of low- to intermediate-mass hot cores in our study. In contrast, the larger sample of massive hot cores reveals that only a small fraction (<20\%) of observations align with model predictions, while the majority exhibit discrepancies. This finding suggests that source-specific physical parameters should be considered for individual objects.

Overall, our analysis reveals that the model overestimates the ratio of acetone to methanol when compared to observations. This discrepancy implies that current chemical networks may inadequately account for acetone destruction pathways or possibly overlook missing physical conditions in the model (e.g., source-specific environmental parameters)

\section{Conclusions}
\label{sec:conclusions}
We conducted a large-sample hot core survey focusing on acetone (CH$_3$COCH$_3$) toward 60 hot cores from the ATOMS project. Rotational temperatures and column densities were derived for 15 sources. We analyzed the correlations between acetone and acetaldehyde (CH$_3$CHO), as well as between acetone and methanol (CH$_3$OH), across different types of sources, and compared the observations with the model results. Our main conclusions are as follows:

1. Acetone has been detected in 15 hot cores, with rotational temperatures ranging from 89 to 176 K. Its column densities are in the range of (0.9-24) $\times$ 10$^{16}$ cm$^{-2}$. 

2. Compared the Spatial distributions of acetone at different upper energies and found that their peak emissions coincide with the continuum peaks, exhibiting no significant differences in spatial profiles.

3. The spatial distributions of acetone exhibit similarities to those of acetaldehyde. The line emission peaks of the two species coincide with the continuum peaks in most regions. The emissions of acetone are concentrated toward the hot core regions and generally exhibit a compact spatial distribution, whereas the emission of acetaldehyde shows a more extended spatial profile. 

4. A moderately positive correlation was found between the column densities  and rotational temperature of acetone, with a Pearson’s coefficient of 0.59 for high-mass hot cores. However, the relatively modest strength of the correlation (R$^2$ $\sim$ 0.35) suggests that rotational temperature alone explains only a fraction of the variance in acetone column density.

5. A positive correlation was found between the column densities of acetone and acetaldehyde, with a Pearson’s coefficient of 0.82. This indicates strong dependence of acetone column density on that of acetaldehyde, reflecting a significant increasing trend. This suggests that there is a chemical relationship between them.

6. Compared these observational results with the three-phase model results, the models overpredict the ratio of acetone to methanol relative to the observational data. This discrepancy suggests that current chemical networks may inadequately account for acetone destruction pathways or potential missing physical conditions in the models.


\section*{Acknowledgements}
This work was supported by the Xinjiang Talent Development Fund (XJRC-2025-KJ-PY-KJLJ-104, XJRC-2025-KJ-YJ-CXPT-067) , the National Natural Science Foundation of China under grants 12473025, the Natural Science Foundation of Xinjiang Uygur Autonomous Region under grants 2024D01E37, 2025D01D49, and the Chinese Academy of Sciences "Light of West China" Program (No. xbzg-zdsys-202410). The authors acknowledge Gleb Fedoseev's contribution to discussions about the desorption temperatures of acetone obtained through TPD experiments. This paper makes use of the following ALMA data: ADS/JAO.ALMA\#2019.1.00685.S. ALMA is a partnership of ESO (representing its member states), NSF (USA), and NINS (Japan), together with NRC (Canada), MOST and ASIAA (Taiwan), and KASI (Republic of Korea), in cooperation with the Republic of Chile. The Joint ALMA Observatory is operated by ESO, AUI/NRAO, and NAOJ.

\section*{Data Availability}

The raw data are available in ALMA archive.



\bibliographystyle{mnras}
\bibliography{mnras_acetone} 




\appendix

\section{Additional tables}
\label{sec:tables}
Table \ref{tab:Iden-trans-acetone} lists the identified transitions of CH$_3$COCH$_3$ focused on in this paper. Table \ref{tab:acetone-obser} gives source parameters from the literature used for comparison in Section \ref{sec:compare} and \ref{sec:chemistry}.

\onecolumn
\begin{longtable}{cccc}
\caption{Identified transitions of acetone and acetaldehyde}. 
\label{tab:Iden-trans-acetone} \\
\hline
Transition & Frequency(MHz) &E$_{up}$ (K) &Log$_{10}$(A$_{ij}$)(s$^{-1}$)\\
\hline
\endfirsthead

\multicolumn{4}{c}
{\tablename\ \thetable\ -- \textit{Continued from previous page.}} \\
\hline
 Transition & Frequency(MHz) &E$_{up}$ (K) &Log$_{10}$(A$_{ij}$)(s$^{-1}$)\\
\hline
\endhead

\hline \multicolumn{4}{r}{\textit{Continued on next page}} \\
\endfoot
\endlastfoot
\multicolumn{4}{c}{CH$_3$COCH$_3$, v=0}\\
\hline 
   17(6,11)-17(5,12)EE &98052.3987 &110.58732  &-4.76738\\
   17(7,11)-17(6,12)EE &98053.5346 &110.58738  &-4.76738\\
   
   5(4,1)-4(3,2)AA &98303.3686 &12.85935 &-4.80446\\
   
   5(5,1)-4(4,0)AA &98455.636 &13.96314  &-4.45585\\
   
   16(5,11)-16(4,12)EE &98600.7203 &95.56728  &-4.80797\\
   16(6,11)-16(5,12)EE &98600.9759 &95.56729  &-4.80797\\
   
   5(5,1)-4(4,1)EE &98651.5138 &14.02564 &-4.50157\\
   
   5(5,0)-4(4,0)EE &98800.8904 &14.09453  &-4.49742\\
   10(8,2)-9(9,0)EA &98800.3974 &47.67666  &-6.35147\\
   
   15(4,11)-15(3,12)EE &99052.5099 &81.41249 &-4.86474\\
   15(5,11)-15(4,12)EE &99052.5592 &81.41249 &-4.86474\\
   
   5(5,0)-4(4,1)AA &99266.4322 &13.97429  &-4.44945\\
   
   14(3,11)-14(2,12)EE &99422.0761 &68.12636 &-4.94731\\
   14(4,11)-14(3,12)EE &99422.084 &68.12636 &-4.94731\\
   
   13(2,11)-13(1,12)AE  &99542.6028   &55.79859    &-5.08093\\
   13(3,11)-13(2,12)AE  &99542.6037   &55.79859    &-5.08091\\
   13(2,11)-13(1,12)EA  &99542.709    &55.7986     &-5.08087\\
   13(3,11)-13(2,12)EA  &99542.71     &55.7986     &-5.08094\\
 
   14(3,11)-14(2,12)AA  &99587.764  &68.05   &-4.94463\\
   14(4,11)-14(3,12)AA  &99587.7721 &68.05   &-4.94468\\
 
   10(8,2)-9(9,1)AE &99718.7728 &47.61283 &-6.3280\\
   13(2,11)-13(1,12)EE &99721.5062 &55.71222 &-5.07815\\
   13(3,11)-13(2,12) EE & 99721.5071 &55.71222 &-5.07815\\
 
   13(2,11)-13(1,12)AA  & 99900.0915  &55.62612 &-5.07538\\
   13(3,11)-13(2,12)AA  & 99900.0925  &55.62612  &-5.07543\\
 
   8(2,6)-7(3,5)EE  &100350.3041  &24.56169  &-4.52252\\
   33(21,13)-33(20,14)EA  &100351.0592  &457.91295 &-4.50904\\ 
   40(24,16)-40(23,17)AA  &100351.9657  &666.04023 &-4.48374\\
 
   8(2,6)-7(3,5)AA  &100389.8619  &24.46618   &-4.52232\\
 \hline
 \multicolumn{4}{c}{CH$_3$CHO, v=0}\\
 \hline
   5(1,4)-4(1,3)E &98863.3135 &16.58889 &-4.50824\\
   5(1,4)-4(1,3)A-- &98900.951 &16.51416 &-4.50766\\
 \hline
\end{longtable}

\begin{landscape}
\begin{longtable}{ccccccc}
\caption{Column densities and rotational temperatures of detected CH$_3$COCH$_3$ derived from interferometric observations toward different astrophysical sources.} 
\label{tab:acetone-obser} \\
\hline
\multicolumn{1}{c}{Source} & \multicolumn{3}{c}{CH$_3$COCH$_3$}  &\multicolumn{1}{c}{CH$_3$OH} &\multicolumn{1}{c}{CH$_3$CHO} & Ref.\\
\cline{2-4}
&\multicolumn{1}{c}{T$_{rot}$ (K)} & \multicolumn{1}{c}{N(cm$^{-2}$)} &  \multicolumn{1}{c}{$f_{\text{CH$_3$OH}}$} & \multicolumn{1}{c}{{N(cm$^{-2}$)}} & \multicolumn{1}{c}{{N(cm$^{-2}$)}} &\\
\hline
\endfirsthead

\multicolumn{7}{c}
{\tablename\ \thetable\ -- \textit{Continued from previous page}} \\
\hline
\multicolumn{1}{c}{Source} & \multicolumn{3}{c}{CH$_3$COCH$_3$} &\multicolumn{1}{c}{CH$_3$OH} &\multicolumn{1}{c}{CH$_3$CHO} & Ref.\\
\cline{2-4}
\hline
\endhead

\hline 
\multicolumn{7}{r}{\textit{Continued on next page}} \\
\endfoot
\bottomrule
\multicolumn{7}{p{\linewidth}}{$^a$:The column density of CH$_3$OH was derived through spectral fitting of CH$_3$OH.

$^b$:The column density of CH$_3$OH was derived based on optically thin transitions.

$^c$: The column density of CH$_3$OH was determined by scaling the observed $^{13}$CH$_3$OH column density by the $^{12}$C/$^{13}$C ratio.

$^d$: The column density of CH$_3$OH was determined by scaling the observed CH$_3^{18}$OH column density by the $^{16}$O/$^{18}$O ratio.} 
\endlastfoot
\hline
Sgr B2 (N-LMH)   &170 (13)   &2.9(3) $\times 10^{16}$  &- &- &-    &\citet{snyder2002}\\
Orion-KL &176 (48)-194 (66) &2.0 (0.3)-8.0 (1.2) $\times 10^{16}$ &- &- &- &\citet{friedel2005}\\
G10.47+0.03 &200 &5$\times 10^{17}$ &5.56$\times 10^{-2}$ &9 $\times 10^{18a}$ &- &\citet{rolffs2011}\\
IRAS 22198(MM2) &120 &5.0$\times 10^{16}$ &2.17$\times 10^{-3}$  &2.3 $\times 10^{19b}$ &6.0 $\times 10^{16}$ &\citet{palau2011}\\
AFGL 5142(MM1) &210 &1.0$\times 10^{17}$ &4.35$\times 10^{-2}$  &2.3 $\times 10^{18b}$ &- &\citet{palau2011}\\
AFGL 5142(MM2) &140 &4.0$\times 10^{16}$ &1.00$\times 10^{-2}$  &4.0 $\times 10^{18b}$ &- &\citet{palau2011}\\
NGC 7129 FIRS 2 &200 &1$\times 10^{15}$ &2.94$\times 10^{-6}$  &3.4 $\times 10^{20c}$ &4.0 $\times 10^{15}$ &\citet{fuente2014}\\
Orion-KL(HC) &126(13) &5.43(0.82)$\times 10^{15}$ &2.15(0.84)$\times 10^{-2}$ &2.53(0.91) $\times 10^{17a}$ &- &\citet{feng2015}\\
Orion-KL(mm2) &112(7) &2.48(0.10)$\times 10^{15}$ &1.51(0.61)$\times 10^{-2}$ &1.64(0.66) $\times 10^{17a}$ &- &\citet{feng2015}\\
Orion-KL(mm3a) &101(5) &5.61(1.14)$\times 10^{14}$ &2.25(0.96)$\times 10^{-3}$ &2.49(0.93) $\times 10^{17a}$ &- &\citet{feng2015}\\
Orion-KL(mm3b) &102(5) &8.16(1.30)$\times 10^{14}$ &2.86(1.57)$\times 10^{-3}$ &2.85(1.49) $\times 10^{17a}$ &- &\citet{feng2015}\\
W51 North &140 &1.4 $\times 10^{17}$ &6.09$\times 10^{-2}$ &2.3 $\times 10^{18b}$ &- &\citet{rong2016}\\
IRAS 16293-2422A &125(25) &2.4(0.4)$\times 10^{16}$ &1.8(0.8) $\times 10^{-3}$ &1.3(0.4) $\times 10^{19b}$ &3.5(1.1) $\times 10^{15}$  &\citet{manigand2020}\\
IRAS 16293-2422B &130(20) &1.6(0.6,0.5)$\times 10^{16}$ &1.95(1.0,0.7) $\times 10^{-3}$ &8.2(2.7,1.4)$\times 10^{18c}$ &3.5(0.2,0.4) $\times 10^{16}$  &\citet{nazari2024}\\
IRAS 20126+4104(disk) &230(30) &2.6$\times 10^{17}$ &1.04$\times 10^{-1}$ &2.5 $\times 10^{18c}$ &- &\citet{palau2017}\\
IRAS 20126+4104(outflow) &190(20) &1.7$\times 10^{16}$ &2.36$\times 10^{-2}$ &7.2 $\times 10^{17c}$ &- &\citet{palau2017}\\
G328.2551-0.5321(shock-A) &180 &4$\times 10^{17}$ &2.50$\times 10^{-2}$ &1.6 $\times 10^{19b}$ &3.2 $\times 10^{16}$ &\citet{csengeri2019}\\
G328.2551-0.5321(shock-B) &180 &5$\times 10^{17}$ &2.50$\times 10^{-2}$ &2.0 $\times 10^{19b}$ &4.0 $\times 10^{16}$ &\citet{csengeri2019}\\
G328.2551-0.5321 (Inner Envelope) &110 &4.1$\times 10^{16}$ &1.00$\times 10^{-2}$ &4.1 $\times 10^{18b}$ &6.83 $\times 10^{15}$ &\citet{csengeri2019}\\
G9.62+0.19(MM4) &113(31) &1.1(0.2)$\times 10^{16}$ &1.47(0.27)$\times 10^{-2}$ &7.5(0.3) $\times 10^{17b}$ &5.0 $\times 10^{15}$ &\citet{peng2022}\\
G9.62+0.19(MM8) &100(38) &1.0(0.5)$\times 10^{16}$ &2.50(1.25)$\times 10^{-2}$ &4.0 $\times 10^{17b}$ &5.3 $\times 10^{15}$ &\citet{peng2022}\\
G25.82+0.17C1  &113(9.1) &4.74 $\times 10^{17}$ &1.33(0.19)$\times 10^{-1}$ &3.57(0.05) $\times 10^{18d}$ &- &\citet{baek2022}\\
G27.36+0.16C1  &213(31.5) &4.3(0.31) $\times 10^{18}$  &8.35(0.61)$\times 10^{-1}$ &5.15(0.05) $\times 10^{18d}$ &- &\citet{baek2022}\\  
G49.49-0.39C4  &238.8(27.2) &1.69(0.73)$\times 10^{18}$  &7.51(3.25)$\times 10^{-1}$ &2.25(0.04) $\times 10^{18d}$ &- &\citet{baek2022}\\
GAL 31.41 + 0.31 &170(21) &5.6(1.3)$\times10^{17}$  &7.00(2.0)$\times 10^{-3}$ &8.0(1.3) $\times 10^{19d}$ &3.4(1.6) $\times 10^{16}$ &\citet{mininni2023}\\
IRAS14498-5856C1 &152(8)  &3.4(1.0)$\times10^{15}$ &3.09(1.45)$\times 10^{-3}$ &1.1(0.4)$\times10^{18c}$ &2.2(0.3)$\times10^{15}$ &\citet{li2025a}\\
IRAS15520-5234C3 &100(4)  &1.6(0.3)$\times10^{15}$ &3.90(0.79)$\times 10^{-4}$ &4.1(0.3)$\times10^{18c}$ &6.5(1.1)$\times10^{14}$ &\citet{li2025a}\\
IRAS15596-5301C1 &113(9)  &2.2(0.3)$\times10^{15}$ &1.00(0.23)$\times 10^{-3}$ &2.2(0.4)$\times10^{18c}$ &3.0(0.9)$\times10^{14}$ &\citet{li2025a}\\ 
IRAS15596-5301C2 &95(8) &2.6(0.3)$\times10^{15}$ &4.26(1.16)$\times 10^{-4}$ &6.1(1.5)$\times10^{18c}$ &8.8(2.0)$\times10^{14}$ &\citet{li2025a}\\
IRAS16060-5146C1 &120(2)  &7.6(1.5)$\times10^{15}$ &6.91(1.50)$\times 10^{-3}$ &1.1(0.1)$\times10^{18c}$ &2.1(0.3)$\times10^{15}$ &\citet{li2025a}\\
IRAS16060-5146C2 &147(12) &1.8(0.4)$\times10^{16}$ &1.39(0.44)$\times 10^{-2}$ &1.3(0.3)$\times10^{18c}$ &6.0(0.9)$\times10^{15}$ &\citet{li2025a}\\
IRAS16060-5146C3 &150(7)  &7.0(0.8)$\times10^{16}$ &1.67(0.25)$\times 10^{-2}$ &4.2(0.4)$\times10^{18c}$ &1.5(0.3)$\times10^{16}$ &\citet{li2025a}\\
IRAS16071-5142C1 &138(2)  &5.2(0.6)$\times10^{16}$ &4.00(0.55)$\times 10^{-3}$ &1.3(0.1)$\times10^{19c}$ &2.0(0.4)$\times10^{16}$ &\citet{li2025a}\\
IRAS16076-5134C2 &163(8)  &1.0(0.2)$\times10^{16}$ &7.69(1.94)$\times 10^{-3}$ &1.3(0.2)$\times10^{18c}$ &1.8(0.4)$\times10^{15}$ &\citet{li2025a}\\
IRAS16272-4837C1 &119(6)  &1.7(0.5)$\times10^{16}$ &4.60(2.01)$\times 10^{-3}$ &3.7(1.2)$\times10^{19c}$  &8.9(1.8)$\times10^{15}$ &\citet{li2025a}\\
IRAS16272-4837C2 &92(4)   &5.2(0.9)$\times10^{15}$ &1.73(0.35)$\times 10^{-3}$ &3.0(0.3)$\times10^{18c}$  &2.8(0.6)$\times10^{15}$ &\citet{li2025a}\\
IRAS16272-4837C3 &145(6)  &2.3(0.3)$\times10^{17}$ &1.00(0.14)$\times 10^{-1}$ &2.3(0.1)$\times10^{18c}$  &4.2(0.7)$\times10^{16}$ &\citet{li2025a}\\
IRAS16351-4722C1 &153(7)   &1.6(0.3)$\times10^{15}$ &5.52(1.10)$\times 10^{-3}$ &2.9(0.2)$\times10^{17c}$  &3.0(0.6)$\times10^{14}$ &\citet{li2025a}\\
IRAS16351-4722C2 &132(12)  &7.8(1.1)$\times10^{16}$ &1.18(0.22)$\times 10^{-2}$ &6.6(0.8)$\times10^{18c}$ &1.4(0.3)$\times10^{16}$ &\citet{li2025a}\\
IRAS17220-3609C1 &125(3) &1.9(0.2)$\times10^{16}$ &1.58(0.31)$\times 10^{-2}$ &1.2(0.2)$\times10^{18c}$ &9.0(1.8)$\times10^{15}$ &\citet{li2025a}\\
IRAS17220-3609C2 &125(3)  &1.7(0.5)$\times10^{16}$ &4.72(1.44)$\times 10^{-3}$ &3.6(0.3)$\times10^{18c}$ &8.0(2.1)$\times10^{15}$ &\citet{li2025a}\\
\multirow{2}{*}{G19.01-0.03} &160(45,30) &2.6(2.0,1.0)$\times10^{16}$ &1.53(1.29,0.69)$\times 10^{-2}$ &1.7(0.6,0.4)$\times10^{18d}$ &- &\citet{chen2023,chen2025a}\\
 &200(25) &4.4(1.3,1.2)$\times10^{16}$ &1.38(0.73,0.64)$\times 10^{-2}$ &3.2(1.4,1.2)$\times10^{18d}$ &8.0(3.0)$\times10^{15}$ &\citet{chen2023,chen2025a}\\
\multirow{2}{*}{G19.88-0.53} &110(15) &7.2(2.8,1.9)$\times10^{15}$ &1.31(0.60,0.41)$\times 10^{-3}$ &5.5(1.3,0.9)$\times10^{18d}$ &4.0(1.5)$\times10^{15}$ &\citet{chen2023,chen2025a}\\
 &95(15)  &4.5(1.6,1.3)$\times10^{15}$ &1.31(0.60,0.41)$\times 10^{-2}$ &5.5(1.3,0.9)$\times10^{18d}$ &- &\citet{chen2023,chen2025a}\\
G22.04+0.22  &145(10) &1.5(0.2)$\times10^{17}$ &7.50(2.12,1.51)$\times 10^{-2}$ &2.0(0.5,0.3)$\times10^{19d}$ &4.5(1.0)$\times10^{16}$ &\citet{chen2023,chen2025a}\\
G23.21-0.37 &105(15) &4.2(0.8,0.7)$\times10^{16}$ &1.50(0.39,0.37)$\times 10^{-3}$ &2.8(0.5)$\times10^{19d}$ &4.0(0.6)$\times10^{16}$ &\citet{chen2023,chen2025a}\\
G34.30+0.20 &120(10) &2.8(0.4)$\times10^{16}$ &2.33(0.67)$\times 10^{-3}$ &1.2(0.3)$\times10^{19d}$ &1.3(0.3)$\times10^{16}$  &\citet{chen2023,chen2025a}\\
\multirow{2}{*}{G34.41+0.24} &138(8) &3.7(0.5)$\times10^{16}$ &2.18(0.83,0.59)$\times 10^{-3}$ &1.7(0.6,0.4)$\times10^{19d}$ &1.2(0.3)$\times10^{16}$ &\citet{chen2023,chen2025a}\\
 &93(15) &9.5(3.0,2.5)$\times10^{15}$ &1.58(6.38,5.75)$\times 10^{-3}$ &6.0(1.5)$\times10^{18d}$ &9.0(2.0)$\times10^{15}$ &\citet{chen2023,chen2025a}\\
G345.5+1.5 &160(30) &7.9(3.8,2.8)$\times10^{15}$ &6.08(3.24,2.57)$\times 10^{-2}$ &1.3(0.3)$\times10^{18d}$ &6.0(2.0)$\times10^{14}$ &\citet{chen2023,chen2025a}\\
G35.03+0.35 &110(20) &5.5(2.5,1.8)$\times10^{15}$ &2.12(1.49,1.13)$\times 10^{-3}$ &2.9(1.4,1.1)$\times10^{18d}$ &2.5(1.0)$\times10^{15}$ &\citet{chen2023,chen2025a}\\
G35.20-0.74N &118(8) &3.8(0.5,0.3)$\times10^{16}$ &3.8(0.91,0.82)$\times 10^{-3}$ &1.0(0.2)$\times10^{19d}$ &5.0(1.0)$\times10^{15}$ &\citet{chen2023,chen2025a}\\    
IRAS16547-4247 &180(20) &6.5(2.1)$\times10^{16}$ &2.6(1.04)$\times 10^{-2}$ &2.5(0.6)$\times10^{18d}$ &2.0(0.4)$\times10^{16}$ &\citet{chen2023,chen2025a}\\
\multirow{2}{*}{NGC6334-38} &110(20) &8.0(2.0)$\times10^{16}$ &- &- &- &\citet{chen2023,chen2025a}\\
 &170(50,30)  &2.7(2.0,0.9)$\times10^{16}$ &3.97(3.23,1.89)$\times 10^{-3}$ &6.8(2.3)$\times10^{18d}$ &1.8(0.5)$\times10^{15}$ &\citet{chen2023,chen2025a}\\
\end{longtable}
\end{landscape}

\section{Additional figures}
\label{sec:figures}

Figure \ref{fig:B1} shows the best-fitting results for sources detected by CH$_3$COCH$_3$. The method for calculating the local noise for each
window was adopted from \citet{loomis2021}.
\begin{figure}
\centering
\includegraphics[width=16cm]{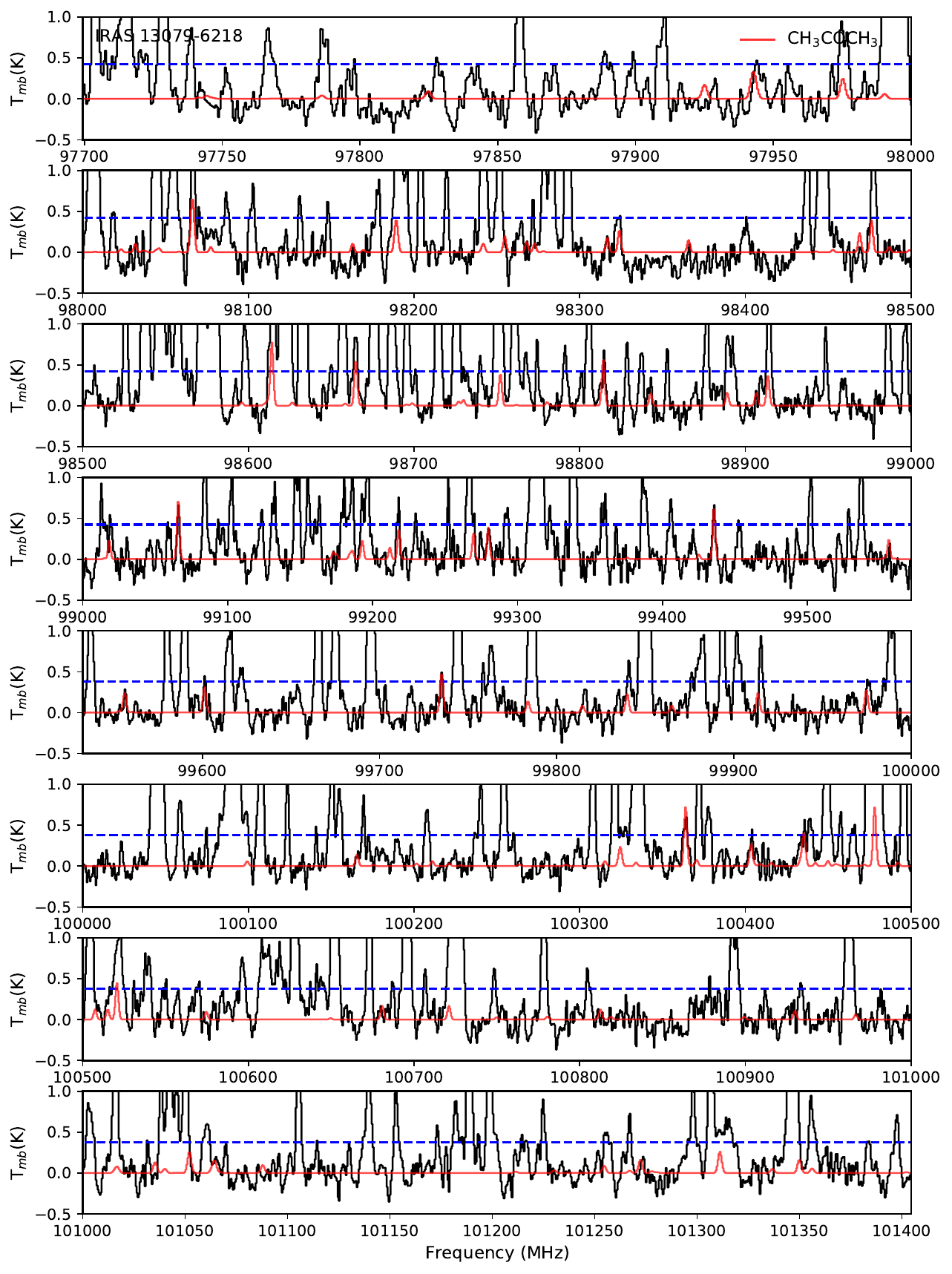}
\caption{The observed spectrum is shown in black, with the synthetic spectrum for acetone in red. The horizontal blue dashed line indicates the 3 $\sigma$ noise level in each window.}
\label{fig:B1}
\end{figure}

\begin{figure}
\centering
\ContinuedFloat
\includegraphics[width=16cm]{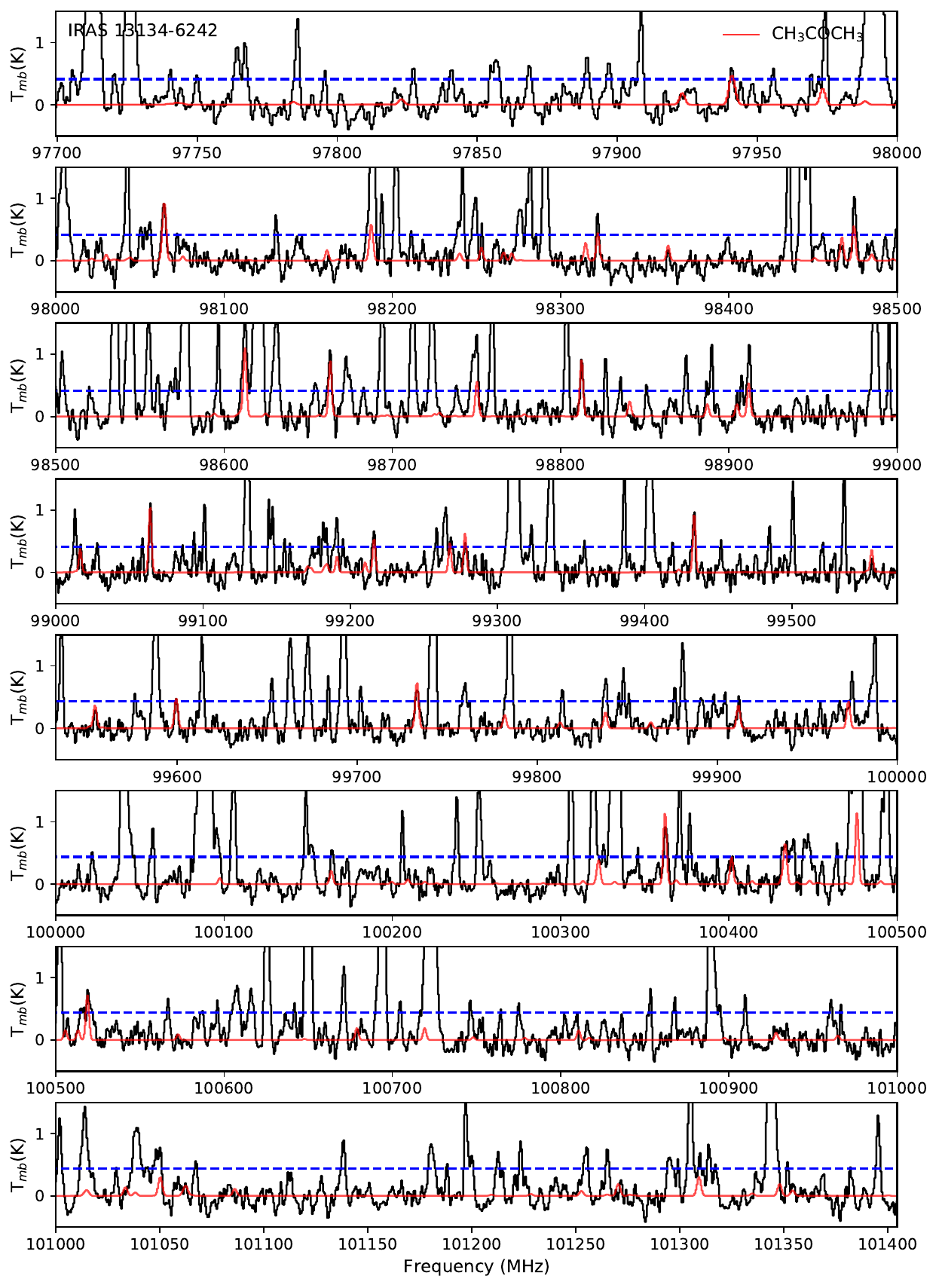}
\caption{Continued.}
\end{figure}
\begin{figure}
\centering
\ContinuedFloat
\includegraphics[width=16cm]{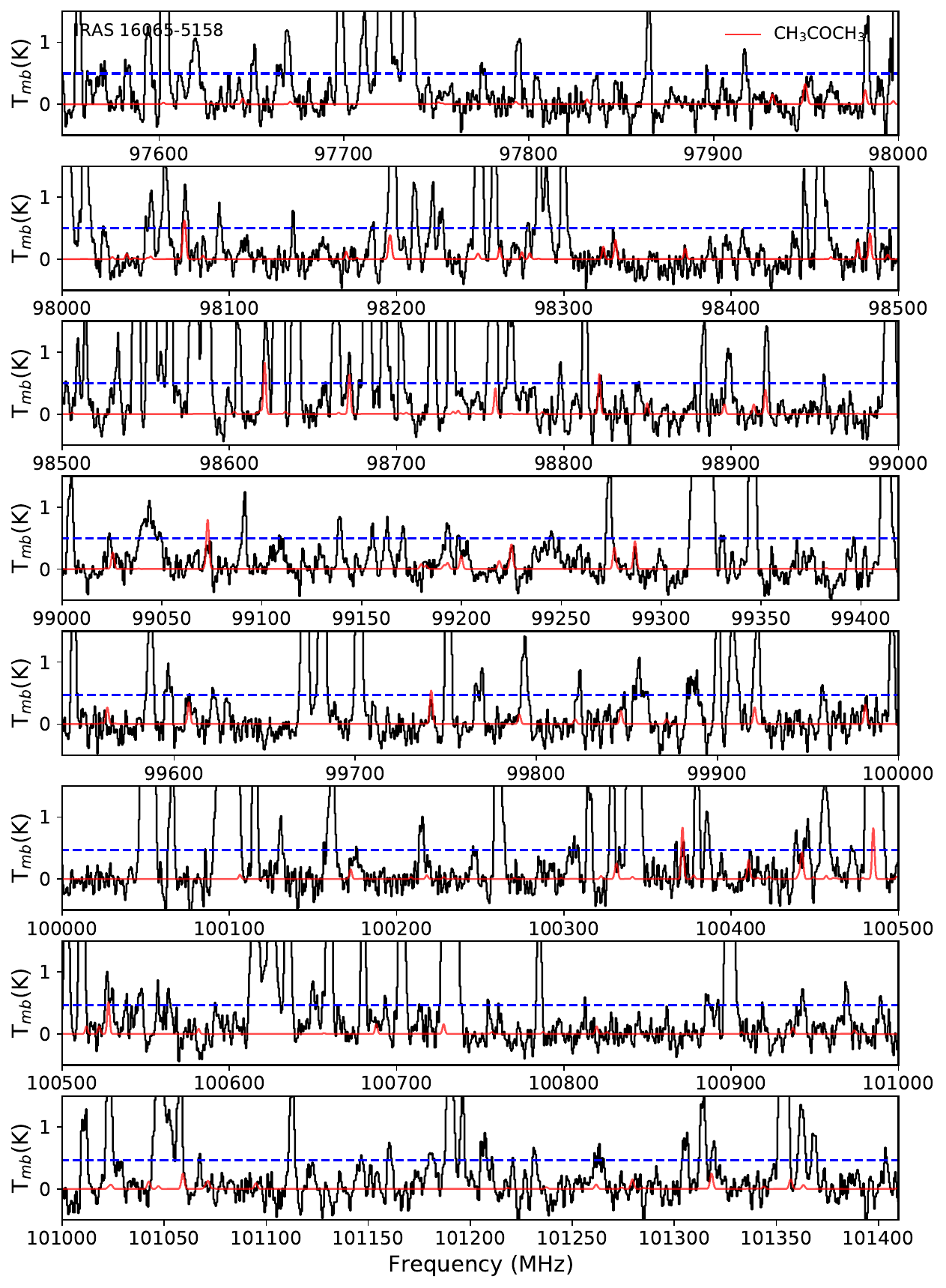}
\caption{Continued.}
\end{figure}

\begin{figure}
\centering
\ContinuedFloat
\includegraphics[width=16cm]{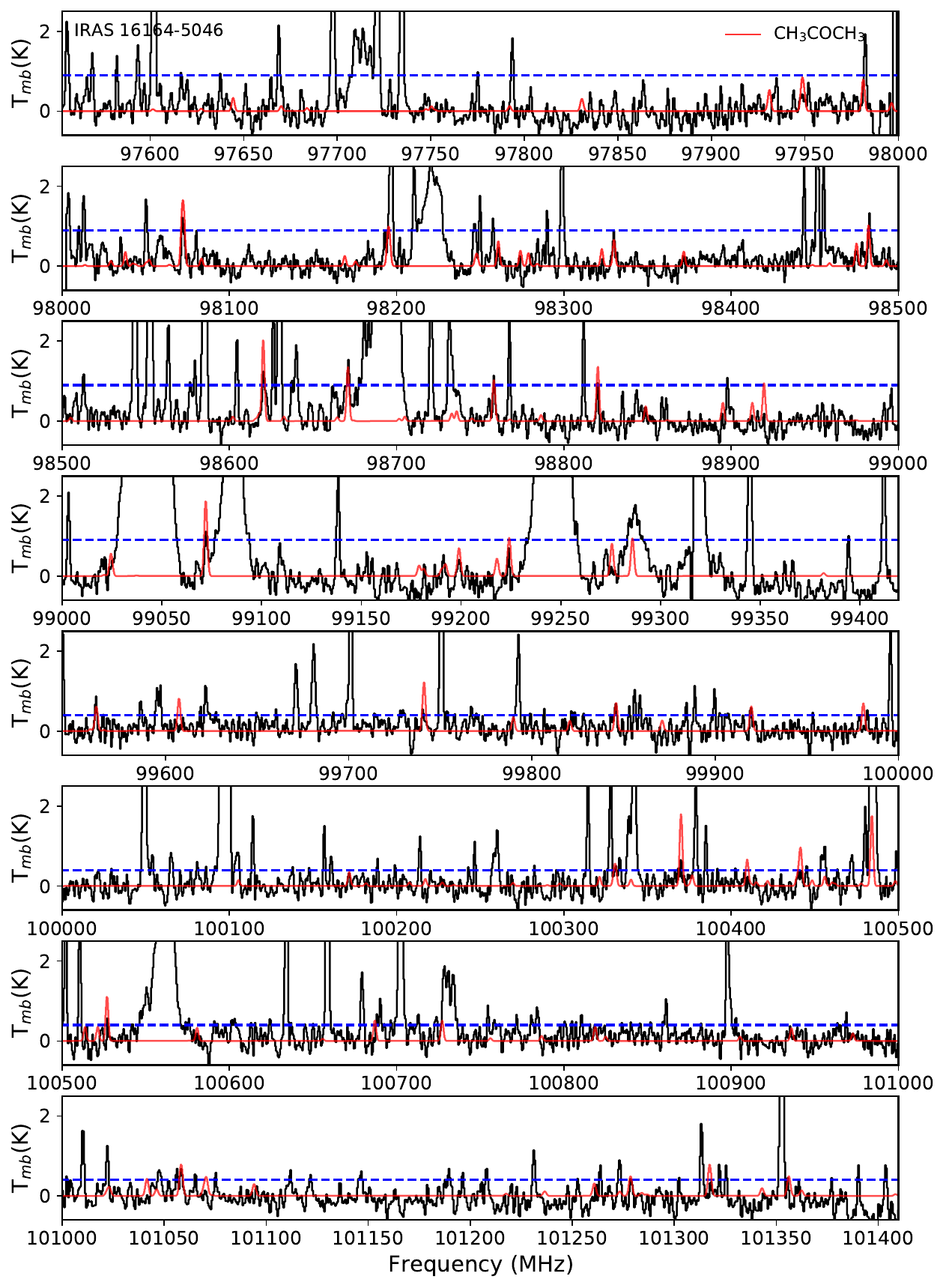}
\caption{Continued.}
\end{figure}
\begin{figure}
\centering
\ContinuedFloat
\includegraphics[width=16cm]{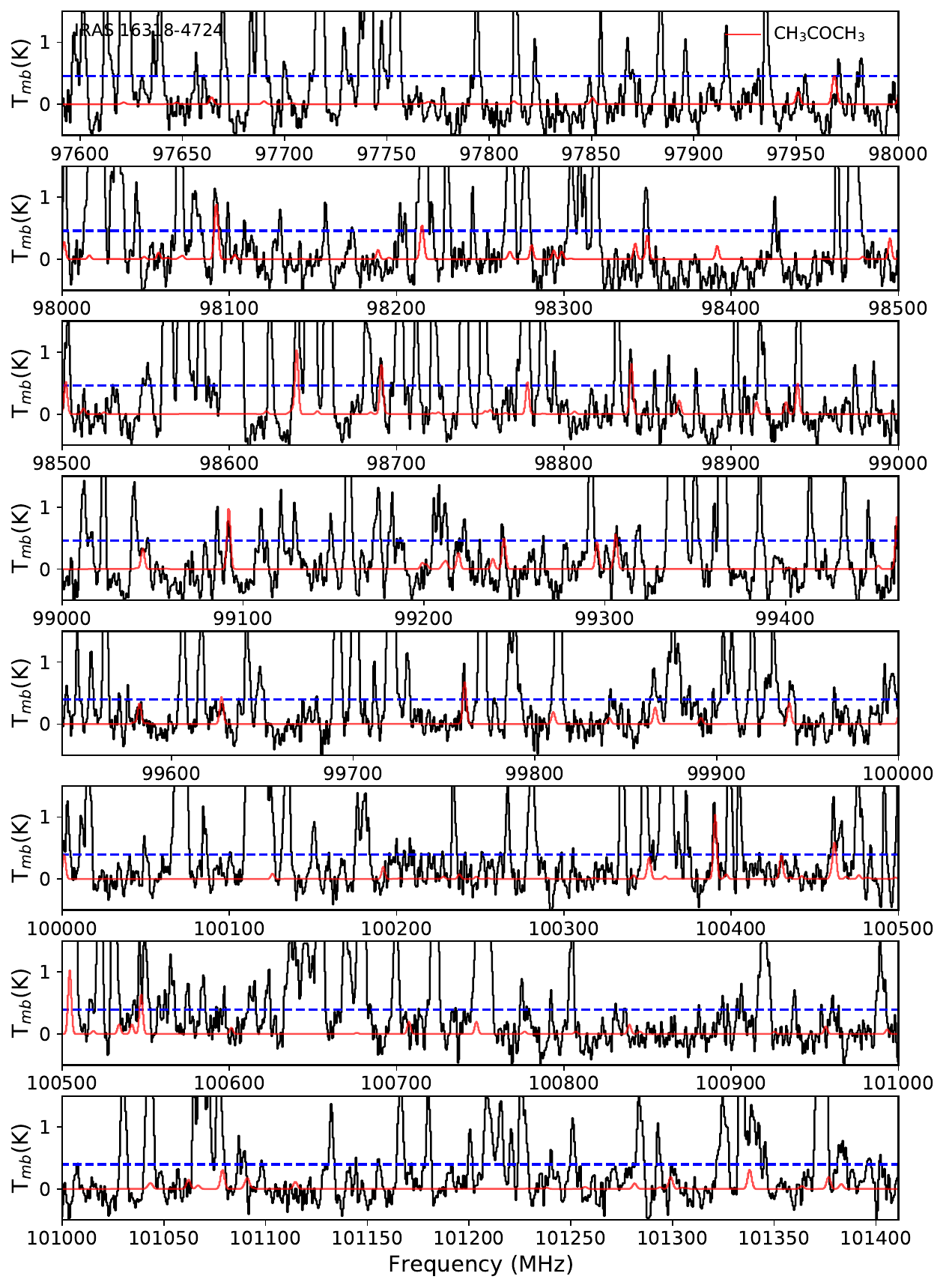}
\caption{Continued.}
\end{figure}

\begin{figure}
\centering
\ContinuedFloat
\includegraphics[width=16cm]{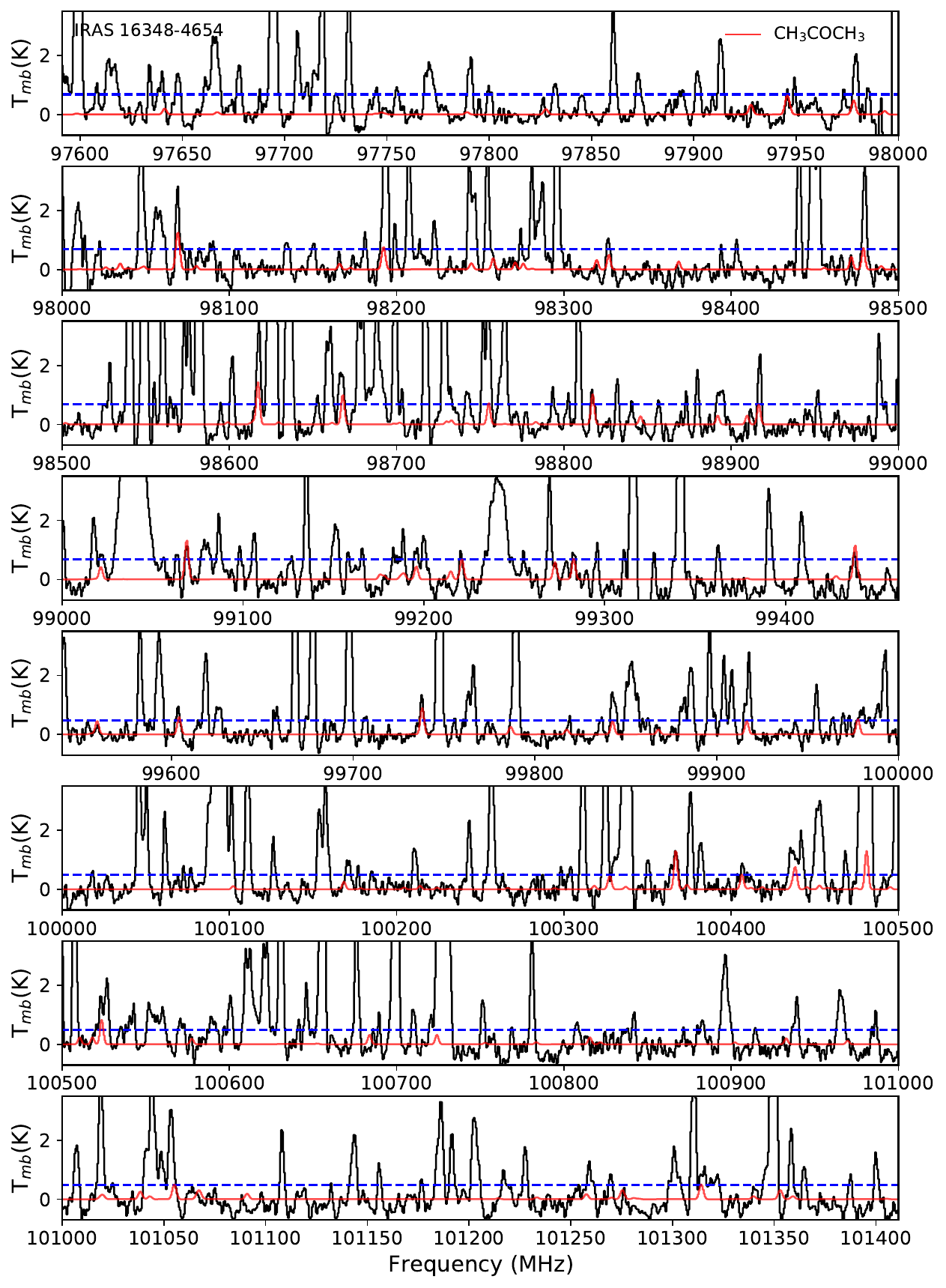}
\caption{Continued.}
\end{figure}
\begin{figure}
\centering
\ContinuedFloat
\includegraphics[width=16cm]{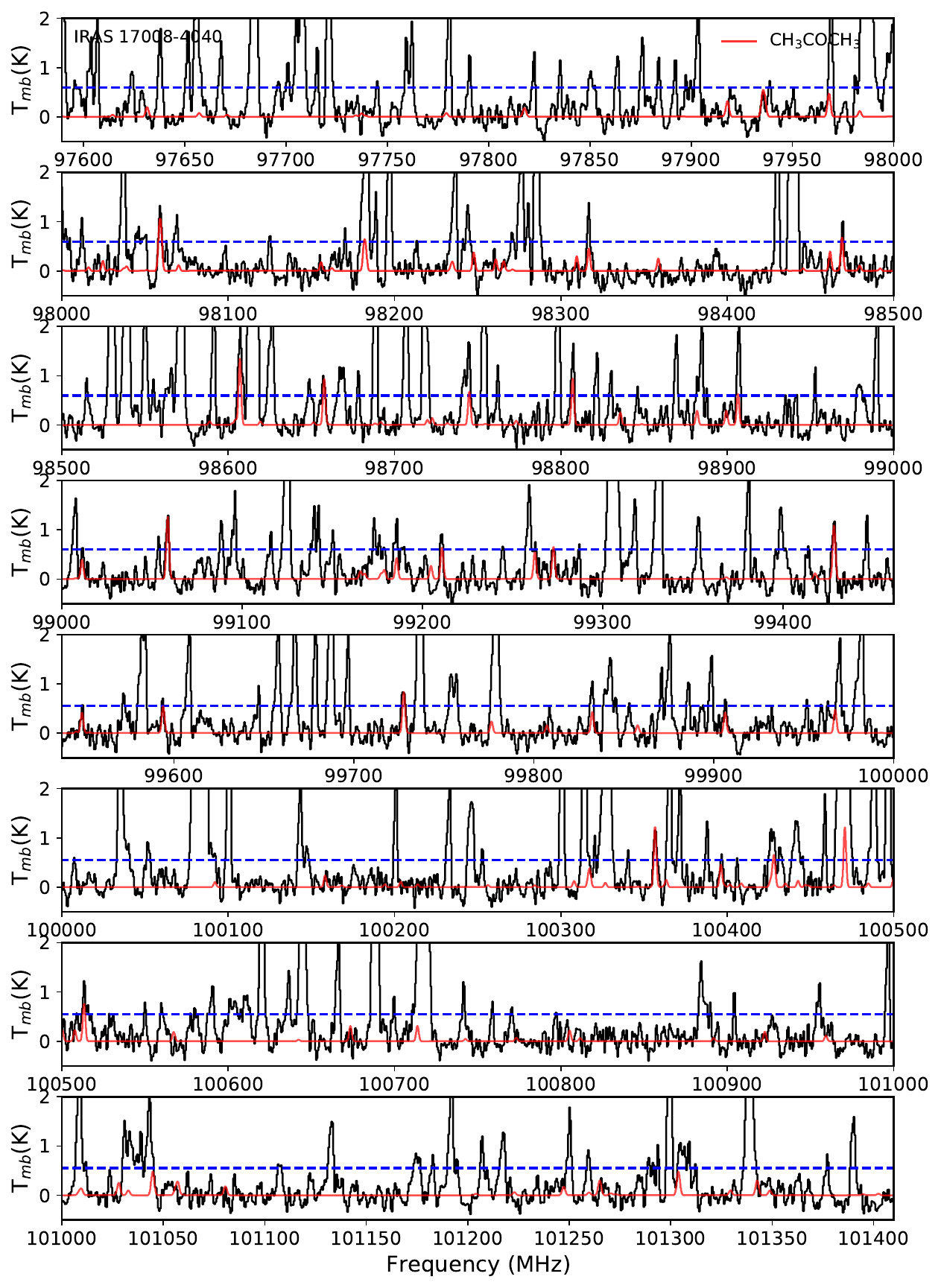}
\caption{Continued.}
\end{figure}

\begin{figure}
\centering
\ContinuedFloat
\includegraphics[width=16cm]{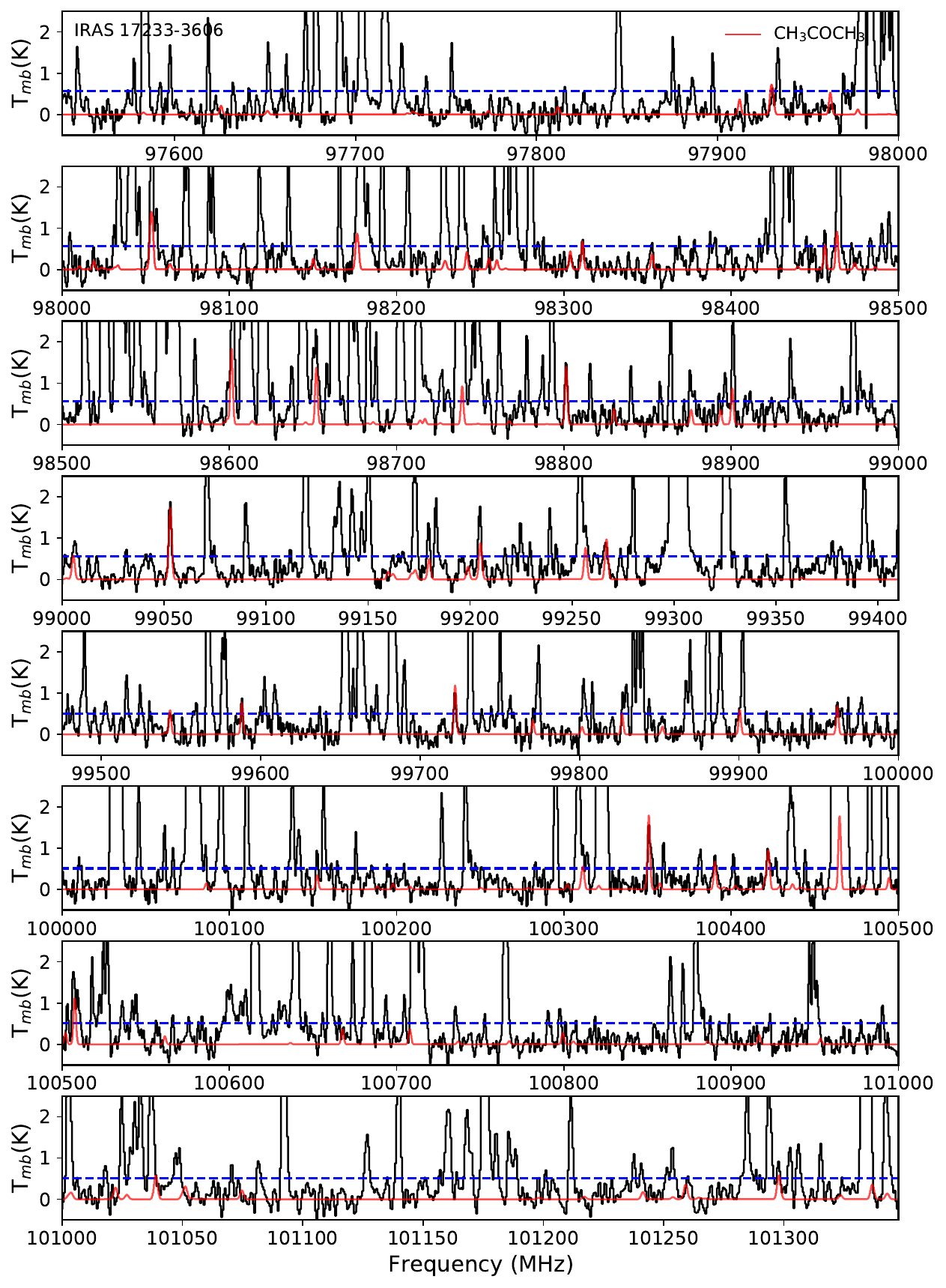}
\caption{Continued.}
\end{figure}
\begin{figure}
\centering
\ContinuedFloat
\includegraphics[width=16cm]{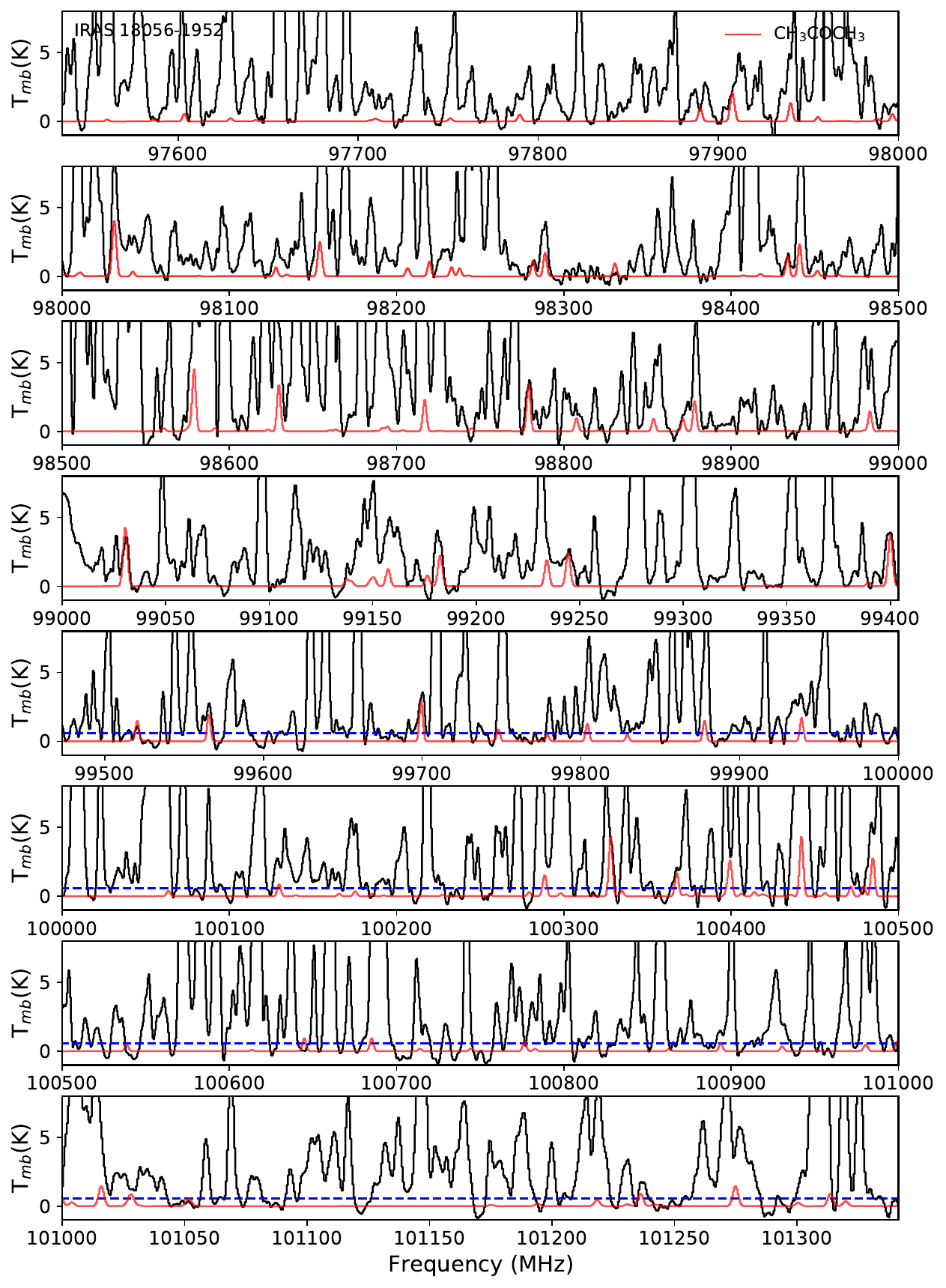}
\caption{Continued.}
\end{figure}
\begin{figure}
\centering
\ContinuedFloat
\includegraphics[width=16cm]{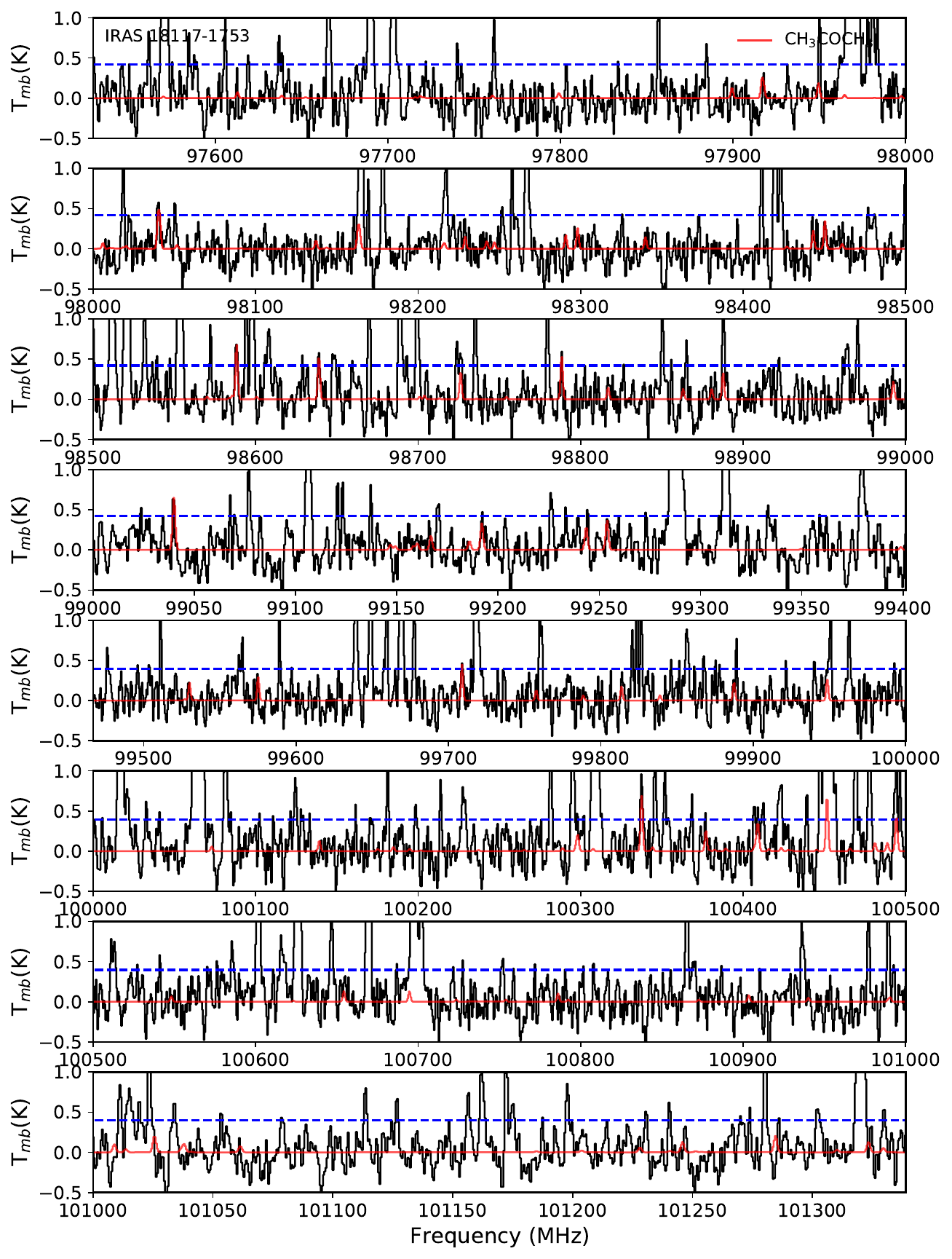}
\caption{Continued.}
\end{figure}

\begin{figure}
\centering
\ContinuedFloat
\includegraphics[width=16cm]{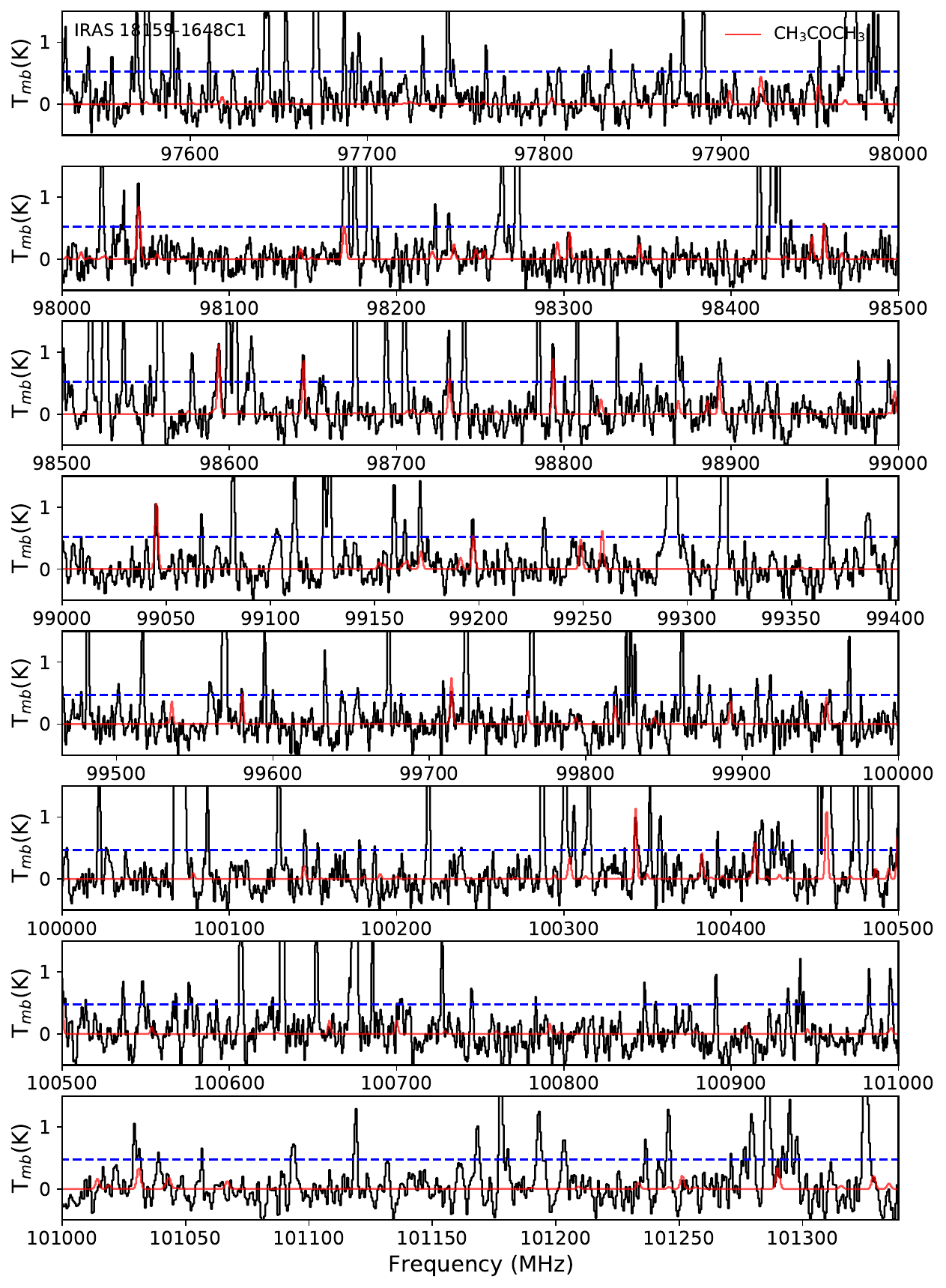}
\caption{Continued.}
\end{figure}
\begin{figure}
\centering
\ContinuedFloat
\includegraphics[width=16cm]{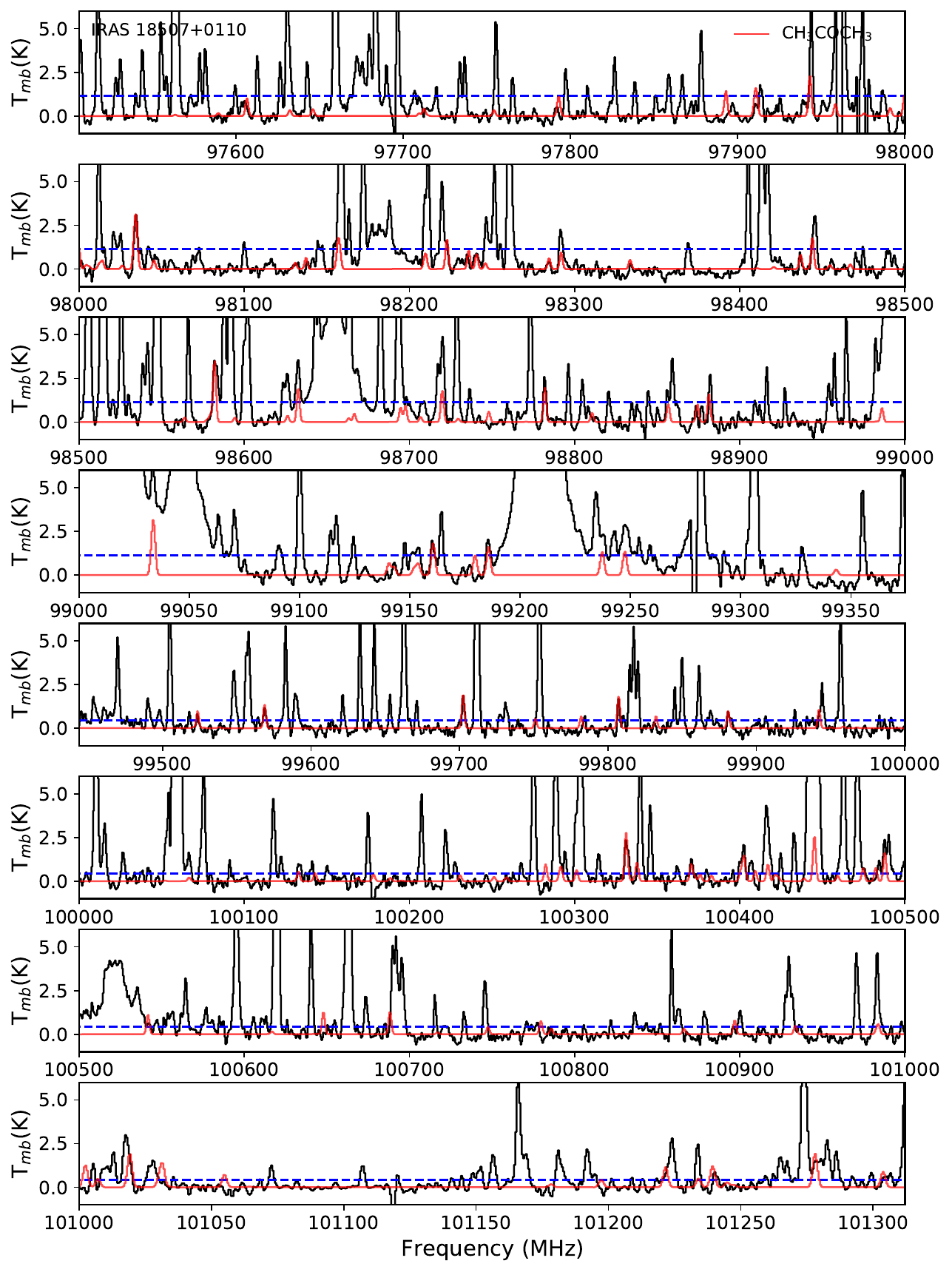}
\caption{Continued.}
\end{figure}

\begin{figure}
\centering
\ContinuedFloat
\includegraphics[width=16cm]{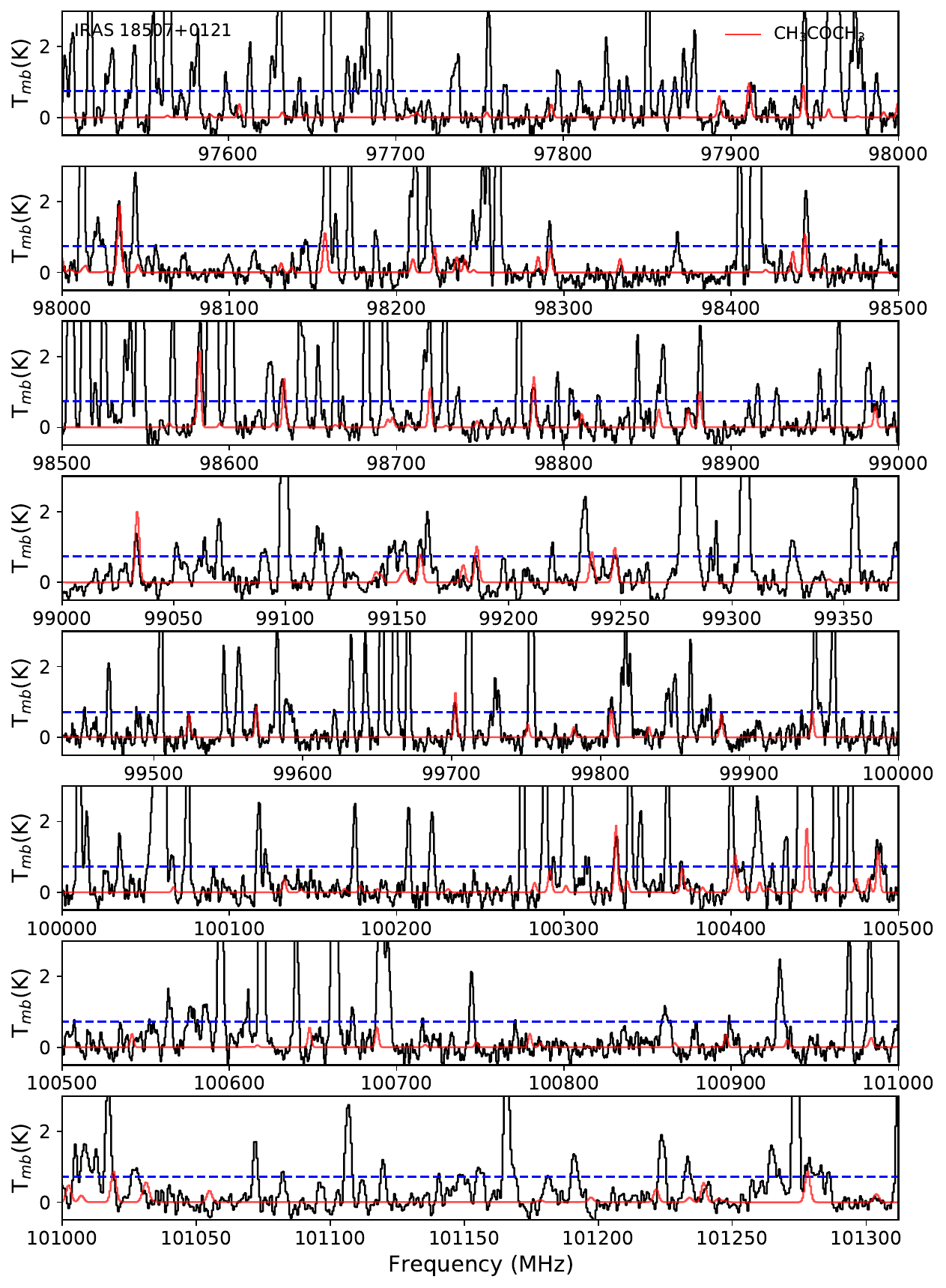}
\caption{Continued.}
\end{figure}
\begin{figure}
\centering
\ContinuedFloat
\includegraphics[width=16cm]{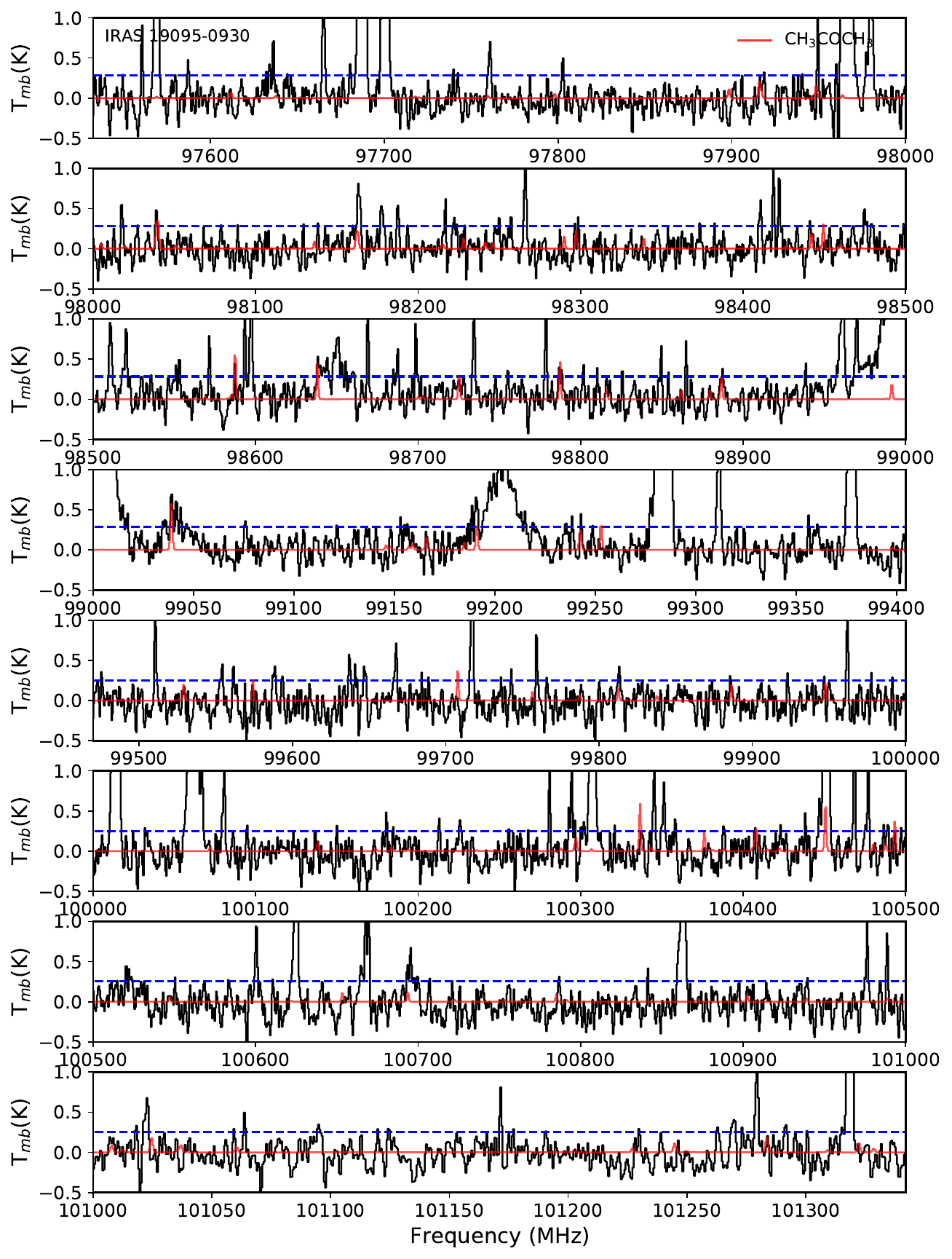}
\caption{Continued.}
\end{figure}


\bsp	
\label{lastpage}
\end{document}